\documentclass{aa}  
\usepackage{graphicx}
\usepackage{txfonts}

\usepackage{array}
\usepackage{ulem}
\newcolumntype{P}[1]{>{\centering\arraybackslash}p{#1}}
\usepackage{subfig}
\usepackage{booktabs}
\usepackage{xcolor}
\usepackage{multirow}
\usepackage{graphicx}	
\usepackage{amsmath}	
\usepackage{amssymb}	
\usepackage{multicol}   
\usepackage[export]{adjustbox}
\usepackage[flushleft]{threeparttable}

\def\mbh{$M_{\rm BH}$\/}

\def\lledd{$L/L_\mathrm{Edd}$}

\def\rfe{$R_{\rm FeII}$}

\def\feiiq{\rm Fe{\sc ii}$\lambda$4570\/}

\def\ltsima{$\; \buildrel < \over \sim \;$}
\def\ltsim{\lower.5ex\hbox{\ltsima}}  
\def\gtsima{$\; \buildrel > \over \sim \;$}

\def\gtsim{\lower.5ex\hbox{\gtsima}}

\def\civ{{\sc{Civ}}$\lambda$1549\/}

\def\cm3{cm$^{-3}$\/}
\def\hb{{\sc{H}}$\beta$\/}

\def\oiiiopt{{\sc{[Oiii]}}$\lambda\lambda$4959,5007\/}

\def\feii{{Fe\sc{ii}}\/}

\def\fe{{\sc{Fe}}\/}

\def\fe76087{{\sc [Fe vii]}$\lambda$6087\/}
\def\oiii{{\sc [Oiii]}$\lambda$5007}

\def\kms{km~s$^{-1}$}

\def\ergss{erg\, s$^{-1}$\/}

\def\rb{$r_{\rm BLR}$\/}

\usepackage{enumerate}
\def\arcsec{\hbox{$^{\prime\prime}$}}
\def\arcmin{\hbox{$^{\prime}$}}

\usepackage[colorlinks=true,linkcolor=blue,citecolor=blue]{hyperref}
\begin{document}

   \title{Investigating the origin of radio emission in candidate super-Eddington accreting black holes}


   \author{M. Gendron-Marsolais\inst{1,}\inst{2}
   \and P. Marziani\inst{3,}\inst{1}
   \and M. Berton\inst{4}
   \and E. Järvelä\inst{5}
   \and A. del Olmo\inst{1}
   \and M. Sargent\inst{6}
   \and M. D'Onofrio\inst{7}
   \and L. Crepaldi\inst{7,}\inst{8}
   \and A. Damas-Segovia\inst{9}
   \and B. Punsly\inst{10}
   \and L. Verdes-Montenegro\inst{1}
   }


   \institute{Instituto de Astrofísica de Andalucía, Glorieta de Astronom\'\i a, Granada, IAA-CSIC, Spain\\
    \email{mlgem@ulaval.ca}
    \and
    Département de physique, de génie physique et d’optique, Université Laval, Québec (QC), G1V 0A6, Canada
    \and
    Istituto Nazionale di Astrofisica, Osservatorio di Padova, Vicolo dell'Osservatorio 5, 35122, Padova, Italy
    \and
    European Southern Observatory (ESO), Alonso de Córdova 3107, Casilla 19, Santiago 19001, Chile
    \and
    Department of Physics and Astronomy, Texas Tech University, Box 41051, Lubbock 79409-1051, TX, USA
    \and
    Institute of Physics, Laboratory of Astrophysics, \'Ecole Polytechnique F\'ed\'erale de Lausanne (EPFL), Observatoire de Sauverny, Versoix CH-1290, Switzerland
    \and
    Dipartimento di Fisica e Astronomia ``G. Galilei", Universit\`a di Padova, Vicolo dell'Osservatorio 3, 35122, Padova, Italy
    \and
    Istituto Nazionale di Astrofisica, Osservatorio di Cagliari, Via della Scienza 5, 09047, Selargius, Italy
    \and
    Max-Planck-Institut für Radioastronomie, Auf dem Hügel 69, 53121 Bonn, Germany
    \and
    ICRA, Physics Department, University La Sapienza, Roma, Italy
        }

   \date{}

 
  \abstract
  {Recent works show that the radio power of quasars accreting at very high rates can reach surprisingly high values. These studies suggest that this radio emission might originate from star formation, but lack of data leaves open the possibility that they could also contain a jetted active galactic nucleus (AGN).}
  {We investigate the origin of the radio emission of  a sample of 18 super-Eddington candidates, over a wide range of redshifts. These sources are expected to have extreme radiative output per unit black hole mass, show high-velocity outflows and are therefore thought to be a prime mover of galactic evolution via radiative and mechanical feedback. }
  {We present new Karl G. Jansky Very Large Array (VLA) observations at L, C and X-band of these sources, which we combine with observations from the LOw-Frequency ARray (LOFAR) Two-metre Sky Survey (LoTSS) and the Very Large Array Sky Survey (VLASS). We also use optical and IR data to derive estimates of accretion and wind parameters, as well as star formation rates to compare with the ones derived from the radio emission.} 
  { Based on the radio variability, luminosity, morphology, radio spectral properties, radio vs IR estimates of star formation rate and radio-to-mid IR flux ratio, we find that 7 of our 18 targets are likely to have their radio emission predominantly coming from SF, and 6 from a combination of SF and AGN-related mechanisms, while only three sources indicate a core or jetted AGN only origin for the detected radio emission. This is consistent with previous studies, and supports the prevalence of lower power radio structures associated with  star-forming activity rather than relativistic jets in the high Eddington ratio regime. In the same sample, however, we find three sources for which the data suggest a concomitant presence of super-Eddington accretion and relativistic ejections. }
  {}


   \maketitle
%

\section{Introduction}

The main sequence (MS) of quasars is a classification scheme based on the spectral properties of quasars, most commonly represented by the plane between the intensity ratio \rfe = W(\feiiq)/W(\hb) (where W(\feiiq) is the equivalent width of the blend of multiplets between 4434~\AA\ and 4684~\AA\ and W(\hb) the equivalent width of the broad component of the \hb\ emission line) and the full width at half maximum (FWHM) of the \hb\ emission line.
It is a powerful analysis tool of the observational properties of type 1 active galactic nuclei (AGN, e.g., \citealt{sulentic_phenomenology_2000,marziani_main_2018}). The relevance of the MS for quasar astronomy is that several observational and physical properties, such as the ones related to the prominence of outflows and the prevalence of jetted sources, change systematically along the MS. There is a growing consensus that the main factors shaping the MS are the Eddington ratio (\lledd) and the viewing angle, defined as the angle between the line-of-sight and the symmetry axis of the AGN (e.g., \citealt{panda_quasar_2019}). 
Quasars can be divided into two broad populations based on their position in the MS : Population A (narrower \hb\ lines, with FWHM $< 4000 \text{ km s}^{-1}$) and B (broad \hb\ lines, with FWHM $> 4000 \text{ km s}^{-1}$). The sub-population of quasars located towards the high \rfe\ end of the MS (extreme Population A, or xA for brevity) are characterized by extreme observational parameters and are linked to high Eddington ratio \citep{marziani_highly_2014,du_fundamental_2016}. They show a distinctive UV spectrum that makes them easily recognizable even at high z \citep{martinez-aldama_extreme_2018}. The highest radiators per unit mass should be preferentially found at high z, as this indeed seems to be the case: the most luminous, intrinsically blue quasar population composite spectra closely resemble the highest radiator UV spectra at low- and moderate z \citep{nardini_most_2019}. They are believed to be in their early stage of evolution \citep{mathur_narrow-line_2000}, and they may constitute the dominant source of ionizing photons in the reionization epoch.

A recent analysis has shown that radio powers can reach surprisingly high values in xA objects \citep{ganci_radio_2019}, some entering the radio-loud (RL) regime, that is with a Kellermann's parameter (defined as the ratio $R_{k}$ between radio flux density at 5 GHz and the optical flux density at 5100 \AA, \citealt{kellermann_vla_1989}), above 10. 
The strong radio emission from RL sources is typically associated with the presence of a relativistic jet, whereas in radio-quiet (RQ) sources the jet is expected to be non-relativistic or intrinsically weaker than in RL. Star formation (SF) is considered an important agent  especially at lower power and lower frequencies: the fraction of star-forming galaxies is increasing at fluxes below $S_{1.4 \text{ GHz}} \simeq 10$ mJy (e.g., \citealt{condon_14_1989,smolcic_vla-cosmos_2017,mancuso_galaxy_2017}), although other mechanisms more closely related to the AGN have been proposed.

However, despite their high radio powers, the sample of strong FeII emitters from \citep{ganci_radio_2019} obeys to the FIR-radio correlation for star-forming galaxies and RQ quasars. Moreover, a large fraction ($75\%$) of radio-loud narrow-line Seyfert 1s (RL NLSy1s) shows mid-IR emission properties consistent with SF being the dominant component also in radio emission \citep{caccianiga_wise_2015,jarvela_unravelling_2022}. Therefore, it is legitimate to suggest that the radio emission from most sample sources in \cite{ganci_radio_2019} is dominated by processes not related to the AGN. Nonetheless, the lack of data leaves open the possibility that they could also harbor a relatively low radio-power relativistic jet or outflow. This would be the case if they were sources with relatively low black hole mass, as jet power scales nonlinearly with black hole mass \citep{heinz_non-linear_2003}. Indeed, in recent years a number of objects characterized by prominent FeII emission and formally RQ have been classified as jetted \citep[e.g.,][]{lahteenmaki_radio_2018,berton_radio-emitting_2018,berton_absorbed_2020,jarvela_narrow-line_2021,marziani_optical_2021,vietri_spectacular_2022}.

Mrk 231, a low-$z$, high-$L$ prototypical xA source, is one of the RL sources from the \citealt{ganci_radio_2019} sample. Mrk 231 has been described by \cite{sulentic_low_2006} as a high-L, highly accreting quasar misplaced at late cosmic epochs. Emission line properties are extreme, in terms of FeII emission, CIV$\lambda$1549 blueshift, and blueward asymmetry of \hb.  It is a broad absorption line (BAL) QSOs suffering some internal extinction and with extreme absorption troughs in  radial velocity \citep{sulentic_low_2006,veilleux_complete_2016}. Mrk 231 is known to possess an unresolved core, highly variable and of high brightness temperature \citep{condon_compact_1991}. Superluminal radio components of Mrk 231 have been detected, and the relation between BALs, radio ejections, and continuum change \citep{reynolds_relativistic_2017}, illustrating the complex interplay of thermal/nonthermal nuclear emission that can be perhaps typical of sources accreting at extremely high rates. Still, according to the location in the diagram of \citealt{ganci_radio_2019}, the dominant emission mechanism is SF. This inference is consistent with the enormous CO luminosity of the AGN host \citep{rigopoulou_molecular_1996}.

Mrk 231 notwithstanding, the wide majority of extreme radiators remains little studied, and this is especially true as far as their radio properties are concerned. Strong FeII emitters are known to have high accretion rate, possibly at a super-Eddington level \citep{du_fundamental_2016}, although their radiative output is expected to remain within a factor $\sim 2$ above the Eddington limit \citep{mineshige_slim-disk_2000}. High \lledd\ can coexist with powerful jetted radio emission (log P$_\nu > 10^{31}$ erg s$^{-1}$ Hz$^{-1}$). There is no physical impossibility in this respect \citep{czerny_astronomical_2018}, and recent works consider the relativistic jet and non-relativistic wide angle outflows as two aspects of a hydromagnetically-driven wind \citep{reynolds_relativistic_2017}. Concomitant high accretion and high radio power are observed in compact steep-spectrum (CSS) radio sources (e.g., \citealt{wu_black_2009}). 
General relativistic magnetohydrodynamic simulations also can reproduce high \lledd\ sources with jets \citep{mckinney_double_2017,liska_formation_2022}.
Our view is likely biased by the relative rarity of this sort of sources at low-z \citep{odea_compact_1998}, and especially by the scarcity of high-quality UV/optical spectra, and their rather sketchy interpretation that, in the past, ignored the MS trends. The basic issue is whether the RL sources in the xA spectral bins of the MS are truly jetted (as RL NLSy1s are, \citealt{foschini_properties_2015}) or dominated by radio emission from SF.

Generally speaking, RL extreme accretors may appear indistinguishable in the \hb\ spectral range from their RQ counterparts, as both RQ and RL strong-FeII show a large blueshift in their [O III] profiles \citep{berton_o_2016,berton_jet-induced_2021}. The radio power may be associated with a relatively small black hole mass. They could be systems in their early stage of evolution for which a spin-up has been made possible by a recent accretion event \citep{mathur_narrow-line_2000,sulentic_phenomenology_2000}, but the possibility of radio emission due to SF calls into question the radio-mode AGN feedback as a cosmologically relevant process in massive galaxy formation. This class of sources is believed to be representative of highly accreting quasars observed at very early cosmic epoch, as indeed proved by the NLS1-like object recently discovered at z = 6.56 \citep{wolf_x-ray_2023}. 
Sources in the re-ionization epoch may be both accreting at super-Eddington rates, and their associated starburst be the primary source of ionizing photons (e.g., \citealt{decarli_rapidly_2017}). 
Several highest-z quasars have been detected in the radio, implying extremely powerful radio emission (e.g., \citealt{belladitta_extremely_2019}). Which is however the main culprit, SF or black hole activity? Establishing their relative contribution in high accretors is expected to provide strong constraints on the ability of nuclear activity to provide the main source of ionizing photons at re-ionization epoch. 

In this work we analyze the observations carried out with the Karl G. Jansky Very Large Array (VLA) of a sample of 18 xA sources with intensity ratio \rfe $  \gtrsim 1$ taken from \cite{marziani_optical_2003,sani_enhanced_2010} and \cite{ganci_radio_2019}.  
We add observations from radio surveys as well as optical and IR data to complete our analysis and determine the origin of the radio emission.
In Section~\ref{sample_data}, we present the sample and the data reduction process, while the analysis of these datasets is presented in section~\ref{results}.  
We discuss our findings in section~\ref{discussion} and provide a summary of our results in section~\ref{conclusion}. Appendix~\ref{optical_spectro} includes an analysis of optical and IR data relevant for the interpretation of the radio observations.
We adopted $H_{0} = 70 \text{ km s}^{-1} \text{Mpc}^{-1}$, $\Omega_{\rm M} = 0.3$ and $\Omega_{\rm vac} = 0.7$.

\section{Sample selection and data reduction}\label{sample_data}

Our complete sample is shown in Table~\ref{table_targets} and contains a total of 18 sources that according to their optical spectra are classified as xA objects in the MS, that is with \rfe $\gtrsim 1$. Throughout the text, we will refer to the sources by their short names as defined in column 2 of Table~\ref{table_targets}. We stress again that this condition is thought to be associated with black holes of highest radiative output per unit black hole mass \citep[e.g.,][]{marziani_searching_2001,sun_dissecting_2015,du_fundamental_2016,panda_quasar_2019,marziani_super-eddington_2025}. Among them, several show multi-frequency evidence of extraordinary activity level which may be ascribed to super-Eddington accretion \citep[][and references therein]{marziani_super-eddington_2025}.  
Five of these sources were taken from radio survey \cite{ganci_radio_2019} and the rest are supplemented by low z sources from \citet{sani_enhanced_2010} and \citet{marziani_optical_2003} detected in FIR.

\begin{figure}
    \centering
    \includegraphics[width=1\linewidth]{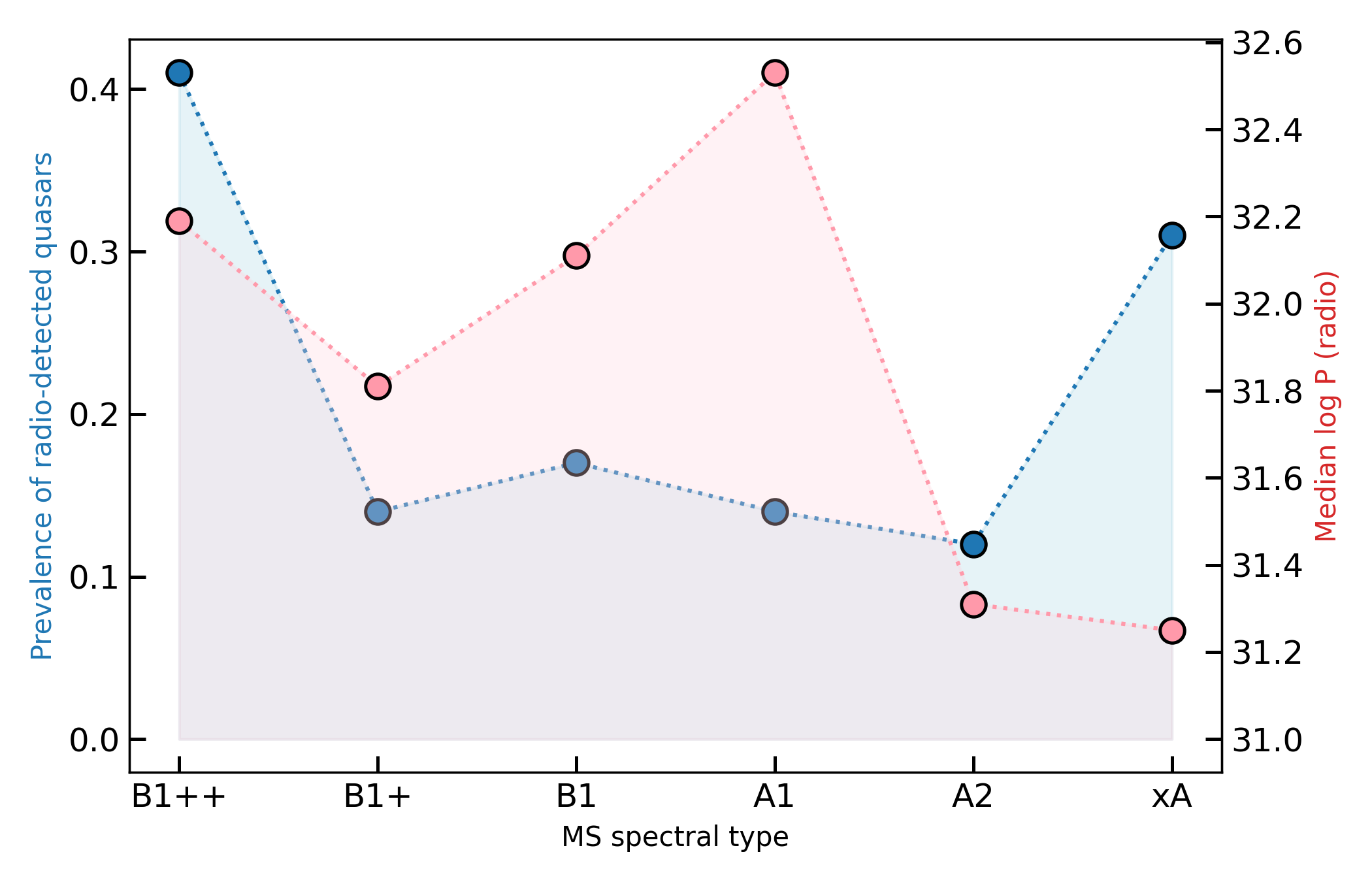}
    \caption{Prevalence of FIRST-detected type-1 AGN \citep{marziani_is_2013} (blue circles and cyan shading) and median radio power (red circles and rose shading) as a function of spectral type along the quasar main sequence.}
    \label{fig:rd}
\end{figure}

In optically selected quasar samples, RL sources tend to concentrate at the two ends of the MS. Figure \ref{fig:rd} illustrates the incidence and the radio power as a function of the spectral types along the quasar MS, for FIRST  detected sources in the sample of \citep{marziani_is_2013}. The B spectral types are where most powerful radio sources are found; if they are observed pole-on, the appear as core-dominated with narrower lines that let them populate part of spectral type A1 \citep{wills_relativistic_1986,jarvis_orientation_2006,ganci_radio_2019}. The surprising result is that in correspondence of the xA spectral type there is a threefold increase in the prevalence of radio detected sources. In this case, however, the radio power is systematically much lower.

The properties of radio emission found in one domain — such as the xA regime defined by \rfe$\gtrsim$1 (spectral types A3 and A4; see \citealt{sulentic_average_2002})  — can be  generalized to other sources with similar spectral characteristics. However, these results do not extend to quasars of markedly different spectral types along the MS, particularly those in the extreme end of Population B. In simple terms, Population B sources may exhibit comparable radio power to xA quasars, yet the origin of that radio emission could be fundamentally different-- most likely jet-driven \citep{zamfir_new_2008,ganci_radio_2019}.

An additional consideration in interpreting these data concerns the spatial origin of the far-infrared emission. Relativistic jets are nuclear phenomena, and the presence of radio lobes corroborates jet activity, whereas star formation can be nuclear, circum-nuclear, or distributed across the host. Because the available measurements do not spatially resolve the FIR emission attributed to star formation, a contribution from a mildly beamed (i.e., weakly Doppler-boosted) or misaligned relativistic jet cannot be ruled out.

\begin{table*}
\centering\tabcolsep=3pt
\caption{Targets of the sample}
\label{table_targets}
\begin{tabular}{lccccll}
\hline
 NED Alias      & Short name  &   z   &  RA (2000)   &  Dec (2000) & Opt. & FIR \\
\hline
Mrk 957 		& J00418+4021 &	0.071	&	00h41m53.420	&	+40d21m17.29s	&	HO87  &	IRAS, ISO, AKARI	\\
PG0043+039 		& J00457+0410 &	0.385	&		  00h45m47.225s 	&	 	 +04d10m23.38s	&	M03	&	IRAS	\\
IZw1 		   	& J00535+1241 &	0.059	&	00h53m34.920s  	 	&	+12d41m35.87s	&	M22	&	IRAS, ISO, MIPS,	\\
& & & & & &  AKARI	\\
Mrk 142 		& J10255+5140 &	0.045	&	10h25m31.279s  	 	&	+51d40m34.85s	&	M03	&	IRAS, MIPS	\\
Mrk 1298 	   	& J11292-0424 &	0.062	&	11h29m16.729s  	 	&	-04d24m07.26s	&	M03	&	IRAS, MIPS, AKARI 	\\
WISEA J114201.86+603029.7   &	J11420+6030  & 0.718	&		  11h42m01.847s 	&	 	 +60d30m30.23s	&	SDSS	&	\ldots	\\
SDSS J120734.63+150643.6	&	J12075+1506  & 0.750	&		  12h07m34.631s 	&	 	 +15d06m43.70s	&	SDSS	&	\ldots	\\
SDSS J120910.62+561109.6    &	J12091+5611  & 0.453	&		  12h09m10.620s 	&	 	 +56d11m09.30s	&	SDSS	&	\ldots	\\
SDSS J123640.33+563021.6	&	J12366+5630  & 0.698	&		  12h36m40.346s 	&	 	 +56d30m21.43s	&	SDSS	&	\ldots	\\
Mrk 231 	   	& J12562+5652	  & 0.042	&	12h56m14.222s  	 	&	+56d52m25.10s	&	M03	&	IRAS, ISO, MIPS,	\\
& & & & & &  AKARI, PACS\\
& & & & & & \\
SBS 1259+593	& J13012+5902	&	0.476	&		  13h01m12.933s 	&	 	 +59d02m06.75s	&	M22	&	IRAS, ISO, MIPS	\\
FBQS J1405+2555	& J14052+2555	&	0.164	&		  14h05m16.218s 	&	 	 +25d55m34.12s	&	M22	&	IRAS, ISO, MIPS	\\
PG 1404+226     & J14063+2223	&	0.098	&		  14h06m21.890s 	&	 	 +22d23m46.51s	&	M22	&	IRAS, MIPS	\\
PG 1415+451	    & J14170+4456	&	0.114	&		  14h17m00.825s 	&	 	 +44d56m06.33s	&	M22	&	IRAS, ISO, MIPS	\\
SDSS J142549.19+394655.0    & J14258+3946 &		0.505	&		  14h25m49.190s 	&	 	 +39d46m55.00s	&	SDSS	&	\ldots	\\
SDSS J144733.05+345506.7	& J14475+3455 &		0.662	&		  14h47m33.052s 	&	 	 +34d55m06.82s	&	SDSS	&	IRAS	\\
PG 1543+489	& J15455+4846 &		0.401	&		  15h45m30.237s 	&	 	 +48d46m09.00s	&	M22	&	IRAS, ISO,	\\
& & & & & & MIPS	\\
PG 1700+518	& J17014+5149 &		0.292	&		  17h01m24.898s 	&	 	 +51d49m20.38s	&	M03	&	IRAS, ISO,	\\
& & & & & & MIPS	\\
\hline
\end{tabular}
\tablefoot{Column 1: NED Alias. 
    Column 2: Short name. 
    Column 3: redshift.
    Column 4 and 5 : coordinates.
    Column 6: Source of optical spectra: HO87: \citealt{halpern_narrow-line_1987};  M03: \citealt{marziani_optical_2003}; M22: \citealt{marziani_intermediate-ionization_2022};  SDSS: Sloan Digital Sky Survey. 
    Column 7: Source of IR. MIPS:  Multiband Imaging Photometer; PACS: Photodetector Array Camera and Spectrometer.} 
\end{table*}

\subsection{VLA data}

The radio observations presented in this paper are from the VLA project 20B-081 (PI: M. Berton). There are 17 datasets in total - each including a single target, except one of them, which includes two sources, J1209+5611 and J1301+5902. The observations were all taken from 2020-12-04 to 2020-12-08, in three bands: L (1.5 GHz), C (5 GHz), and X (10 GHz). Due to the COVID pandemic, the data were taken in a non-standard configuration, with two arms fully extended in A-configuration and one arm in the more compact B-configuration except for one antennae moved to the end position of the arm. Fig.~\ref{fig_plotants_MRK957} shows the antenna positions  during the observations of J00418+4021 on 2020-12-04, and the associate \textit{uv} plane coverage of the L-band scan. Such non-uniform configuration might have impacted the quality of some of the images (slightly elongated synthesized beam), but overall we do not note a significant impact: the rms noise is consistent with the expected values.
For L-band, we used the 8-bit samplers with center frequencies 1.25 and 1.75 GHz, for C-band, the 8-bit samplers centered at 4.72 and 5.75 GHz, and for X-band, two 3-bit samplers centered at 9 and 11 GHz. Each target was observed for a total of 10 minutes per band, which accounting for amplitude and phase calibration and overheads correspond to 20 minutes per band.

We applied the CASA \citep{casa_team_casa_2022} pipeline version 6.4.1.12 to each dataset. The pipeline outputs were visually inspected to make sure the pipeline ran well on all datasets. Once calibrated, the target observations were split away for each band in individual measurement sets. We did not perform any (time or frequency) averaging. Each was imaged individually using \textsc{tclean} from CASA (version 6.5.2.26). We used 2 Taylor coefficients to model the emission's spectral structure across the band and applied W-projection corrections with 14, 4 and 2 planes (for the L, C and X-band, respectively) to correct for the wide-field, non-coplanar baseline effect \citep{cornwell_noncoplanar_2008}. We chose cell sizes of $0.2\arcsec$, $0.07\arcsec$ and $0.03\arcsec$ for the L, C and X-band images, respectively, and image sizes of $9000 \text{ pixels} \times 9000 \text{ pixels} = 30 \arcmin \times 30\arcmin$, $7500 \text{ pixels} \times 7500 \text{ pixels} = 8.75 \arcmin \times 8.75\arcmin$ and $9000 \text{ pixels} \times 9000 \text{ pixels} = 4.5 \arcmin \times 4.5\arcmin$ (approximate sizes of the full width at half maximum of the field of view at the respective frequencies). Due to the presence of very bright sources near J11292-0424, J15455+4846 and J17014+5149, we increased the size of the L-band image which improved significantly the image quality. We used the multi-scale \textsc{clean} algorithm from \cite{cornwell_multiscale_2008} to probe the different scales of the structures in the images: 0, 5 and 15 pixels. Briggs weighting is used with a default robust factor of 0.5 (which corresponds to an intermediate between natural and uniform weighting). 
We run a round of phase self-calibration to improve the overall dynamic range in two cases: for J11292-0424, whose field encompassed the brightest source, and for 12562+5652, which is the brightest target.
Finally, wideband primary beam corrections were calculated with the CASA task \textsc{widebandpbcor}, listing all the spectral windows and the middle channel. Cutouts of the final images are shown in Fig.~\ref{fig:fig_VLA_QSO}, while the obtained beam sizes, rms and dynamic ranges (DR) are listed in Table~\ref{table_images}. The noise level in each image has been evaluated by taking the average of the rms noise in the four corners of the cutouts, so near the sources, but in regions devoid of other bright radio sources. On average, the rms reached are $\sim 30$, 10 and $8 \mu$Jy/beam for the L, C and X-band images (which are close to the noise expected for such observations based on the VLA exposure calculator), except for the case of J12562+5652, whose brightness limits the noise reachable. 
Flux densities encompassed in 5 sigma contours (or peak flux densities for compact sources) are listed in Table~\ref{table_flux}. To test the robustness of these measurements, we compared them with integrated flux densities derived from 2D Gaussian fits. We used the CASA task \textsc{imfit} to test this, and obtained values within $\pm5\%$ from the 5 sigma contours method for the sources with an elliptical morphology, confirming the validity of this approach.
For flux calibration we assume a relative error of $5\%$ for all VLA bands\footnote{The Flux density scale calibration accuracy is based on the VLA Observational Status Summary 2020B.}. The uncertainties is then given by the sum in quadrature of the rms and the calibration error. Of the 18 sources observed with the VLA, 17 are detected at X-band and 16 at L and C-band. For undetected sources, we indicate a 5 sigma upper limit in Table~\ref{table_flux}.

\subsection{Complementary radio data}

In order to complement our VLA observations, we added to the analysis radio observations from the LOw-Frequency ARray (LOFAR) Two-metre Sky Survey (LoTSS) DR2 \citep{shimwell_lofar_2022} and the Very Large Array Sky Survey (VLASS) \citep{lacy_karl_2020}.

We looked for our sources in LoTSS DR2 by first identifying the full bandwidth Stokes I continuum mosaic (with a central frequency of 144 MHz) that better covers each targets and downloaded the fits file of the full resolution image. Cutouts around the sources are presented in Fig.~\ref{fig:fig_VLA_QSO}, while the beam sizes, rms and DR of the mosaic are listed in Table~\ref{table_images}. We determined the noise level in the same way as for the VLA images, by taking the average of the rms noise in the four corners of the cutouts. LOFAR images have beam sizes of $6\arcsec \times 6\arcsec$ and an average noise of about 0.1 mJy/beam. All 10 sources covered by LoTSS DR2 are detected. Flux densities encompassed in 5 sigma contours (or peak flux densities for spatially unresolved sources) are listed in Table~\ref{table_flux}. We assumed a relative error of $10\%$ on the flux calibration \citep{shimwell_lofar_2022} and the uncertainties on the fluxes are then given by the sum in quadrature of the rms and the calibration error.

We obtained the VLASS images (with central frequency of 3 GHz) encompassing our sources through the Canadian Astronomy Data Centre (CADC) portal. All sources are covered by the survey, either in Quick Look (QL) or Single Epoch (SE) continuum images, in at least two epochs. We selected SE images when available to extract the flux, as they are intended to be the reference continuum images for each epoch of VLASS, and otherwise the QL image from the epoch closest in time to the VLA observations (see column 8 in Table~\ref{table_images}). The images have beam sizes of about $2-4\arcsec$ and an average noise of about 0.1 mJy/beam.
Flux densities encompassed in 5 sigma contours (or peak flux densities for spatially unresolved sources) are listed in Table~\ref{table_flux}.
We assumed a relative error of 3 and $10\%$ on the flux calibration for SE and QL images, respectively, and the uncertainties on the fluxes are then given by the sum in quadrature of the rms and the calibration error.
We additionally extracted the flux from all epochs and plotted for each source their flux variation on Fig.~\ref{fig_VLASS_time_series}.
For sources with detections in at least two epochs, we computed the variability index using the formula adopted from \cite{aller_pearson-readhead_1992}:
 \begin{equation}
    V = \frac{(S_{max}-\sigma_{S_{max}})-(S_{min}+\sigma_{S_{min}})}
    {(S_{max}-\sigma_{S_{max}})+(S_{min}+\sigma_{S_{min}})},
\end{equation}
where $S_{min}$ and $S_{max}$ are the maximum and minimum values of the flux density over all epochs of observations, while $\sigma_{S_{min}}$ and $\sigma_{S_{max}}$ are their associated errors.
These are shown on Fig.~\ref{fig_VLASS_time_series}. Negative values of $V$ correspond to cases where the error is greater than the observed scatter of the data. Three of our sources exceed the value of $V =0.1$, which makes them potential candidates for variable objects (e.g. \citealt{kunert-bajraszewska_vlass-based_2025}): J12091+5611, J12562+5652 and J15455+4846. For these sources, the flux extracted from the SE image has been selected for the rest of the analysis, and we verified that it also correspond to the epoch closest in time with the VLA observations.

\subsection{Optical and IR data}

The optical data come from three main sources: the \citet{marziani_optical_2003} atlas of low-$z$\ type-1 AGN covering the \hb\ spectral range; the \citet{marziani_intermediate-ionization_2022} comparative analysis, where the \hb\ spectral range was paired to the rest frame UV blend at $\lambda1900$, and the SDSS. In addition, the spectrum of J00418+4021 was digitized from a published plot \citep{halpern_narrow-line_1987}. Continuum S/N is $\gtrsim 20$\ excluding J00418+4021. Spectral resolution is $\lambda/\delta\lambda \sim 2000$\ for the SDSS data, $\lambda/\delta\lambda \sim 1000$\ for the other spectra. 
The SDSS spectra and \citet{marziani_optical_2003} were fit for this work using the same multi-component, non-linear analysis employed in \citet{marziani_intermediate-ionization_2022}. A special attention was devoted in this work to the fitting of the \oiiiopt\ lines, as well as in the detection of a possible blueshifted (outflow) component in \hb. Appendix \ref{optical_spectro} reports the results of a multicomponent nonlinear analysis of the \hb\ and \oiii\ lines.  Accretion parameters (Eddington ratio and black hole mass estimates derived from optical spectra) are also reported in Appendix \ref{optical_spectro}.

We  used IR fluxes extracted from the Wide-field Infrared Survey Explorer (WISE, \citealt{wright_wide-field_2010}), as well as FIR fluxes from the Infrared Astronomical Satellite (IRAS) Survey \citep{neugebauer_infrared_1984} -- except for source J14063+2223, which did not have IRAS data at 100 $\mu$m, so we used the flux from the Infrared Space Observatory (ISO) survey \citep{kessler_infrared_1996}. IR and FIR fluxes used during the analysis are reported in table \ref{tab:IR-FIR}. We considered also AKARI \citep{murakami_infrared_2007}, ISO, Spitzer MIPS, and Herschel data. The AKARI fluxes at 60$\mu m$ are available for only 4 objects (listed in Table \ref{table_targets}), and are about $\approx 20$\%\ lower than the IRAS fluxes, as expected from the comparison  to more precise data from newer telescopes  \citep{serjeant_evolution_2009}. We therefore expect that the SFR should be consistent with the ones derived from IRAS data, for the same normalization used in Section \ref{sf}.

\section{Results}\label{results}

\subsection{Radio loudness}

We start by calculating the Kellermann's parameter $ R_\mathrm{K}$\  \citep{kellermann_vla_1989}, using the radio flux density at 1.4 GHz  and the optical flux density at 5100 \AA\ from column 2 in Table~\ref{tab:meashb}. The observed radio flux density $f_\mathrm{\nu,o}$ reported in Table \ref{table_flux} has been converted to rest frame by applying a K-correction: $f_\mathrm{\nu,e} = f_\mathrm{\nu,o}[(1 + z)^{(-\alpha+1)}]$  \citep{ganci_radio_2019}, where the spectra index $\alpha$ is reported in sixth column of Table \ref{table_alpha}. 
We use the following criteria to distinguish between radio loud and radio quiet sources:
\begin{enumerate}[-]
    \item Radio loud (RL): sources with $ R_\mathrm{K} \geq 10 $, from the radio fluxes at 1.4GHz reported in Table ~\ref{table_flux}, and from the optical fluxes at 5100 \AA\ (Appendix \ref{optical_spectro});
    \item Radio quiet (RQ): sources with $R_\mathrm{K} < 10$.
\end{enumerate}
This criterion applied  to our sample yields a large fraction of  RL sources, 8 out of 18. We note however that $ R_\mathrm{K} \geq 10 $\ is not by itself a sufficient condition to demonstrate the presence of a relativistic jet, especially among xA quasars \citep{zamfir_new_2008,ganci_radio_2019}. If a more restrictive condition $\log R_\mathrm{K} \geq 1.8 $ (corresponding to $ R_\mathrm{K} \geq 70 $) is applied as suggested by \cite{zamfir_new_2008} for the lower boundary of the RL phenomenon when analysing FRII sources, only J00418+4021 and J14258+3946 would be classified as RL. We also note that the optical continuum of J00418+4021 might have been underestimated, making J14258+3946 the sole bona-fide RL source of the sample.

\subsection{Radio morphology}

Following \cite{berton_radio-emitting_2018}, we define three morphological classes reflecting the properties of the sources seen in the maps, using the ratio ($\mathcal{R}$) of the peak to the total flux density of each source: 
\begin{enumerate}[-]
    \item Compact: sources with $\mathcal{R} \geq 0.95$, labeled as C in Table~\ref{table_flux}
    \item Intermediate: sources with $ 0.75 \leq \mathcal{R} < 0.95 $, labeled as I in Table~\ref{table_flux}.
    \item Extended: sources with $ \mathcal{R} < 0.75 $, labeled as E in Table~\ref{table_flux}.
\end{enumerate}
Most sources of our sample are compact, but at the lowest frequency band (LOFAR data) a much higher fraction of the sources are extended ($60\%$ at 144 MHz, vs $11\%$, $17\%$, $11\%$ and $11\%$ at 1.5, 3, 5 and 10 GHz respectively). 

Overall, only three targets show clear extended emission at most frequencies: J00418+4021, J12562+5652 and J17014+5149.
J00418+4021 and J12562+5652 have faint diffuse emission, especially visible at lower frequencies, surrounding their bright core. The case of J17014+5149 is different: its radio emission resolves into two distinct point-like sources at high frequency.

\begin{figure}
    \center
	\includegraphics[width=0.52\textwidth]{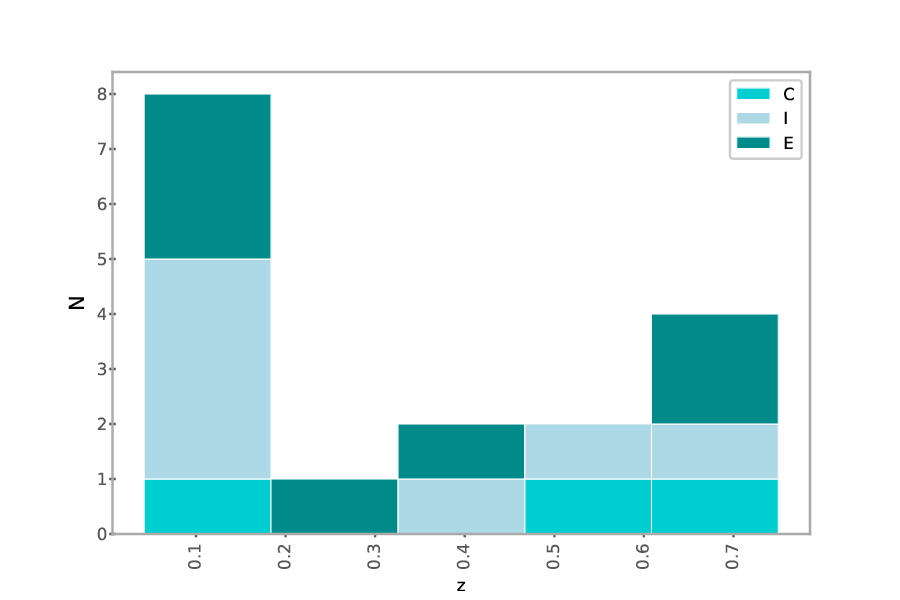}
    \caption{Radio morphology distribution across the redshifts of the sources. C indicates compact at all frequencies observed, I is intermediate morphology detected in at least one band, and E indicates the presence of diffuse emission in at least one band.}
    \label{fig_z_morph}
\end{figure}

We present the distribution of radio morphology detected across the various redshifts of our sources in Figure~\ref{fig_z_morph}. Extended emission is detected even in our highest-redshift sources. At the highest redshift (0.75, J12075+1506), our highest resolution images resolve structures down to 2-3 kpc, which should be enough to resolve relativistic jets or SF.

\subsection{Radio spectral index} \label{Radio spectral index}

Multi-frequency radio observations allow us to precisely measure the spectral index, helping us to understand the origin of the radio emission. We adopt $S_{\nu} \propto \nu^{\alpha}$, where $\alpha$ is the spectral index. We measure the spectral index between bands such as:
\begin{equation}
    \alpha_{\nu_{1}-\nu_{2}} = \frac{\log(S_{2}/S_{1})}{\log(\nu_{2}/\nu_{1})},
\end{equation}
where $S_{1}$ and $S_{2}$ are the flux densities at the observing frequencies $\nu_{1}$ and $\nu_{2}$, respectively. To measure the global spectral index across the whole frequency range available, we fitted the fluxes across the bands using \textsc{numpy.polyfit} for sources with detection in at least three bands. Uncertainties in the spectral indices were calculated using standard propagation of errors.
All the spectral indices are reported in Table~\ref{table_alpha}. We plot the spectrum of each source detected in at least three bands in Figure~\ref{fig_spectrun}. From the three sources identified with significant variability in VLASS observations, we note that the flux from J12562+5652 non-simultaneous observations at 144 MHz and 3 GHz fall outside of the fitted spectral slope.

For the few sources with extended emission at lower frequencies, the large mismatch in resolution with higher frequencies images prevented a spatially resolved analysis of the spectral index. In the case of J17014+5149, where two distinct point-like sources are resolved at C and X bands, we calculated the spectral indices of each point-like source. We used the CASA task \textsc{imsmooth} to smooth the X-band image to the C-band beam (which has the lowest spatial resolution) and then apply the CASA task \textsc{imregrid} to regrid the X-band image using the C-band image as reference. We extracted fluxes from both point-like sources and obtained identical values (within the uncertainties) of $\alpha_{5-10}= -1.0 \pm 0.1$. This is consistent with the spectral index calculated from the integrated flux of the sources.

\begin{figure*}
	\includegraphics[width=\textwidth]{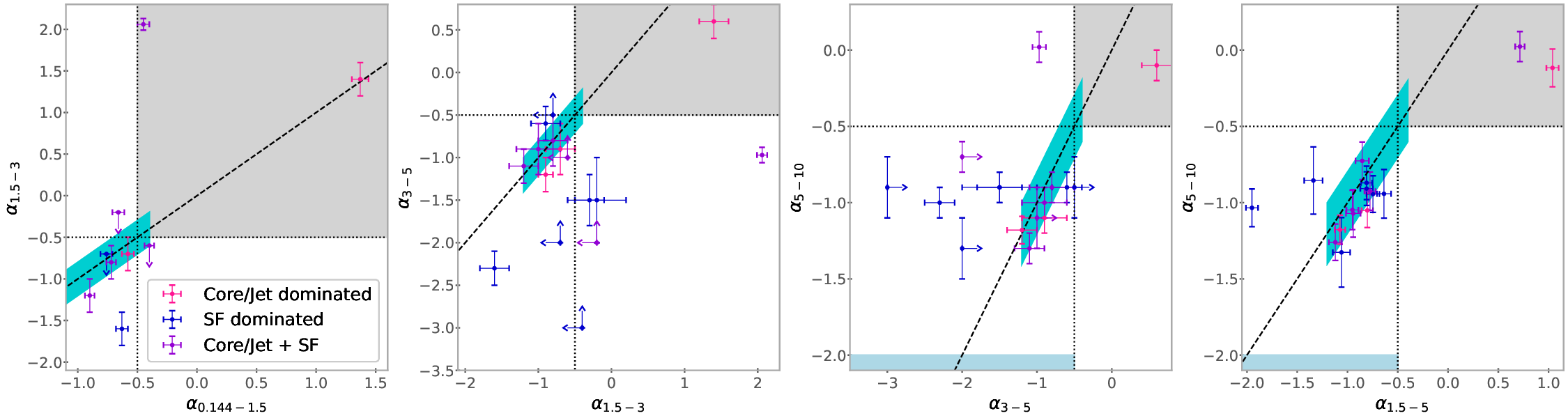}
    \caption{The spectral indices 
    $\alpha_{1.5-5}$ versus $\alpha_{0.144-1.5}$,
    $\alpha_{3-5}$ versus $\alpha_{1.5-3}$, 
    $\alpha_{5-10}$ versus $\alpha_{3-5}$ and 
    $\alpha_{5-10}$ versus $\alpha_{1.5-5}$. Markers are colored depending on the identified dominating mechanism behind the detected radio emission (Table~\ref{table_classification}).
    The dashed line is the $1:1$ ratio line of equal slopes, while the horizontal and vertical dotted lines are $\alpha = -0.5$. 
    The area showing flat or inverted spectra, which may be associated with a core-dominated jet or a corona, is filled in gray. 
    The area with a spectral index $\alpha = -0.8 \pm 0.4$ with a scatter of 0.2, which may be related to SF, is filled in turquoise. 
    The area showing very steep spectral slopes with $\alpha_{5-10} < -2$ in the right-hand panel, which may imply possible relic emission, is filled in blue.}
    \label{fig_alpha-alpha_plot}
\end{figure*}

A common approach to characterize radio spectra is the use of radio color-color diagrams. We show the spectral indices 
$\alpha_{0.144-1.5}$ versus $\alpha_{1.5-3}$, 
$\alpha_{1.5-3}$ versus $\alpha_{3-5}$ and 
$\alpha_{3-5}$ versus $\alpha_{5-10}$ in Figure~\ref{fig_alpha-alpha_plot}, top panels.
To remove the impact of non-simultaneous observations as well as the missing LOFAR images and the non-detections from VLASS images, we also show a color-color diagram including only the L, C and X-band VLA observations: $\alpha_{1.5-5}$ versus $\alpha_{5-10}$ in Figure~\ref{fig_alpha-alpha_plot}.
We do not identify any sources with an ultra steep drop-off (see blue area in Figure~\ref{fig_spectrun}-right), indicative of relic emission, when significant cooling happened since the last AGN activity, leading to a spectrum with exceptional steepening ($\alpha<-2$) at high frequencies. This is consistent with \cite{fraix-burnet_phylogenetic_2017}, where xA sources are interpreted as AGN in an early evolutionary stage.

We use these spectral indices to calculate the radio luminosity of the detected sources:
\begin{equation}\label{radio_lum}
    L_{1.4 GHz} = 4 \pi D_{L}^{2} S_{1.4 GHz} (1 + z)^{-\alpha-1}, 
\end{equation}
where $L_{1.4 GHz}$ is the radio luminosity (W Hz$^{-1}$) at 1.4 GHz, $D_{L}$ is the luminosity distance (Mpc), calculated with Astropy using the redshifts from Table~\ref{table_targets}, $\alpha$ is the spectral index calculated by measuring the fitted spectral index, and $S_{1.4 GHz}$ is the integrated radio flux density at 1.4 GHz extrapolated using the fitted spectral index. These luminosities are reported in Table~\ref{table_alpha}.

\begin{figure}
    \center
	\includegraphics[width=0.4\textwidth]{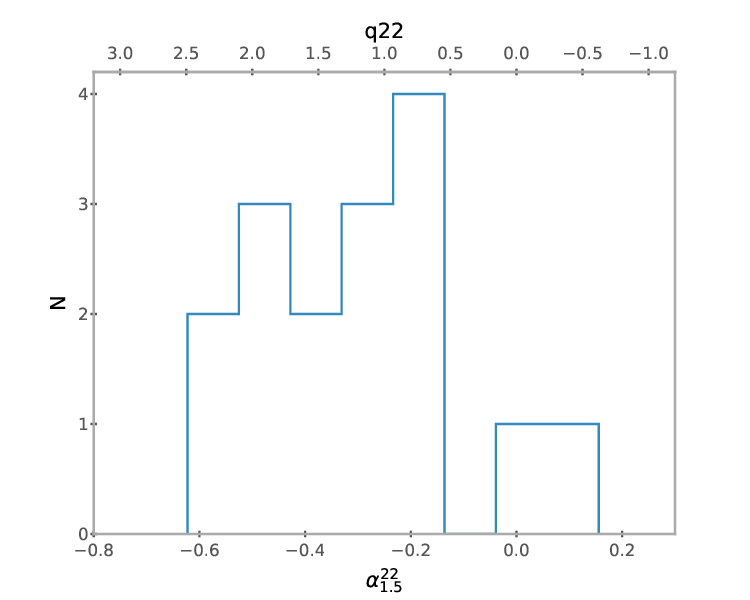}
    \caption{Distribution of the two-point spectral index between radio (1.4 GHz) and mid-IR ($22\mu m$). All sources have $\alpha^{22}_{1.5}$ below $\sim-0.1$ except J12091+5611 and J14258+3946.}
    \label{fig_alpha-IR_plot}
\end{figure}

Radio-to-mid IR flux ratio is a useful tool to identify the source of radio emission since it differs significantly for jet-dominated AGN (i.e. blazars) and star-forming galaxies. We follow \cite{caccianiga_wise_2015} and use the two point spectral index between 1.5GHz and $22\mu m$, defined as:
\begin{equation}
    \alpha^{22}_{1.5} = -
    \frac{\log(S_{1.5}/S_{22})}{\log(1.5 \text{GHz}/ (c/ 22 \mu m))},
\end{equation}
where $S_{22}$ is the flux densities at $ 22\mu m$, as well as the $Q22$ parameter, defined as:
\begin{equation}
    Q22 = \log(S_{22}/S_{1.5}).\label{eq:q22}
\end{equation}
Blazars, for which the radio emission is dominated by the relativistic jet, have steep $\alpha^{22}_{1.5} > 0.2$ indices, while star forming infrared galaxies have $\alpha^{22}_{1.5} < -0.25$, with SF as the main contributor to the radio emission \cite{caccianiga_wise_2015}.
The distribution of radio-to-mid IR flux ratios of the detected sources in our sample is shown on Fig.~\ref{fig_alpha-IR_plot}: all sources have $\alpha^{22}_{1.5}$ below $\sim-0.1$ except two, J12091+5611 ($\alpha^{22}_{1.5}$ = 0.02) and J14258+3946 ($\alpha^{22}_{1.5}$ = 0.16). The distribution does not span a range as wide as in \cite{caccianiga_wise_2015}. Instead we find only a few objects with intermediate values: J12091+5611 ($\alpha^{22}_{1.5}$ = 0.02), J12366+5630 ($\alpha^{22}_{1.5}$ = -0.14) and J14475+3455 ($\alpha^{22}_{1.5}$ = -0.16). For these sources, the radio emission most likely origin from a combination of jet and SF.

The flux densities at $ 22\mu m$ measured for our sources and used to derived the $Q22$ parameter could be affected by the suite of broad emission bands from polycyclic aromatic hydrocarbons (PAHs) molecules. The PAH bands are tracers of star formation activity, and the 11.3$\mu$m\ feature is believed to remain so even in the presence of an AGN \citep{diamond-stanic_relationship_2012}.
Considering the redshift of our sample, the $Q22$ parameter could be affected mainly by the relatively weak feature at $\approx 17 \mu$m \citep{smith_mid-infrared_2007}. With a typical equivalent width in the observed frame of $\sim 0.5 $\AA, it is expected to contribute only to slightly more than $\sim 10$\%\ to the total flux in the $S_{22}$ band. PAH features are also expected to be fainter in type-1 AGN, either because of dilution or destruction by the AGN radiation field \citep{xie_ionization_2022}. A systematic analysis of the MIR properties of the present sample would be in order, considering the mixed nature of these sources, but is beyond the scope of the present work. However, we do not use $Q22$ as diagnostic of star formation, but rather as diagnostic for the presence of an excess radio emission that lowers $Q$22 and increases $\alpha^{22}_{1.5}$. We therefore classify the 2 sources that fall in the blazar domain (i.e. with $\alpha^{22}_{1.5} > 0$) as core/jet dominated.

\begin{figure}
	\includegraphics[width=0.5\textwidth]{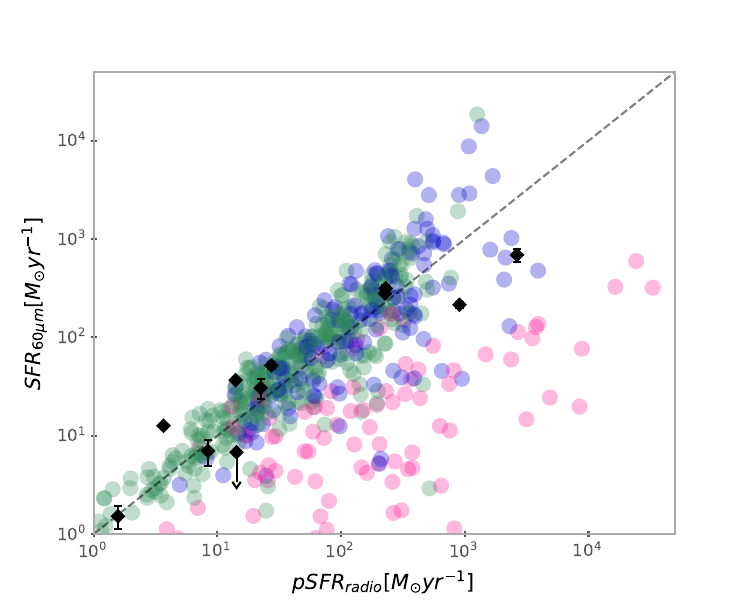}
    \caption{SFR estimate based on FIR luminosity vs. radio pseudo-SFR for all sources radio-detected and with IRAS $60\mu m$ data (black diamonds). Star-forming galaxies (green circles), RQ quasars (blue circles) and RL quasars (pink circles) are from \protect\cite{bonzini_star_2015}, with their $p\text{SFR}_{\text{radio}}$ converted using eq. 22 of \protect\cite{molnar_non-linear_2021}. The dashed line is the 1 : 1 ratio line of equal SFR. }
    \label{fig_SFR_radio-vs-IR}
\end{figure}

\subsection{Star formation estimates}
\label{sf}

We obtain star formation estimates from the IR and radio luminosity following \cite{ganci_radio_2019}. Using the starburst calibration derived by \cite{kennicutt_star_1998}, we estimate the SFR based on the IR luminosity with:
\begin{equation}
    \text{SFR}_{60\mu m} \approx 4.5 \times 10^{-44} L_{60\mu m},
\end{equation}
where $\text{SFR}_{60\mu m}$ is in $M_{\odot} yr^{-1}$ and $L_{60\mu m}$ in erg/s is derived from $60\mu m$ flux densities from IRAS and ISO data  as:
\begin{equation}\label{eqn:L60}
   L_{60\mu m} = 4 \pi D_{L}^{2} \nu S_{60\mu m} / K_{FIR},
\end{equation}
with $K_{FIR}$ being the $K$ correction in the FIR that we estimated as in equation 10 of \cite{ganci_radio_2019}. It is of order unity, varying within the range $\sim 1-1.4$. 
The SFR derived from $60-70 \mu$ m   samples  roughly the peak of the emission from warmer dust (50 - 60 K). Galaxies with hotter dust (e.g.\ very compact starbursts) will have a higher $60\mu$m/$100\mu$m  ratio than cooler, more quiescent disks.  Conversely, cold cirrus tails will depress $L_{60\mu m}$ relative to the total FIR emission. We also considered $L$(FIR) from IRAS data at 60$\mu$m and 100$\mu$m ($L$(FIR) defined from a linear combination of the 60$\mu$m and 100$\mu$m fluxes according to \citealt{helou_thermal_1985}, see Tab. \ref{tab:IR-FIR}).
$L$(FIR) {is comparable} to $L_{60\mu m}$, with some scatter associated with differences in the 100  $\mu$m/60$\mu$m ratio  ranging from $\approx 0.8$ to 2. The different origin of the FIR emission at 60$\mu$m and 100$\mu$m\ make it preferable that the 60$\mu$m scaling is used for comparison with radio emission, as it is expected to be more closely connected to recent star formation. However, for sources at $z\gtrsim 0.5$, the wavelength entering into the $60\mu$m band is $\lambda \lesssim 60/(1+z) \lesssim 40 \mu$m. Even at $60 \mu$m, there is a danger of contamination from the hot dust emission in the AGN torus \citep{mor_hot_2012,netzer_star_2016,fuller_sofiaforcast_2019}.  Nonetheless, the highest redshift sources for which there usable FIR data are J15455+4846 and J17014+5149. In this case, the $60\mu$m scaling, and the $L$(FIR) yield SFR that are within a factor 2 from the SFR derived by ISO including the flux at 150 $\mu$m\ following the scaling by \citet{stickel_isophot_2000}.

 We compare the $L_{60\mu m}$\  estimates with the radio pseudo-SFR for each object following eq. 22 of \cite{molnar_non-linear_2021}:
\begin{equation}
   \text{log}( p\text{SFR}_{\text{radio}}) = (0.823 \pm 0.009) \times \text{log}(L_{1.4 GHz}) - (17.5 \pm 0.2),
\end{equation}
where $p\text{SFR}_{\text{radio}}$ is in $M_{\odot} yr^{-1}$ and $L_{1.4 GHz}$ in W Hz$^{-1}$. 
These are displayed in Table~\ref{table_alpha}. $p\text{SFR}_{\text{radio}}$ assumes that all the radio emission detected is synchrotron emission associated with supernova remnants - which is incorrect for sources with an AGN-related contribution. 
Fig. \ref{fig_SFR_radio-vs-IR} shows the FIR SFR estimate vs. radio pseudo-SFR for all sources radio-detected and with IRAS $60\mu m$ data. For comparison we added star-forming galaxies (green), RQ quasars (blue) and RL quasars (red) from \cite{bonzini_star_2015}, with their $p\text{SFR}_{\text{radio}}$ converted using eq. 22 of \cite{molnar_non-linear_2021}. All of the 10 sources with measured FIR SFR and radio pseudo-SFR more or less follow the 1 : 1 ratio line of equal SFR.

\begin{table*}
\caption{Indicators of the dominating mechanism behind the detected radio emission of the sources and overall classification}
\label{table_classification}
\begin{center}
\begin{tabular}{P{1.6cm}P{1.7cm}P{1.5cm}P{1.7cm}P{1.6cm}P{1.6cm}P{1.6cm}P{0.6cm}P{2.4cm}}
\hline
Object 
& $L_{1.4 \text{GHz}}$  
& VLASS variability 
& Radio spectra 
& Radio morphology 
& FIR-radio correlation 
& $\alpha^{22}_{1.5}$ 
& $R_{k}$
& Overall Classification \\
\midrule
 J00418+4021 & - & - & Linear/Curved & Extended & SF & SF & RL & SF dominated \\
 J00457+0410 & - & - & - & - & - & - & - & - \\
 J00535+1241 & - & - & Linear/Curved & C/I & SF & SF & RQ & SF dominated \\
 J10255+5140 & - & - & Linear/Curved & C/I & SF & SF & RQ & SF dominated  \\
 J11292-0424 & - & - & Linear/Curved & C/I & SF & SF & RQ & SF dominated \\
 J11420+6030 & - & - & Linear/Curved & C/I & - & Combination & RQ & Core/Jet + SF \\
 J12075+1506 &Core/Jet& - & Linear/Curved & C/I & - & SF & RQ & Core/Jet + SF \\
 J12091+5611 & - & Core/Jet & Linear/Curved & C/I & - & Core/Jet & RL & Core/Jet dominated \\
 J12366+5630 & - & - & Linear/Curved & C/I & - & Combination & RL  &  Core/Jet + SF \\
 J12562+5652 & - &Core/Jet&Flat/Inverted&Extended& SF & SF & RL & Core/Jet + SF \\
 J13012+5902 & - & - & - & - & - & - & - & - \\
 J14052+2555 & -  & - & Linear/Curved & C/I & SF & SF & RQ & SF dominated \\
 J14063+2223 & - & - & Linear/Curved & C/I & - & SF & RQ & SF dominated \\
 J14170+4456 & - & - & Linear/Curved & C/I & SF & SF & RQ & SF dominated \\
 J14258+3946 &Core/Jet& - & Flat/Inverted & C/I & -& Core/Jet & RL &  Core/jet dominated \\
 J14475+3455 &Core/Jet& - & Linear/Curved & C/I & SF & Combination & RL & Core/Jet + SF \\
 J15455+4846 & - &Core/Jet& Linear/Curved & C/I & SF & SF & RL & Core/Jet + SF \\
 J17014+5149 &Core/Jet& - & Linear/Curved & Jets & SF & Combination & RL & Jet dominated \\
 \hline
\end{tabular}
\tablefoot{
    Column 1: Object coordinate name.
    Column 2: Sources with $\log L_{1.4 \text{ GHz}} > 31.6 $ erg s$^{-1}$ Hz$^{-1}$ certainly have an AGN-related contribution to their radio emission.
    Column 3: A variability index $V > 0.1$ (based on VLASS observations) indicate core/jet origin.
    Column 4: Shape of the radio spectra.
    Column 5: Particular morphology from radio observations - extended, compact or intermediate (C/I), or presence of jets.
    Column 6: Sources for which $p\text{SFR}_{\text{radio}} \simeq \text{SFR}_{60\mu m}$ are likely dominated by SF, while sources below the correlation are likely a combination.
    Column 7: Sources with $\alpha^{22}_{1.5} < -0.25$ are SF, those with $\alpha^{22}_{1.5} > 0$ are core/jet-dominated and the rest are combination of both. 
    Column 8: For reference, the calculated Kellermann’s parameter $R_{k}$ (same as in last column of Table~\ref{table_flux}).
    Column 9: Overall origin of the emission mechanism. 
}
\end{center}
\end{table*}

\begin{figure}
	\includegraphics[width=0.47\textwidth]{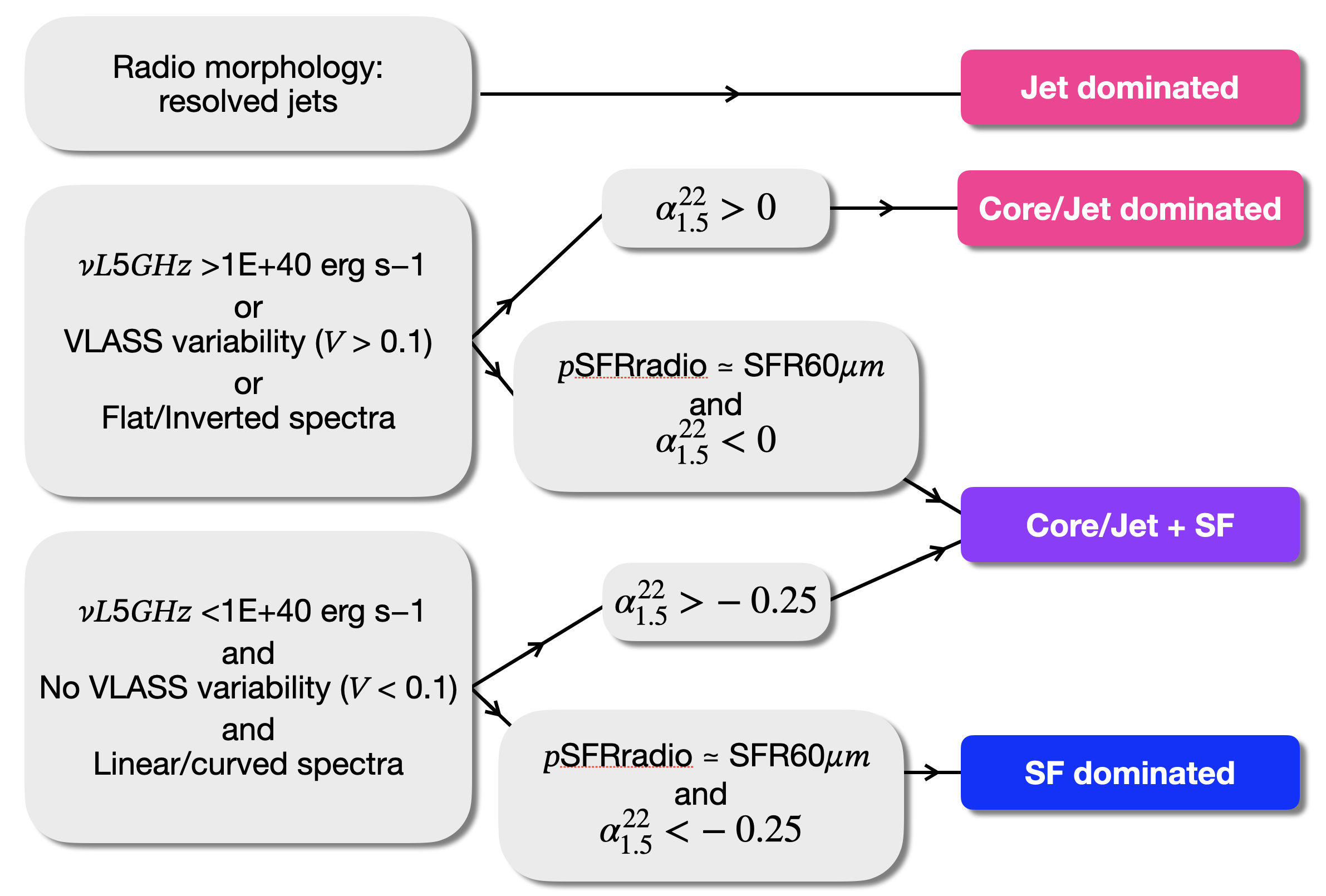}
    \caption{Summary of the classification scheme used to identify the radio emission dominating mechanisms for each sources of our sample based on the combination of indicators compiled in Table~\ref{table_classification}.}
    \label{schema_QSO}
\end{figure}

\subsection{Classification of the dominating radio mechanism}

The radio emission from the sources of our sample may be coming from a combination SF, AGN-driven winds/outflows, relativistic jets, and/or accretion disc coronal emission \citep{chen_radio_2022}. 
For each source, we now intend to identify the dominating mechanisms based on the combination of several indicators, all compiled in Table~\ref{table_classification}. The calculated Kellermann’s parameter $R_{k}$ are shown for reference but were not used for the classification. The classification scheme we decided to follow is shown on Figure~\ref{schema_QSO}, with details given below.

We first use the radio luminosity, which for starbursts typically do not exceed $\nu L_{\nu,int}$ = 1E+40 erg s$^{-1}$ \citep{sargsyan_star_2009}. 
We adopted the limit from \cite{zamfir_new_2008}, $\log L_{1.4 \text{ GHz}} = 31.6 $ erg s$^{-1}$ Hz$^{-1}$.
This indicates that the radio emission from at least 4 of our 18 sources must have an AGN-related component (core or jet). Then, we note the three sources of our sample with significant radio variability at 3 GHz ($V >0.1$), J12091+5611, J12562+5652 and J15455+4846, indicating a core or jet-related origin.
We also consider the shape of the radio spectra obtained in section \ref{Radio spectral index}.
SF produces a synchrotron spectrum with a $\alpha = -0.8 \pm 0.4$, no significant curvature from $1-10$ GHz and a spectral turnover at around a few hundreds MHz. 
Radio emission from AGN jets has steeper spectral slopes with increasing frequencies, indicating the aging of the relativistic electron population due to inverse-Compton, synchrotron, ionization, or bremsstrahlung energy losses.
Core-dominated jet or accretion disk corona produces optically thick synchrotron emission, characterized by a flat or inverted spectrum. 
From the 16 detected sources in at least three bands, 2 have flat or inverted spectra between 144 MHz (or 1.5 GHz) to 10 GHz: J12562+5652 and J14258+3946, with fitted spectral indices of $0.5\pm0.3$ and $1.0\pm0.2$, respectively, both RL sources.
J12562+5652 shows some extended emission (especially at 144 MHz), but, as it is the brightest source, the limited dynamical range reachable with the short exposure of the VLA observations might have prevented the detection of faint extended emission at higher frequencies (note the unusually high rms values from Table~\ref{table_images} for this target). This explains why the flux density measured at 144 MHz appears way above the fitted spectral slope.
J14258+3946 is the most luminous source of our sample and also has the highest $\alpha^{22}_{1.5}$ = 0.16. 

The non-simultaneity of the observations (LOFAR and VLASS) and the mismatch in resolution between frequencies prevent us to clearly differentiate between SF vs jet origin for the rest of the sources with linear or curved spectrum.

J17014+5149 is the only source where we resolve two symmetrical radio jets. Indeed, the two distinct point-like sources resolved at higher frequencies have identical (within the uncertainties) $\alpha_{5-10}= -1.0 \pm 0.1$. It also has the second highest radio luminosity of all our sample. Due to this morphology, we directly classify it as jet-dominated. For the rest of the sources, their morphologies or just the presence or lack-of extended emission do not allow us to constrain more the classification.

While correlation is not causation, we consider that the 10 sources having $p\text{SFR}_{\text{radio}} \simeq \text{SFR}_{60\mu m}$ are likely to have their radio emission from SF origin (see Figure~\ref{fig_SFR_radio-vs-IR}). None of these sources are located below the correlation, indicating that the entire detected radio emission can be resulting from SF. In addition to this criteria, we consider the two-point spectral index between radio and mid-IR (see Figure~\ref{fig_alpha-IR_plot}). We divided sources with $\alpha^{22}_{1.5} < -0.25$, typical of star forming infrared galaxies, from those with $\alpha^{22}_{1.5} > 0$ (more typical of core or jets), while for the rest of the sources, the radio emission likely origin from a combination of the AGN and SF. Only one source, J12075+1506, for which no IR data is available, cannot be classified following exactly the classification scheme from Figure~\ref{schema_QSO}. However, since it has $\log L_{1.4 \text{ GHz}} > 31.6 $ erg s$^{-1}$ Hz$^{-1}$ and $\alpha^{22}_{1.5} < -0.25$, the origin of its emission has been interpreted as a combination of the AGN and SF.

Overall, we find that 7 of our 18 targets are likely to have their radio emission dominantly coming from SF, and 6 from a combination of SF and AGN-related mechanism. Only three sources (all RL) indicate a core or jetted AGN only origin for the detected radio emission.  
Among the 8 RL sources, one is classified as SF dominated and four as a combination of SF and AGN-related mechanism.

\subsection{Bremsstrahlung radio emission}
\label{radiowind}

Free-free (Bremsstrahlung) emission arises from hot, ionized gas. The mechanism itself operates independently of bulk motion, while a BLR or a NLR wind could contribute to or enhance such emission—particularly by providing a more extended or structured ionized medium. The vast majority of super-Eddington sources in our sample exhibit evidence of \oiiiopt\ outflows (see the results of the multicomponent nonlinear analysis of the \oiiiopt\ line and estimation of wind parameter in Appendix~\ref{optical_spectro}), which in some cases are powerful enough to dominate the line emission. For sources with high Eddington ratios, free-free emission associated with a wind is a plausible contribution to their radio output \citep{blundell_origin_2007,chen_windy_2024}. 

As shown in the previous sections (Sect. \ref{Radio spectral index}), in our sample the radio spectral indices are {in general}  not consistent with sole optically thin or thick Bremsstrahlung emission \citep{baskin_radiation_2021} nor with emission from a corona. The energetics of the \oiii\ outflows also disfavors Bremsstrahlung emission as the origin for the radio power.

The mass of the gas emitting the \oiii\ lines can be computed by using a conversion between line luminosity and gas mass \citep{cano-diaz_observational_2012,carniani_ionised_2015,marziani_blue_2016,marziani_quasar_2017,fiore_agn_2017}: 
\begin{equation} 
M_\mathrm{[OIII]} \sim 1\cdot 10^6 L_{44}\left(\frac{Z}{5Z_\odot}\right) ^{-1} n_{4}^{-1} \mathrm{M}_\odot
\end{equation}
where  the line luminosity is expressed in units of 10$^{44}$ \ergss, the metallicity is scaled to five times solar, and the density  is normalized to $10^4$ cm$^{-3}$. Given the free-free emissivity $j_{\rm ff}$ \ \citep{rybicki_radiative_1979}, the radio power expected from the emitting gas is:

\begin{eqnarray}
P_{\nu,\mathrm{[OIII]}} \sim \frac{M_\mathrm{[OIII]}}{n \mu m_\mathrm{p}} j_\mathrm{ff} &\sim &  2.9 \cdot 10^{29} 
\bar g_{10} \, T_\mathrm{
e,10000}^{-1/2}  Z_{5} ^{-1} \cdot \\
& & n_\mathrm{4}  
L_{\mathrm{[OIII]},44}\,\mathrm{erg \, s^{-1} Hz^{-1}} \nonumber
\end{eqnarray}

where $\bar g_{10}$\ is value for the average Gaunt factor  normalized to $ 10$\ \citep{van_hoof_accurate_2014,van_hoof_accurate_2015,chluba_improved_2020}.    
The typical density of the spatially resolved NLR is $n \gtrsim 10^3$cm$^{-3}$ \citep[][c.f. \citealt{kakkad_spatially_2018}]{singha_close_2022};  an upper limit could be provided  density of the innermost part of the NLR of NGC 5548 has been found to be $n \sim 10^5$  \ cm$^{-3}$ \citep{peterson_size_2013}. We therefore scaled the equation for the radio-power to $n \sim   10^4$  \ cm$^{-3}$. If so, the observed median \oiii\ luminosity of $\sim 10^{42}$ \ergss\ (Appendix \ref{optical_spectro})  is not sufficient to explain the observed radio power, especially if $L_{\rm 1.4 GHz \gtrsim 10^{31}}$ \ergss\ Hz$^{-1}$. In the case of a compact NLR \citep{zamanov_kinematic_2002}, however, a higher density $\sim 10^5$ cm$^{-3}$\ of the \oiii\ emitting region  could account for powers comparable tp $L_{\rm 1.4 GHz} \sim\ 10^{29}$ \ergss\ Hz$^{-1}$.

\begin{figure*}
    \center
	\includegraphics[width=0.9\textwidth]{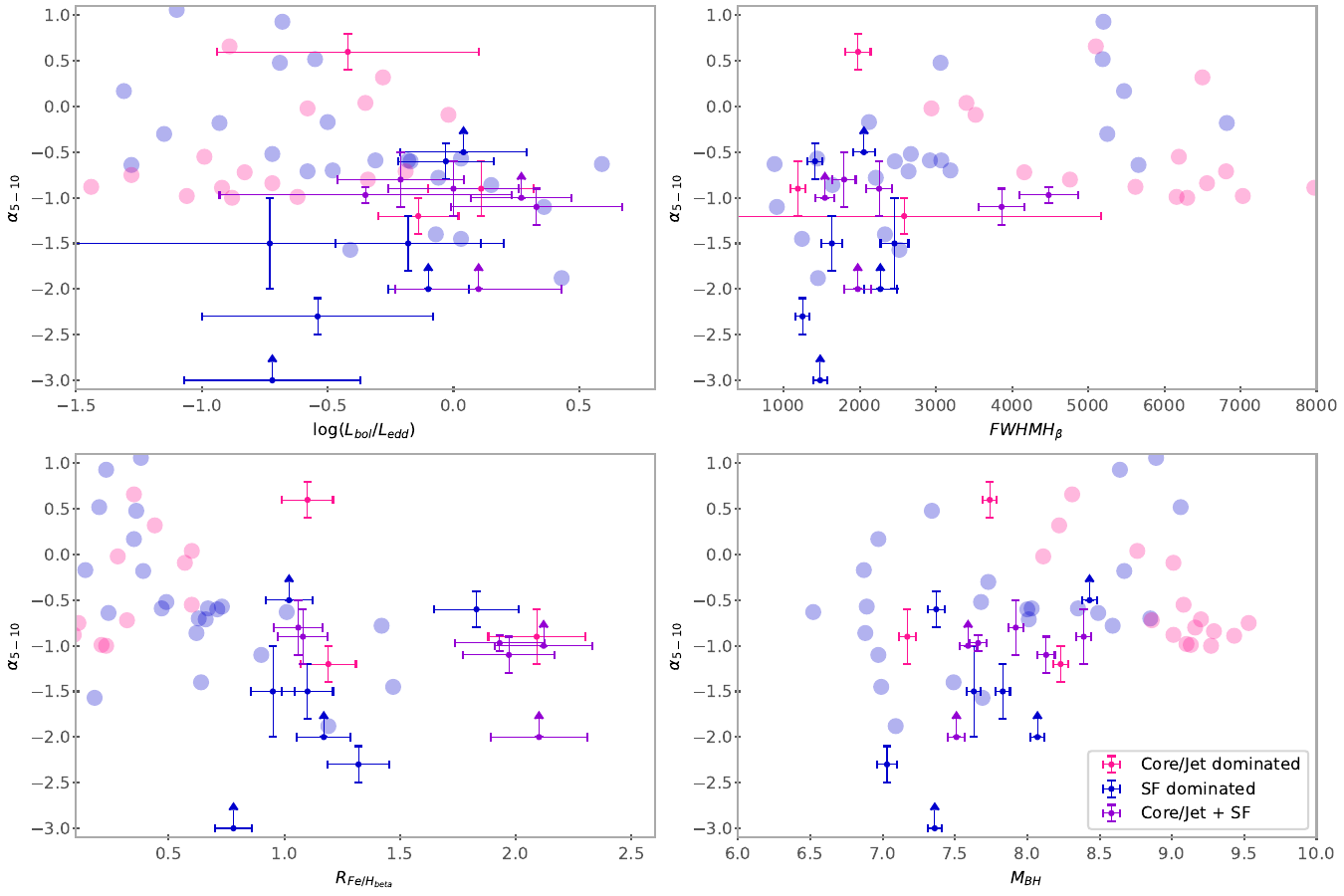}
    \caption{The spectral index $\alpha_{5-10}$ versus the Eddington ratio, the FWHM of the \hb\ line (in km/s), the flux ratio \rfe\ and \mbh. Markers are colored depending on the identified dominating mechanism behind the detected radio emission (Table~\ref{table_classification}).
    Blue (pink) filled circles represents the radio quiet (loud) quasars respectively from \protect\cite{laor_what_2019}.}
    \label{fig_alpha-accretion}
\end{figure*}

\subsection{Correlation analysis}\label{correlation}

\subsubsection{Accretion parameters vs radio spectral indices}
\cite{laor_what_2019} compared the radio spectral index of a sample of 25 RQ Palomar–Green (PG) quasars with the Eddington ratio, the FWHM of the \hb\ line, the flux ratio \rfe\ and \mbh. 
They found that radio quiet and radio loud quasars were showing different trends. 
The radio spectral indices of the radio quiet quasars in their sample correlate with the FWHM of the \hb\ line as well as with the Eddington ratio : high Eddington ratio sources have steeper radio spectrum (indicative of an extended optically thin synchrotron source, i.e. possibly a weak jet or wind component) while lower Eddington ratio sources have flatter radio spectrum (indicative of a compact optically thick synchrotron source, i.e. compact core, possibly a weak jet base or an accretion disc corona).
Radio loud quasars do not show such a correlation but their radio spectral indices correlate with \mbh: most quasars with $M_{BH} > 10^9 M_{\odot}$ have a steeper ($<-0.5$) spectral indices (radio is dominated by the extended lobe emission) while \mbh\ $< 10^9 M_{\odot}$ quasars have spectral indices $>-0.5$, and their radio emission is unresolved.

We compare these results with our sample in Fig. \ref{fig_alpha-accretion}. The distribution of our sources mainly overlap the radio quiet quasars from \cite{laor_what_2019}. Indeed, most of the xA data points (those classified SF-dominated and Core/Jet+SF, all with $\alpha_{5-10} \lesssim$ -0.5) are consistent with the high \lledd\ branch of \cite{laor_what_2019}.
A single source in our sample (J12562+5652 $\equiv$ Mark 231) is special in that it has an inverted spectrum ($\alpha_{5-10}>0$); this object may correspond more to the RL regime of \cite{laor_what_2019}.
An important difference is that two core-jet dominated objects show \rfe$>1$, while all RL sources of \cite{laor_what_2019} show \rfe$\lesssim$ 0.5,  and that those core-jet dominated sources in our sample have Eddington ratio close to 1.

\subsubsection{Correlation   of radio, IR and optical data}

The data for the xA sources allow for a correlation analysis between radio, IR, and optical spectral properties. Fig. \ref{fig:corrmatrix} shows the results for most parameters measured in the present papers (from Tables \ref{table_alpha},  \ref{tab:meashb}, \ref{tab:measoiii} and \ref{tab:valuesmbh}, or computed as explained in the previous sections). Apart from some predictable correlations (for example, \mbh\ and \lledd\ with luminosity), correlations between \rfe\ and \lledd\ have been found or predicted in several past works \cite[][and references therein]{marziani_searching_2001,sun_dissecting_2015,du_fundamental_2016,marziani_super-eddington_2025}. This correlation is not especially strong since almost all sources of the sample are strong \feii-emitters, which implies that all have low equivalent width $W$\ albeit broad \oiiiopt\ and blueward asymmetries in both \hb\ and \oiii. 
The anti-correlation between \oiii\ BLUE peak shift and its FWHM is typical of compact, wind dominated narrow-line regions \citep[NLR, ][]{zamanov_kinematic_2002,marziani_blue_2016,coatman_c_2016,deconto-machado_high-redshift_2023}.
An important result is the lack of a strong correlation  ($r_\mathrm{P}\lesssim 0.3$) between radio-power and \oiii\ and \hb\ mass outflow rate  $\dot{M}$\ \hb.

\subsubsection{Principal component analysis}

\begin{figure}\hspace{-1cm}
    \center
	\includegraphics[width=0.45\textwidth]{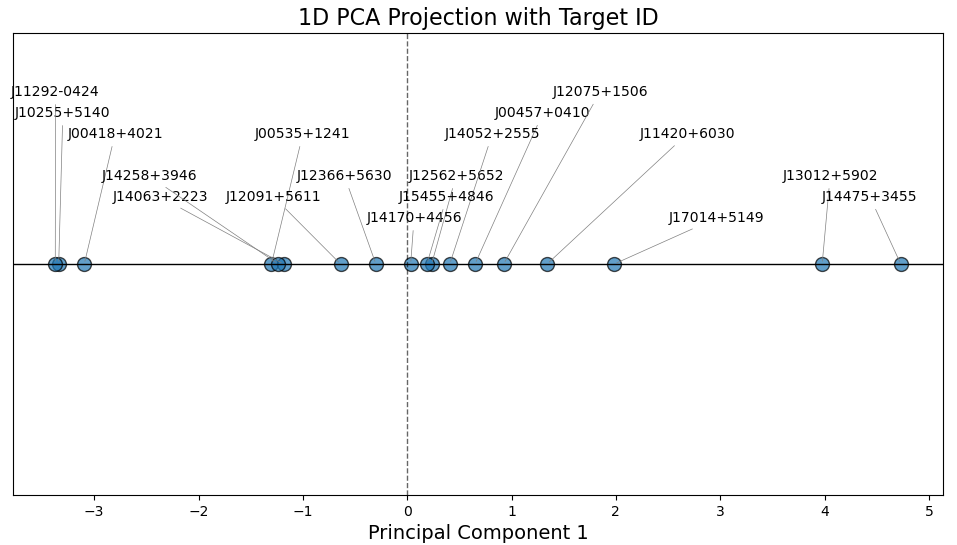} 
    \center
	\includegraphics[width=0.45\textwidth]{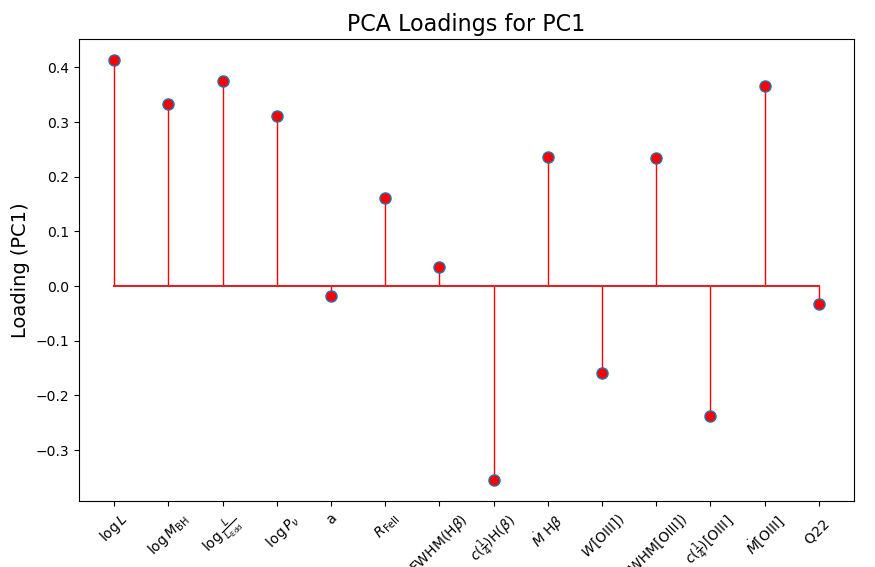}
    \caption{Results of the PCA analysis: top, 1D projection of sources along the PC1; bottom, vector loading along the PC1.  }
    \label{fig:pca}
\end{figure}

The Eigenvector 1 of quasars (E1) has been retrieved in samples where the \rfe\ range is $\approx 0 - 2$ \citep{boroson_emission-line_1992}, while in the present sample, the median value is \rfe$\approx 1.19$, with a 1st quartile at \rfe$\approx 1.09$, and only one object (J10255+5140) with \rfe $\approx 0.8$.
In lieu of the correlation between \rfe\ and \lledd\ \citep{du_fundamental_2016,panda_quasar_2019}, and of  \lledd\ varying little for \rfe$\gtrsim 1$ \citep{marziani_super-eddington_2025}, we cannot expect to detect the Eigenvector 1 in our sample. 
The PCA applied to our sample reveals only 1 eigenvector with a clear physical meaning. Fig. \ref{fig:pca} shows the projection of the targets, and the loads of each vector along principal component 1 (PC1). The PC1 carries $\approx$ 34 \%\ of the sample variance (a bit more than the original E1, but it is associated with trends due to luminosity, i.e., with the original E2 of \citet{boroson_emission-line_1992}. The left panel of Fig. \ref{fig:pca} shows the sources of our sample aligned along the PC1: the least and the most luminous ones are located at the opposite ends of the vector.

\section{Discussion}\label{discussion}

\subsection{Significance of the high \rfe\ and jetted super-Eddington candidates}

In our sample, we have found three
sources (J12091+5611, J14258+3946 and J17014+5149, all RL) for which their radio emission is dominantly coming from a core or jetted AGN.
As shown by Fig.~\ref{fig:rd}, the interpretation of the prevalence of radio-detected and genuinely jetted sources---starting from an optically selected sample---depends on the location along the MS. A rigorous assessment of the incidence of jetted sources would require homogeneous datasets and well-defined parent samples that are not readily available. As a pragmatic proxy, one may consider NLSy1s, even though only a minority of them satisfy the condition \rfe$~\gtrsim 1$. While the most powerful radio sources belong to the B spectral types, a population of lower-power, yet jetted, objects has been uncovered among NLSy1s \citep{komossa_radio-loud_2006,foschini_properties_2015,foschini_jetted_2020,Ojha_intra-night_2024}. About 7 per cent of NLSy1s are radio loud, with an even smaller fraction ($\sim$2.5 per cent) reaching very high radio loudness ($R_K>100$) \citep{orienti_investigating_2015}. Sources detected in the $\gamma$-ray domain are far fewer: $\lesssim 30$ are currently known \citep{foschini_jetted_2020,foschini_new_2022,barba_revised_2025}. Moreover, objects with reliable measurements (i.e., high S/N optical spectra) that simultaneously meet \rfe$~\gtrsim 1$ and are demonstrably jetted---as indicated by a high radio-loudness parameter, $\log R_\mathrm{K} \gtrsim 1.8$ \citep{zamfir_new_2008}---are extremely rare. A search in the recent catalog by \citet{paliya_narrow-line_2023} returns only a handful of candidates among $\sim$28,000 NLSy1s. Only a handful of the $\gamma$-ray emitters would be classified as xA. A couple of additional jetted candidates is reported in the sample of \citet{garnica_spectral_2025}, defined by \rfe$~\gtrsim 0.9$. In addition, some observational studies are reporting compact or jet-like radio emission in high-Eddington systems among eROSITA detected X-ray quasars
\citep[e.g.,][]{khorunzhev_discovery_2021,wolf_first_2021,Obuchi_discovery_2025}.

Super-Eddington accretors with relativistic ejections imply that thick, wind-dominated accretion flows can still generate and sustain magnetically driven, collimated jets. Coexistence is physically expected in a magnetically-arrested disk (MAD) regime because super-Eddington disks are geometrically thick and tend to create a low-density polar funnel while simultaneously launching a wide-angle wind from the disk surface \citep{mckinney_efficiency_2015,yang_properties_2023}.  The main threat to a highly relativistic jet is mass loading (baryons and radiation) in/near the funnel, which can reduce jet magnetization and terminal Lorentz factor \citep{mckinney_efficiency_2015}. The identification of jetted super-Eddington candidates is therefore of special relevance for a full understanding of the accretion process, and the sources identified in the the present work should be considered for dedicated studies.

\subsection{High-rate accretion and star formation}    
 {The analysis presented in the previous sections indicates that SF is among the emission mechanisms responsible for the radio emission of many sources of our sample. Indeed, we find that most have a radio spectral index corresponding with either SF or optically thin jet emission, but when considering the correlation between IR and radio properties, we determine a high prevalence of systems that have composite origin, or that are even SF dominated, showing the importance of a multi-wavelength analysis based on several indicators. }

We selected our sample on the basis on an optical selection criterion \rfe $\gtrsim 1$ which isolates spectral type A3/A4 in the 4DE1 formalism \citep{sulentic_eigenvector_2000}. This selection is independent of radio properties, avoiding any explicit bias toward radio-loud or radio-quiet AGN. Sources in the A3/A4 bins are generally interpreted as radiating at high Eddington ratios, a key driver of Eigenvector 1, and are known to show strong \feii\ emission, narrow Balmer lines, and weak but semi-broad — all hallmarks of high accretion states -- cognate to the coexistence of a virialized low-ionization BLR coexisting with an outflowing, mildly ionized system involving both broad and narrow lines \citep{collin-souffrin_broad-line_1988,elvis_structure_2000,marziani_broad-line_2010}. Our high Eddington ratio sample reveals a high prevalence of systems that have composite radio or are even SF dominated. Within this framework, our results confirm previous SDSS-based statistical findings: a significant fraction of the high-\rfe\ sources exhibit either composite radio emission or emission dominated by star formation  \citep{bonzini_star_2015,ganci_radio_2019}. Consistent results were also found by the analysis of NLSy1 samples \citep{sani_enhanced_2010,caccianiga_wise_2015}, as such sample involve a large fraction of highly accreting sources (even if not all NLSy1s are highly accreting sources, \citealt{marziani_main_2018}). The evidence supports the view that in the high Eddington ratio regime, the prevalence of powerful relativistic jets diminishes, while other forms of radio emission—such as compact or diffuse low-power radio structures associated with winds, disk instabilities, or star-forming activity—become more common.  The relative paucity of high radio power sources for low \mbh\ AGN  may ultimately support a mechanism in which the radio power scales with the \mbh, such as extraction of  energy from a spinning black hole threaded by magnetic fields \citep{blandford_electromagnetic_1977,Foschini_unification_2014,Foschini_power_2024}.

The E1 main sequence, even as it is observed at low redshifts, allows for an evolutionary interpretation \citep{fraix-burnet_phylogenetic_2017}: from the highly-accreting low mass AGN, to the evolved, massive black holes accreting at low rates down to minimum needed to sustain radiative-efficient accretion \citep{giustini_global_2019}.  Figure 7 of \citet{yue_novel_2025} presents cumulative \mbh\ distributions for AGN whose low-frequency radio is AGN- and star formation-dominated, broken down across a grid in luminosity and redshift.  A key result is the consistent offset: AGN-dominated quasars have systematically higher black hole masses than SF-dominated ones, across all bins.   In the context of a  sample selected via high \rfe $  \gtrsim 1$, which maps to spectral types A3/A4 and thus to high Eddington ratio sources, the results of \citet{yue_novel_2025} are especially informative:   sources with lower \mbh\ are expected to have relatively higher \lledd\  at fixed luminosity due to the inverse relation between $L/L_{\mathrm{Edd}}$ and $M_{\mathrm{BH}}$ for a given bolometric output. The star-formation-dominated quasars  likely corresponding to high-$L/L_{\mathrm{Edd}}$ systems in our sample  tend to be less massive than their jetted  counterparts. This behavior consistently extends from low-$z$\ to beyond the cosmic noon, and suggests a common pattern in the evolution of quasars and growth of their black holes, in which star formation, possibly associated with an enhancement of the amount of gas available in proximity of the active nucleus induced by a merging, is  coexisting in the early phases of the growth of the black hole. As the quasars evolves from xA to the extreme population B, the black hole mass increases and the circum-nuclear stellar system evolves, with some stars following accretion-modified stellar evolution pattern \citep[e.g.,][]{wang_molecular_2010,wang_estimating_2009,cantiello_stellar_2021,fabj_eccentric_2024}. In the end, the most massive black holes are expected to be surrounded by a stellar graveyard \citep{marziani_super-eddington_2025}.

\section{Conclusions}\label{conclusion}

We present new VLA observations at L, C and X-band of a sample of 18 quasars accreting at very high rates, which we combine with LoTSS and VLASS observations.
We identify the dominating mechanism behind the detected radio emission of these sources based on a combination of several indicators : radio variability, luminosity, morphology, radio spectral properties and radio-to-mid
IR flux ratio.
We find that 7 of our 18 targets are likely to have their radio emission dominantly coming from SF, and 6 from a combination of SF and AGN-related mechanism. Only three sources indicate a core or jetted AGN only origin for the detected radio emission. This is consistent with previous studies and suggests that, in the high–Eddington-ratio regime, radio emission is more commonly associated with lower-power structures related to star formation than with relativistic jets.
We discuss how this fits into a common evolutionary pattern of quasars, in which star formation is coexisting in the early phases of the growth of the black hole. A PCA analysis confirms the tendency toward homogeneous properties of the sample sources —consistent with a common spectral type —  with the first eigenvector being associated with luminosity-dependent effects. 

We add to the analysis a multi-component nonlinear decomposition of \hb\ and \oiiiopt\ lines, from which we derive estimation of accretion and wind parameters. While these sources are mostly candidate super-Eddington quasars, their  NLR and BLR winds only produce the modest feedback, especially at low-luminosity. We also estimate the Bremsstrahlung emission associated with the condition needed for formation of the \oiiiopt\ lines in the context of photoionization from the active nucleus, but find no clear evidence of its contribution to the total radio power in the present sample.

The upcoming Square Kilometre Array Observatory, with its unparalleled sensitivity and angular resolution, will allow to conduct similar studies and probe the evolution of these objects at much earlier times (e.g. \citealt{latif_radio_2024}).


\begin{acknowledgements}
MLGM and LVM acknowledge financial support from the grant CEX2021-001131-S funded by MICIU/AEI/10.13039/501100011033, from the grant PID2021-123930OB-C21 funded by MICIU/AEI/10.13039/501100011033 and by ERDF/EU. MLGM acknowledges financial support from NSERC via the Discovery grant program and the Canada Research Chair program. AdO and PM acknowledge financial support from the Spanish MCIU through project PID2022-140871NB-C21 by “ERDF A way of making Europe”, and from the Severo Ochoa grant CEX2021-515001131-S funded by MCIN/AEI/10.13039/501100011033.

LOFAR is the Low Frequency Array designed and constructed by ASTRON. It has observing, data processing, and data storage facilities in several countries, which are owned by various parties (each with their own funding sources), and which are collectively operated by the ILT foundation under a joint scientific policy. The ILT resources have benefited from the following recent major funding sources: CNRS-INSU, Observatoire de Paris and Université d'Orléans, France; BMBF, MIWF-NRW, MPG, Germany; Science Foundation Ireland (SFI), Department of Business, Enterprise and Innovation (DBEI), Ireland; NWO, The Netherlands; The Science and Technology Facilities Council, UK; Ministry of Science and Higher Education, Poland; The Istituto Nazionale di Astrofisica (INAF), Italy. This research has made use of the NASA/IPAC Extragalactic Database (NED),
which is operated by the Jet Propulsion Laboratory, California Institute of Technology,
under contract with the National Aeronautics and Space Administration. This research has made use of the NASA/IPAC Infrared Science Archive, which is funded by the National Aeronautics and Space Administration and operated by the California Institute of Technology.
\end{acknowledgements}

%
%

\bibliographystyle{aa}
\bibliography{aa57025-25} 

\begin{appendix}
\onecolumn
\section{Radio images, VLA observations setup and variability}

\vspace{-0.7cm}
\begin{figure*}[h!]
    \centering
    \includegraphics[width=0.96\textwidth]{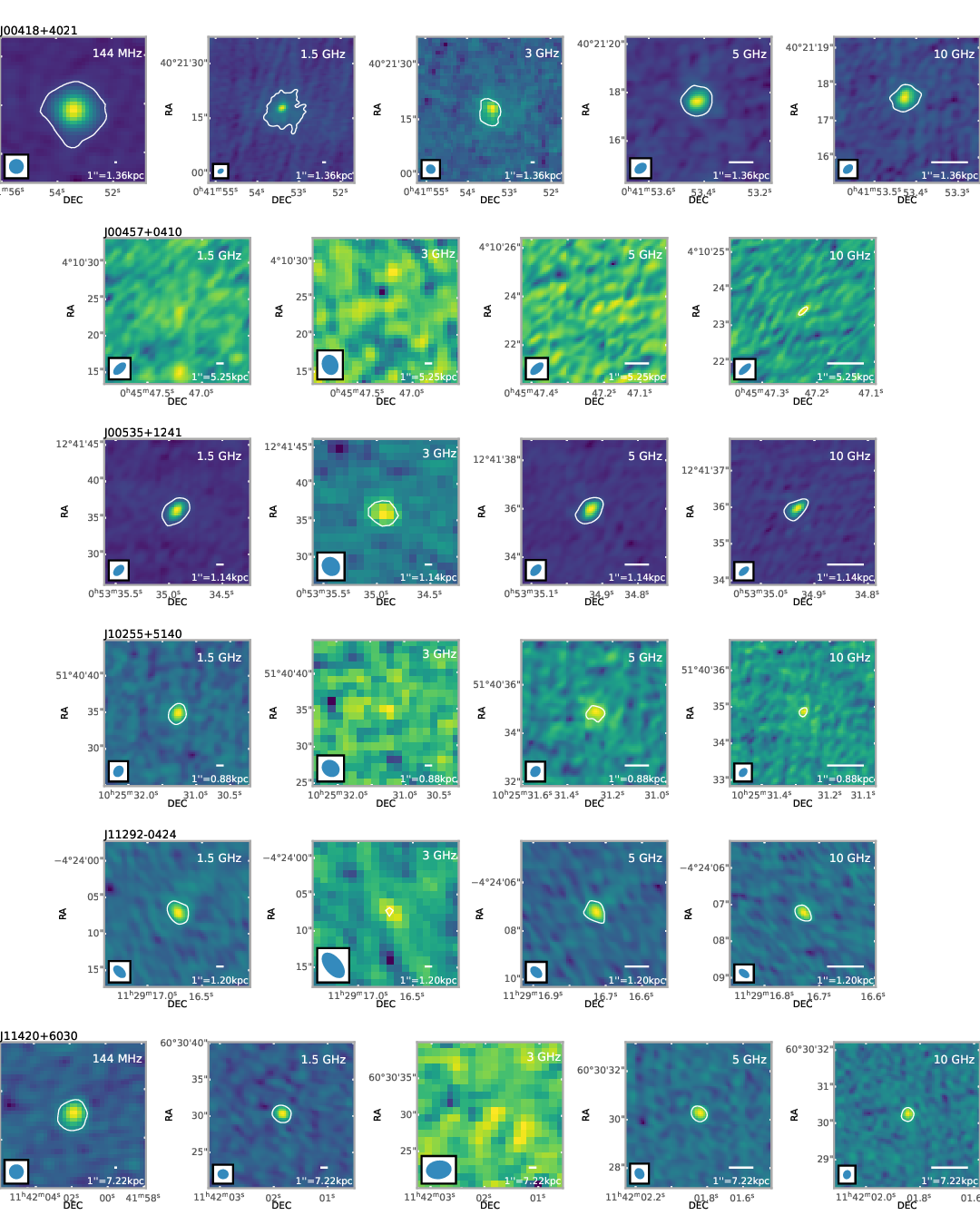}
    \caption{LOFAR (144 MHz) - when available - as well as our VLA L (1.5 GHz), C (5 GHz) and X (10 GHz) band images and VLASS (S-band, 3 GHz) of the 18 targets of the sample, each with a $5\sigma$ contour encompassing the main target in white. The sizes of the beams are shown in the bottom-left corner, while the scales are in the bottom-right corner. The cutouts have sizes of $1\arcmin\times1\arcmin$, $20\arcsec\times20\arcsec$,
    $20\arcsec\times20\arcsec$,
    $6\arcsec\times6\arcsec$ and $4\arcsec\times4\arcsec$ at 144 MHz, 1.5 GHz, 3 GHz, 5 GHz and 10 GHz respectively - except for sources with larger detected diffuse emission (J00418+4021, J12562+5652), which cutouts have been increased to encompass it. }
    \label{fig:fig_VLA_QSO}
\end{figure*}
\clearpage
\begin{figure*}
    \ContinuedFloat
    \centering
    \includegraphics[width=0.99\textwidth]{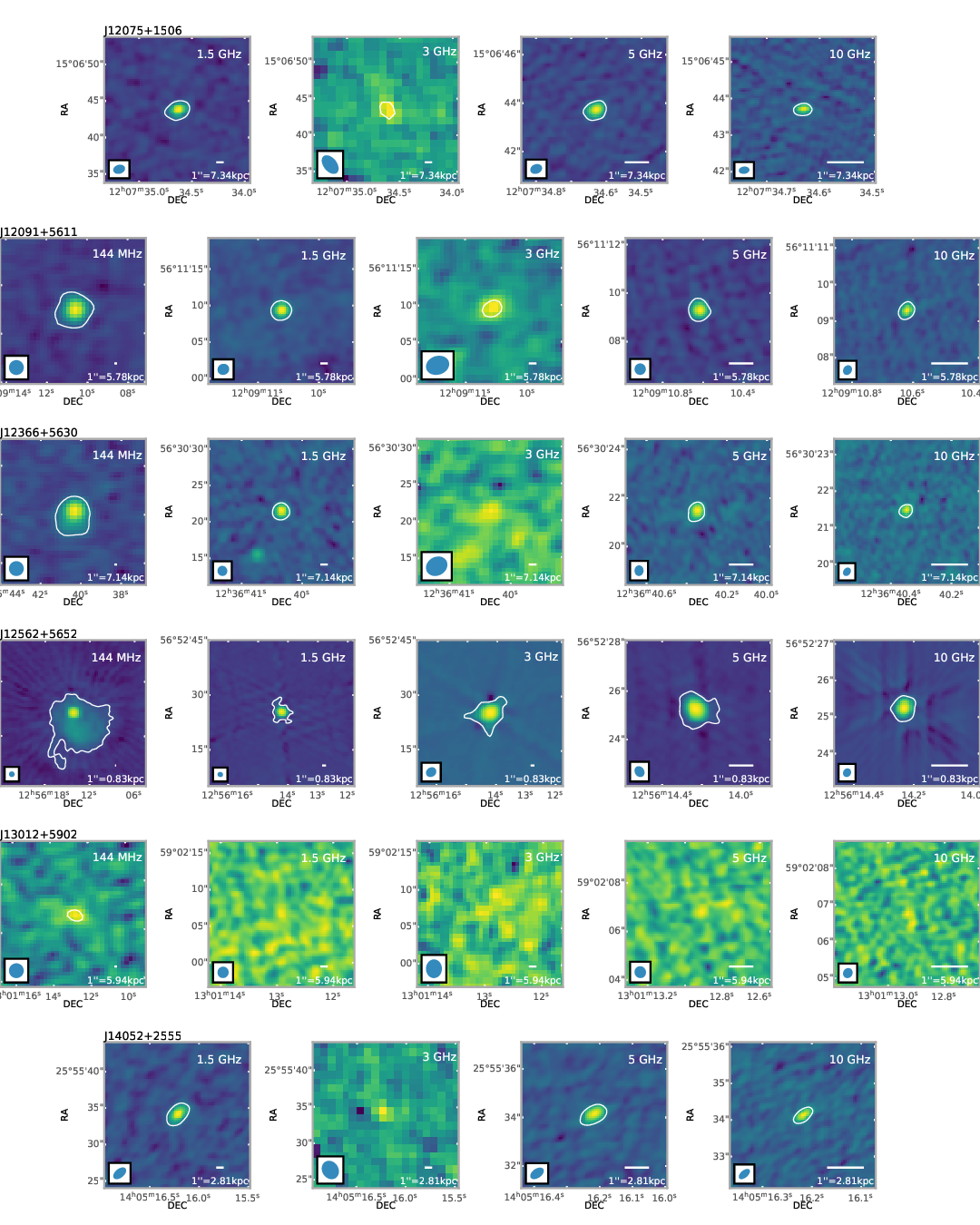}
    \caption{Continuation of Figure \ref{fig:fig_VLA_QSO}.}
\end{figure*}
\clearpage
\begin{figure*}
    \ContinuedFloat
    \centering
    \includegraphics[width=0.99\textwidth]{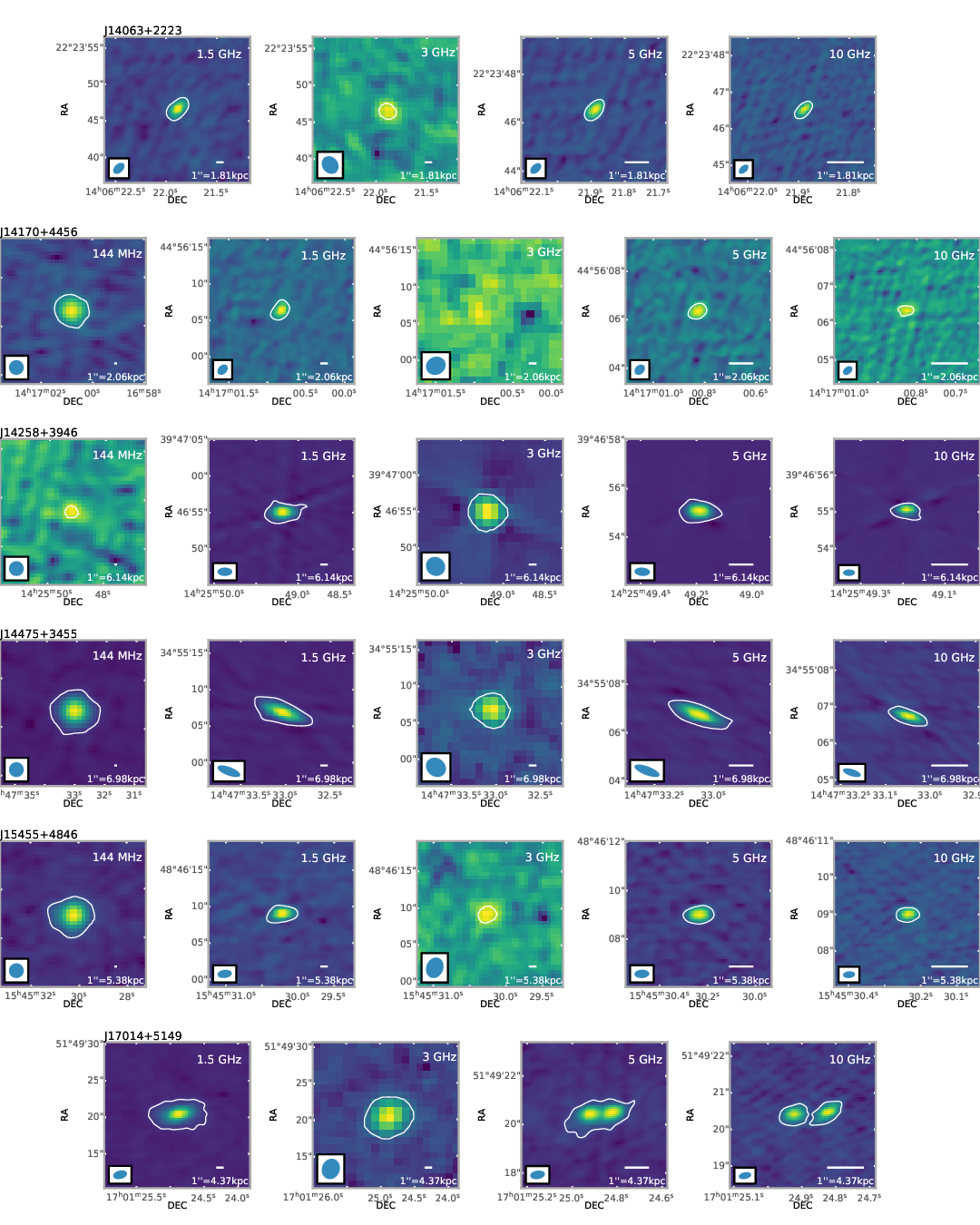}
    \caption{Continuation of Figure \ref{fig:fig_VLA_QSO}.}
\end{figure*}
\clearpage

\begin{figure*}[h!]
    \centering
	\includegraphics[width=.45\linewidth]{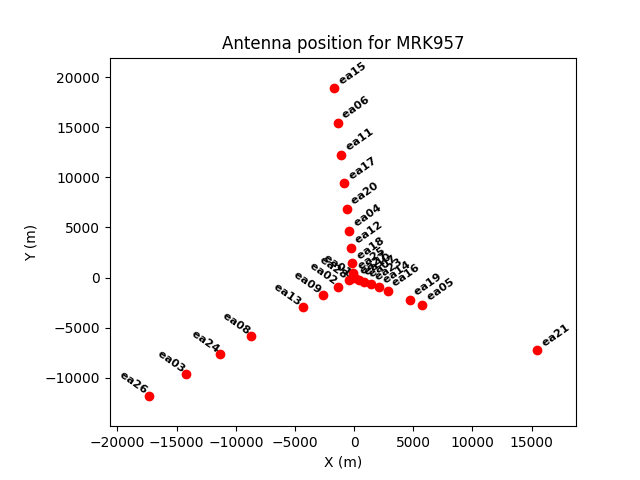}
    \includegraphics[width=.33\linewidth]{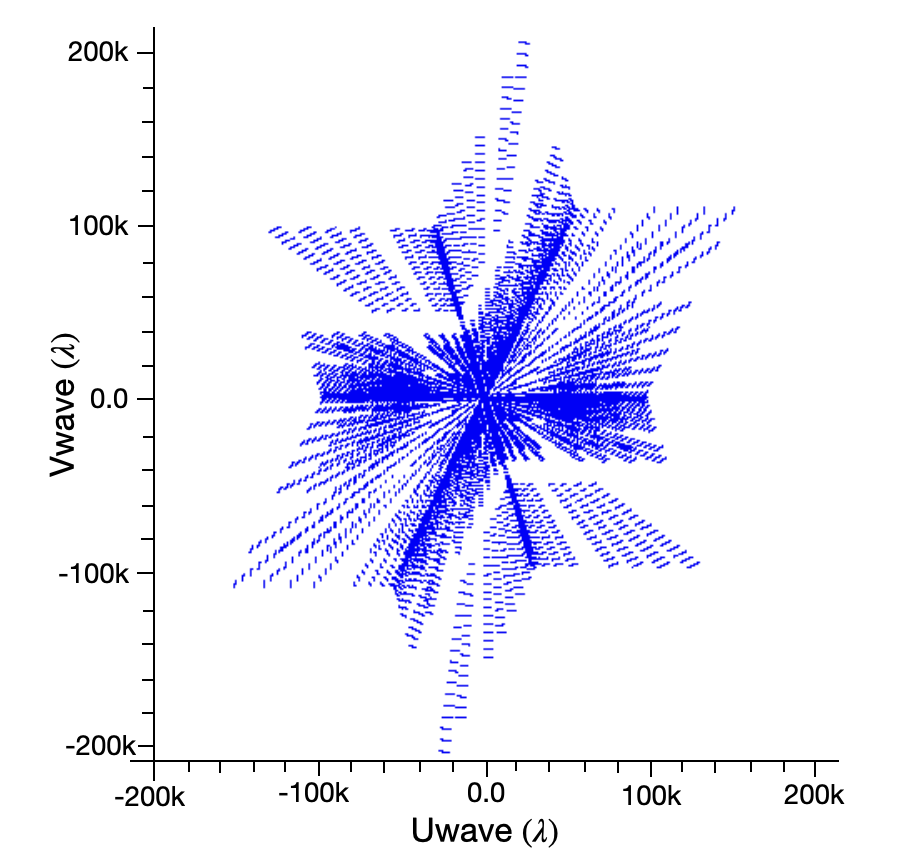}
    \caption{Antenna positions during the observations of J00418+4021 on 2020-12-04, and \textit{uv} plane coverage of the L-band scan (with an averaging on 128 channels).
    }
    \label{fig_plotants_MRK957}
\end{figure*}

\begin{figure*}[h!]
	\includegraphics[width=\textwidth]{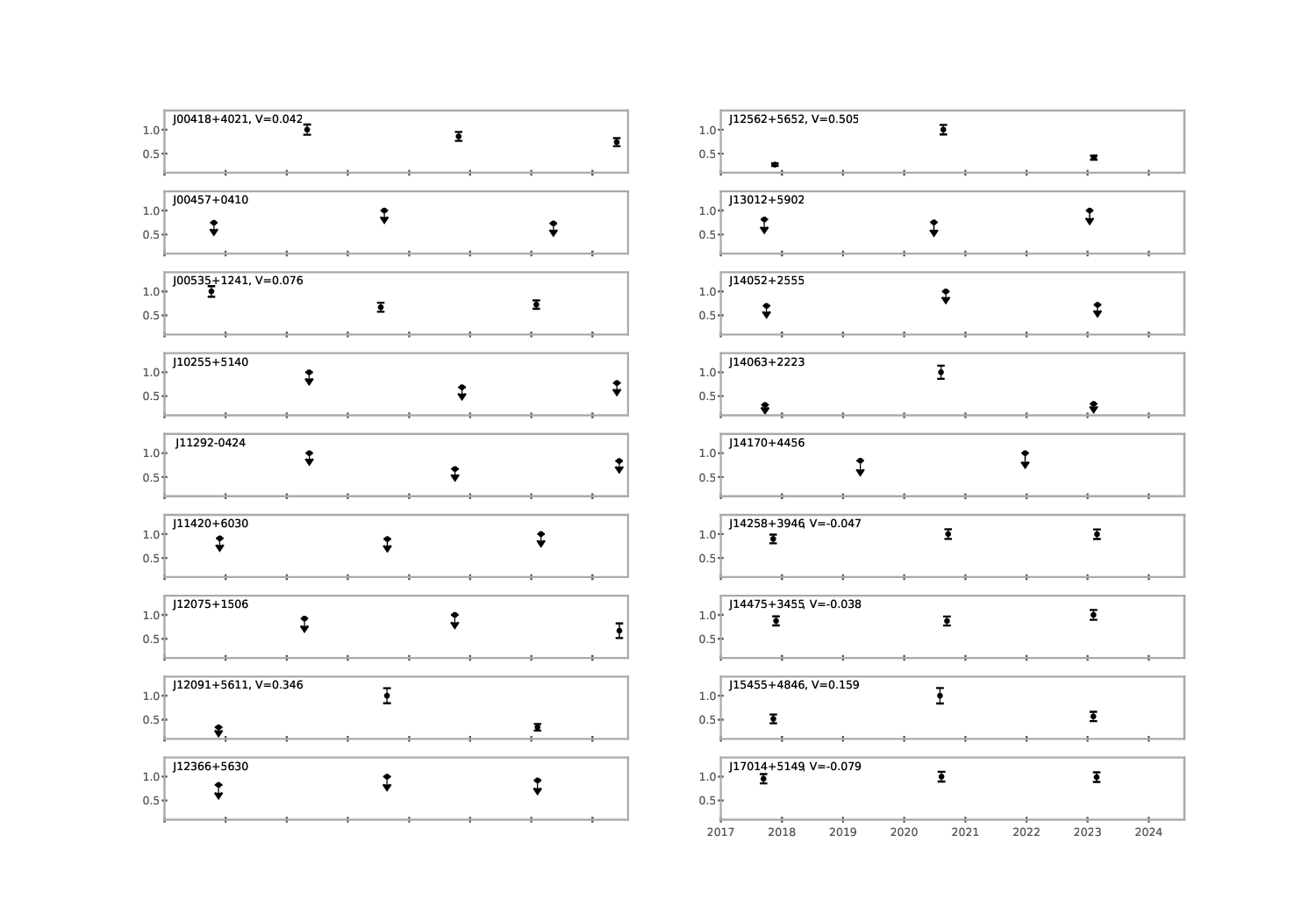}
    \vspace{-1.3cm}
    \caption{Sources variability across VLASS observations. Fluxes have been normalized by the maximum flux observed with VLASS, and the variability index $V$ is shown for sources with detections in at least two epochs.}
    \label{fig_VLASS_time_series}
\end{figure*}

\section{Radio images properties, measured radio flux densities and spectrum, spectral indices and radio luminosity, IR and FIR fluxes}

\begin{table*}
\centering\tabcolsep=3pt
\caption{Final radio images properties$^{*}$}
\label{table_images}
\begin{tabular}{lcccccccccccccccc}
\hline
\multirow{3}{*}{Object} &
\multicolumn{3}{c}{LOFAR 144 MHz} &
\multicolumn{3}{c}{VLA L-band 1.5 GHz} &
\multicolumn{4}{c}{VLASS S-band 3 GHz} &
\multicolumn{3}{c}{VLA C-band 5 GHz} &
\multicolumn{3}{c}{VLA X-band 10 GHz} \\
& {Beam} & {rms} & {DR} 
& {Beam} & {rms} & {DR} 
& {Epoch} & {Beam} & {rms} & {DR} 
& {Beam} & {rms} & {DR}  
& {Beam} & {rms} & {DR} \\
\midrule
 J00418+4021 & 6x6 & 100 &   20000    & 2x1 & 50 &    10000    &       2.2 \text{QL**}        & 3x2 & 100 &    2000    & 0.5x0.4 & 20  &     100     & 0.3x0.2 &  7  &     60      \\
 J00457+0410 &  -  &  -  &     -      & 2x1 & 30 &     800     &       2.1 \text{SE}        & 3x2 & 100 &    300     & 0.7x0.3 & 10  &     100     & 0.4x0.2 &  7  &     20      \\
 J00535+1241 &  -  &  -  &     -      & 2x1 & 30 &     100     &       2.1 \text{QL}        & 3x2 & 200 &    200     & 0.6x0.3 & 10  &     200     & 0.3x0.2 &  7  &     100     \\
 J10255+5140 &  -  &  -  &     -      & 2x1 & 20 &    1000     &       2.2 \text{QL}        & 3x2 & 100 &    1000    & 0.5x0.4 & 10  &     100     & 0.3x0.2 &  7  &     20      \\
 J11292-0424 &  -  &  -  &     -      & 2x1 & 30 &    5000     &       2.2 \text{QL}        & 4x2 & 100 &    3000    & 0.6x0.4 & 10  &     300     & 0.3x0.2 &  9  &     100     \\
 J11420+6030 & 6x6 & 50  &   100000   & 1x1 & 20 &    2000     &       2.1 \text{QL}        & 3x3 & 100 &    300     & 0.5x0.4 & 10  &     30      & 0.2x0.2 &  7  &     20      \\
 J12075+1506 &  -  &  -  &     -      & 2x1 & 30 &     800     &       2.2 \text{QL}        & 3x2 & 100 &    800     & 0.5x0.4 & 10  &     60      & 0.3x0.2 &  8  &     40      \\
 J12091+5611 & 6x6 & 60  &   70000    & 2x1 & 50 &     600     &       2.1 \text{SE}        & 3x3 & 200 &    700     & 0.5x0.5 & 10  &     100     & 0.3x0.2 &  8  &     50      \\
 J12366+5630 & 6x6 & 40  &   50000    & 1x1 & 30 &     500     &       2.1 \text{SE}        & 3x2 & 200 &    200     & 0.4x0.4 &  8  &    2000     & 0.2x0.2 &  8  &    2000     \\
 J12562+5652 & 6x6 & 200 &   60000    & 1x1 & 70 &    4000     &       2.1 \text{SE}        & 3x2 & 300 &    4000    & 0.5x0.4 & 200 &    4000     & 0.2x0.2 & 500 &    1000     \\
 J13012+5902 & 6x6 & 50  &   30000    & 2x2 & 30 &     200     &       2.1 \text{SE}        & 3x2 & 90  &    300     & 0.5x0.5 & 10  &     20      & 0.3x0.2 &  7  &     20      \\
 J14052+2555 &  -  &  -  &     -      & 2x1 & 20 &     600     &       2.1 \text{QL}        & 3x2 & 100 &    600     & 0.6x0.4 & 10  &     40      & 0.4x0.2 &  9  &     30      \\
 J14063+2223 &  -  &  -  &     -      & 2x1 & 30 &    4000     &       2.1 \text{SE}        & 3x2 & 100 &    400     & 0.5x0.3 & 10  &     60      & 0.3x0.2 &  7  &     60      \\
 J14170+4456 & 6x6 & 70  &   20000    & 2x1 & 20 &     200     &       2.2 \text{QL}        & 3x2 & 90  &    1000    & 0.5x0.4 & 10  &     200     & 0.3x0.2 &  8  &     60      \\
 J14258+3946 & 6x6 & 90  &   10000    & 2x1 & 50 &     300     &       2.1 \text{QL}        & 3x2 & 600 &    200     & 0.6x0.3 & 80  &     700     & 0.3x0.2 & 100 &     400     \\
 J14475+3455 & 6x6 & 200 &   20000    & 3x1 & 30 &     400     &       2.1 \text{QL}        & 3x2 & 100 &    200     &  1x0.3  & 10  &     300     & 0.5x0.2 & 10  &     100     \\
 J15455+4846 & 6x6 & 60  &   80000    & 2x1 & 30 &    5000     &       2.1 \text{SE}        & 3x2 & 100 &    800     & 0.6x0.3 &  9  &     100     & 0.3x0.2 &  7  &     60      \\
 J17014+5149 &  -  &  -  &     -      & 2x1 & 30 &    1000     &       2.1 \text{QL}        & 3x2 & 100 &    600     & 0.6x0.3 & 10  &     200     & 0.3x0.2 & 10  &     90      
 \\\hline
\end{tabular}
\tablefoot{* Beams are given in arcseconds, while the rms of the images are given in $\mu$Jy/beam. ** For VLASS images, QL stands for a Quick Look image data product, while SE stands for a Single Epoch continuum image data product.
}
\end{table*}

\begin{table*}
\centering\tabcolsep=3pt
\caption{Measured radio flux densities of the sources in mJy from the original images at their respective resolutions based on $5\sigma$ contours, their respective morphology type$^{*}$ and the calculated Kellermann’s parameter $R_{k}$.}
\label{table_flux}
\begin{tabular}{lccccccccccc}
\hline
\multirow{2}{*}{Object} &
\multicolumn{2}{c}{LOFAR 144 MHz} &
\multicolumn{2}{c}{VLA L-band 1.5 GHz} &
\multicolumn{2}{c}{VLASS S-band 3 GHz} &
\multicolumn{2}{c}{VLA C-band 5 GHz} &
\multicolumn{2}{c}{VLA X-band 10 GHz} &
\multirow{2}{*}{$R_{k}$} \\
& {Flux} & {Type}
& {Flux} & {Type}
& {Flux} & {Type} 
& {Flux} & {Type} 
& {Flux} & {Type} \\
\midrule
 J00418+4021 &   $84\pm8$    &     E     &   $19\pm1$    &     E      & $6.1\pm0.7$$^{\text{QL**}}$ &     E      & $1.66\pm0.09$ &     E      &  $0.85\pm0.05$  &     E    & 1.83  \\
 J00457+0410 &       -       &     -     &    $<0.1$     &     -      &   $<0.7$$^{\text{SE}}$    &     -      &    $<0.05$    &     -      & $0.050\pm0.008$ &     C      & $\lesssim$-0.50  \\
 J00535+1241 &       -       &     -     &  $5.2\pm0.3$  &     I      & $2.7\pm0.3$$^{\text{QL}}$ &     C      & $1.89\pm0.10$ &     I      &  $1.05\pm0.05$  &     C      & 0.07  \\
 J10255+5140 &       -       &     -     & $0.64\pm0.05$ &     I      &   $<0.5$$^{\text{QL}}$    &     -      & $0.12\pm0.01$ &     I      & $0.069\pm0.008$ &     C      & -0.24  \\
 J11292-0424 &       -       &     -     & $0.89\pm0.05$ &     C      & $0.8\pm0.2$$^{\text{QL}}$ &     C      & $0.35\pm0.02$ &     C      &  $0.19\pm0.01$  &     C      & -0.11  \\
 J11420+6030 &  $2.3\pm0.2$  &     I     & $0.91\pm0.05$ &     C      &   $<0.6$$^{\text{QL}}$    &     -      & $0.28\pm0.02$ &     C      &  $0.14\pm0.01$  &     C      & 1.15  \\
 J12075+1506 &       -       &     -     &  $2.2\pm0.1$  &     C      & $1.1\pm0.2$$^{\text{QL}}$ &     C      & $0.67\pm0.04$ &     C      &  $0.34\pm0.02$  &     C      & 0.81  \\
 J12091+5611 &  $8.2\pm0.8$  &     I     &  $2.1\pm0.1$  &     C      & $1.3\pm0.2$$^{\text{SE}}$ &     I      & $0.77\pm0.04$ &     C      &  $0.39\pm0.02$  &     C      & 1.71  \\
 J12366+5630 &  $4.4\pm0.4$  &     E     & $0.93\pm0.05$ &     C      &   $<0.8$$^{\text{SE}}$    &     -      & $0.32\pm0.02$ &     C      &  $0.20\pm0.01$  &     C      & 1.52  \\
 J12562+5652 &  $770\pm80$   &     E     &  $270\pm10$   &     I      & $1130\pm30$$^{\text{SE}}$ &     E      &  $660\pm30$   &     I      &   $670\pm30$    &     C      & 1.53  \\
 J13012+5902 & $0.38\pm0.06$ &     C     &    $<0.1$     &     -      &   $<0.5$$^{\text{SE}}$    &     -      &    $<0.05$    &     -      &     $<0.04$     &     -      & $\lesssim$-1.88  \\
 J14052+2555 &       -       &     -     & $1.02\pm0.06$ &     C      &   $<0.6$$^{\text{QL}}$    &     -      & $0.46\pm0.03$ &     I      &  $0.25\pm0.02$  &     C      & -1.14  \\
 J14063+2223 &       -       &     -     & $1.79\pm0.09$ &     C      & $1.5\pm0.2$$^{\text{SE}}$ &     I      & $0.65\pm0.03$ &     C      &  $0.37\pm0.02$  &     C      & -0.34  \\
 J14170+4456 &  $4.7\pm0.5$  &     E     & $0.79\pm0.04$ &     C      &   $<0.5$$^{\text{QL}}$    &     -      & $0.21\pm0.02$ &     I      & $0.089\pm0.010$ &     C      & -0.94  \\
 J14258+3946 &  $0.6\pm0.1$  &     C     & $15.0\pm0.8$  &     C      &  $40\pm4$$^{\text{QL}}$   &     I      &   $55\pm3$    &     C      &    $51\pm3$     &     C      & 2.99  \\
 J14475+3455 &  $110\pm10$   &     E     & $13.4\pm0.7$  &     C      & $6.0\pm0.6$$^{\text{QL}}$ &     C      &  $3.3\pm0.2$  &     C      &  $1.46\pm0.07$  &     C      & 1.34  \\
 J15455+4846 &   $13\pm1$    &     E     &  $2.4\pm0.1$  &     C      & $1.4\pm0.2$$^{\text{SE}}$ &     I      & $0.91\pm0.05$ &     C      &  $0.50\pm0.03$  &     I      & 1.08  \\
 J17014+5149 &       -       &     -     &   $23\pm1$    &     E      &  $12\pm1$$^{\text{QL}}$   &     I      &  $6.0\pm0.3$  &     E      &   $2.8\pm0.1$   &     E      & 1.24  \\

\hline
\end{tabular}
\tablefoot{* C indicates compact, I is intermediate morphology, and E indicates the presence of diffuse emission. ** For VLASS images, QL stands for a Quick Look image data product, while SE stands for a Single Epoch continuum image data product. The parameter $R_\mathrm{K}$ has been computed as the ratio between the K-corrected flux at 1.4GHz and the optical one at $5100$ \AA. 
}
\end{table*}

\begin{table*}
\centering\tabcolsep=4pt
\caption{IR and FIR fluxes used during the analysis.} 
\begin{tabular}{l c c c c c c c c } 
\hline
\multicolumn{1}{c}{Object} &  
\multicolumn{1}{c}{$S_{60\mu m}$} & 
\multicolumn{1}{c}{$L_{60\mu m}$}   & 
\multicolumn{1}{c}{$S_{100\mu m}$} & 
\multicolumn{1}{c}{$L_{100\mu m}$}   & 
\multicolumn{1}{c}{$L$(FIR)}   & 
\multicolumn{1}{c}{$M_{W4}$} & 
\multicolumn{1}{c}{$S_{22}$}  \\ 
\multicolumn{1}{c}{} &
\multicolumn{1}{c}{[mJy]} &
\multicolumn{1}{c}{[erg s$^{-1}$]} &
\multicolumn{1}{c}{[mJy]} &
\multicolumn{1}{c}{[erg s$^{-1}$]} &
\multicolumn{1}{c}{[erg s$^{-1}$]} &
\multicolumn{1}{c}{} &
\multicolumn{1}{c}{[mJy]}   \\
 \hline
J00418+4021 & 2100 $\pm$ 170  & 1.16 $\pm$ 0.09 E+45  & 3200 $\pm$ 1000  & 1.8 $\pm$ 0.5 E+45  & 5 $\pm$ 2 E+45  & 3.96 $\pm$ 0.02  & 219 $\pm$ 4  \\
J00457+0410 & < 200  & < 3 E+45  & < 300  & < 6 E+45  & < 4 E+45  & 6.19 $\pm$ 0.09  & 28 $\pm$ 2  \\
J00535+1241 & 2160 $\pm$ 50  & 8.2 $\pm$ 0.2 E+44  & 2960 $\pm$ 50  & 1.12 $\pm$ 0.2 E+44  & 3.9 $\pm$ 0.2 E+45  & 2.37 $\pm$ 0.02  & 940 $\pm$ 20  \\
J10255+5140 & 150 $\pm$ 40  & 3.4 $\pm$ 0.9 E+43  & 200 $\pm$ 100  & 4 $\pm$ 2 E+43  & 2 $\pm$ 2 E+44  & 5.57 $\pm$ 0.03  & 50 $\pm$ 2  \\
J11292-0424 & 670 $\pm$ 30  & 2.8 $\pm$ 0.1 E+44  & 1200 $\pm$ 100  & 4.9 $\pm$ 0.6 E+44  & 1.4 $\pm$ 0.2 E+45  & 3.71 $\pm$ 0.02  & 275 $\pm$ 6  \\
J11420+6030 & - & - & - & - & - & 7.7 $\pm$ 0.2  & 7 $\pm$ 1  \\
J12075+1506 & - & - & - & - & - & 6.46 $\pm$ 0.06  & 22 $\pm$ 1  \\
J12091+5611 & - & - & - & - & - & 9.2 $\pm$ 0.5  & 1.8 $\pm$ 0.9  \\
J12366+5630 & - & - & - & - & - & 8.5 $\pm$ 0.2  & 3.5 $\pm$ 0.7  \\
J12562+5652 & 32000 $\pm$ 2000  & 6.2 $\pm$ 0.3 E+45  & 30000 $\pm$ 1000  & 5.9 $\pm$ 0.2 E+45  & 3.5 $\pm$ 0.5 E+46  & 0.27 $\pm$ 0.01  & 6530 $\pm$ 70  \\
J13012+5902 & 30 $\pm$ 50  & 1 $\pm$ 2 E+45  & - & - & 8 $\pm$ 200 E+44  & 6.40 $\pm$ 0.05  & 23 $\pm$ 1  \\
J14052+2555 & 230 $\pm$ 50  & 7 $\pm$ 2 E+44  & - & - & 1$\pm$ 4 E+45  & 4.65 $\pm$ 0.03  & 115 $\pm$ 3  \\
J14063+2223 & < 100  & < 1.5 E+44  & < 100  & < 1.3 E+44  & < 4.0 E+44  & 6.04 $\pm$ 0.05  & 32 $\pm$ 1  \\
J14170+4456 & 110 $\pm$ 30  & 1.6 $\pm$ 0.5 E+44  & 150 $\pm$ 70  & 2$\pm$ 1 E+44  & 4$\pm$ 4 E+44  & 5.28 $\pm$ 0.03  & 65 $\pm$ 2  \\
J14258+3946 & - & - & - & - & - & 8.4 $\pm$ 0.3  & 3.6 $\pm$ 0.9  \\
J14475+3455 & < 200  & < 1.5 E+46  & < 600  & < 4.1 E+46  & < 1.4 E+46  & 5.37 $\pm$ 0.03  & 59 $\pm$ 1  \\
J15455+4846 & 350 $\pm$ 30  & 7.0 $\pm$ 0.5 E+45  & 370 $\pm$ 80  & 7.0 $\pm$ 2 E+45  & 7$\pm$ 2 E+45  & 4.72 $\pm$ 0.02  & 109 $\pm$ 3  \\
J17014+5149 & 480 $\pm$ 40  & 4.8 $\pm$ 0.4 E+45  & 400 $\pm$ 100  & 4 $\pm$ 1 E+45  & 6 $\pm$ 2 E+45  & 4.09 $\pm$ 0.02  & 193 $\pm$ 3  \\
\\\hline
\end{tabular}
\tablefoot{Column 1: Object coordinate name. 
Column 2: $60 \mu m$ flux densities from IRAS or ISO data.
Column 3: $60 \mu m$ luminosity derived from equation \ref{eqn:L60}.
Column 4: $100 \mu m$ flux densities from IRAS or ISO data.
Column 5: $100 \mu m$ luminosity derived from equation \ref{eqn:L60}.
Column 6: $L$(FIR) defined from a linear combination of the 60$\mu$m and 100$\mu$m fluxes according to \citealt{helou_thermal_1985}.
Column 7: WISE band 4 ($22 \mu m$) magnitude. 
Column 8: $22 \mu m$ flux densities derived from Column 7 magnitudes, using $F_{\nu_{0}} \times 10^{-M_{W4}/2.5}$, where $F_{\nu_{0}} = 8.363$Jy \citep{jarrett_Spitzer-WISE_2011} is the zero magnitude flux density in band 4 corresponding to the constant that gives the same response as that of Vega.
} 
\label{tab:IR-FIR}
\end{table*}

\begin{table*}
\begin{center}
\caption{Measured spectral indices and radio luminosity of the sources}
\label{table_alpha}
\begin{tabular}{lccccccc}
\hline 
 Object & $\alpha_{0.144-1.5}$ & $\alpha_{1.5-3}$ & $\alpha_{3-5}$ & $\alpha_{5-10}$ & $\alpha$  & $L_{1.4 GHz}$ & $p\text{SFR}_{\text{radio}}$ \\
        &                      &                  &                  &                 &            & [W Hz$^{-1}$] & $M_{\odot} yr^{-1}$\\
\hline
 J00418+4021 & $-0.63\pm0.05$  & $-1.6\pm0.2$  &  $-2.3\pm0.2$  &  $-1.0\pm0.1$  &  $-1.1\pm0.2$  &      1.02E+23       & 2.72E+01 \\
 J00457+0410 &        -        &       -       &       -        &     $>0.0$     &       -        &          -          & -        \\
 J00535+1241 &        -        & $-0.9\pm0.2$  &  $-0.6\pm0.2$  &  $-0.9\pm0.1$  & $-0.84\pm0.03$ &      4.55E+22       & 1.40E+01 \\
 J10255+5140 &        -        &    $<-0.4$    &     $>-3$      &  $-0.9\pm0.2$  &  $-1.2\pm0.1$  &      3.18E+21       & 1.57E+00 \\
 J11292-0424 &        -        & $-0.2\pm0.4$  &  $-1.5\pm0.5$  &  $-0.9\pm0.1$  & $-0.81\pm0.05$ &      8.88E+21       & 3.66E+00 \\
 J11420+6030 & $-0.40\pm0.04$  &    $<-0.6$    &     $>-1$      &  $-1.1\pm0.2$  &  $-0.7\pm0.1$  &      1.50E+24       & 2.49E+02 \\
 J12075+1506 &        -        & $-1.0\pm0.3$  &  $-0.9\pm0.3$  &  $-1.0\pm0.1$  & $-0.98\pm0.02$ &      6.24E+24       & 8.06E+02 \\
 J12091+5611 & $-0.58\pm0.05$  & $-0.7\pm0.2$  &  $-0.9\pm0.3$  &  $-1.1\pm0.1$  & $-0.74\pm0.06$ &      1.46E+24       & 2.43E+02 \\
 J12366+5630 & $-0.66\pm0.05$  &    $<-0.2$    &     $>-2$      &  $-0.7\pm0.1$  & $-0.74\pm0.03$ &      1.65E+24       & 2.69E+02 \\
 J12562+5652 & $-0.45\pm0.05$  & $2.06\pm0.07$ & $-0.97\pm0.09$ & $0.02\pm0.10$  &  $0.5\pm0.3$   &      1.33E+24       & 2.26E+02 \\
 J13012+5902 &     $<-0.6$     &       -       &       -        &       -        &       -        &          -          & -        \\
 J14052+2555 &        -        &    $<-0.8$    &    $>-0.5$     &  $-0.9\pm0.2$  & $-0.72\pm0.07$ &      8.01E+22       & 2.24E+01 \\
 J14063+2223 &        -        & $-0.3\pm0.2$  &  $-1.5\pm0.3$  &  $-0.9\pm0.1$  & $-0.83\pm0.06$ &      4.64E+22       & 1.43E+01 \\
 J14170+4456 & $-0.76\pm0.05$  &    $<-0.7$    &     $>-2$      &  $-1.3\pm0.2$  & $-0.92\pm0.09$ &      2.43E+22       & 8.38E+00 \\
 J14258+3946 &  $1.37\pm0.07$  &  $1.4\pm0.2$  &  $0.6\pm0.2$   &  $-0.1\pm0.1$  &  $1.0\pm0.2$   &      6.80E+24       & 8.66E+02 \\
 J14475+3455 & $-0.90\pm0.04$  & $-1.2\pm0.2$  &  $-1.1\pm0.2$  &  $-1.3\pm0.1$  & $-1.04\pm0.05$ &      2.64E+25       & 2.64E+03 \\
 J15455+4846 & $-0.72\pm0.04$  & $-0.8\pm0.2$  &  $-0.8\pm0.3$  &  $-0.9\pm0.1$  & $-0.77\pm0.02$ &      1.35E+24       & 2.28E+02 \\
 J17014+5149 &        -        & $-0.9\pm0.1$  &  $-1.2\pm0.2$  & $-1.18\pm0.09$ & $-1.11\pm0.03$ &      7.19E+24       & 9.06E+02 \\
\hline
\end{tabular}
\tablefoot{Column 1: Object coordinate name.
    Column 2: spectral slope at 0.144–1.5 GHz. 
    Column 3: spectral slope at 1.5–3 GHz. 
    Column 4: spectral slope at 3–5 GHz. 
    Column 5: spectral slope at 5–10 GHz. 
    Column 6: spectral slope fit over the whole frequency range available for sources with detection in at least three bands. 
    Column 7: Calculated radio luminosity in W Hz$^{-1}$ at 1.4 GHz using equation~\ref{radio_lum}.
    Column 8: Pseudo radio SFR in $M_{\odot} yr^{-1}$.
}
\end{center}
\end{table*}

\begin{figure*}
	\includegraphics[width=0.98\textwidth]{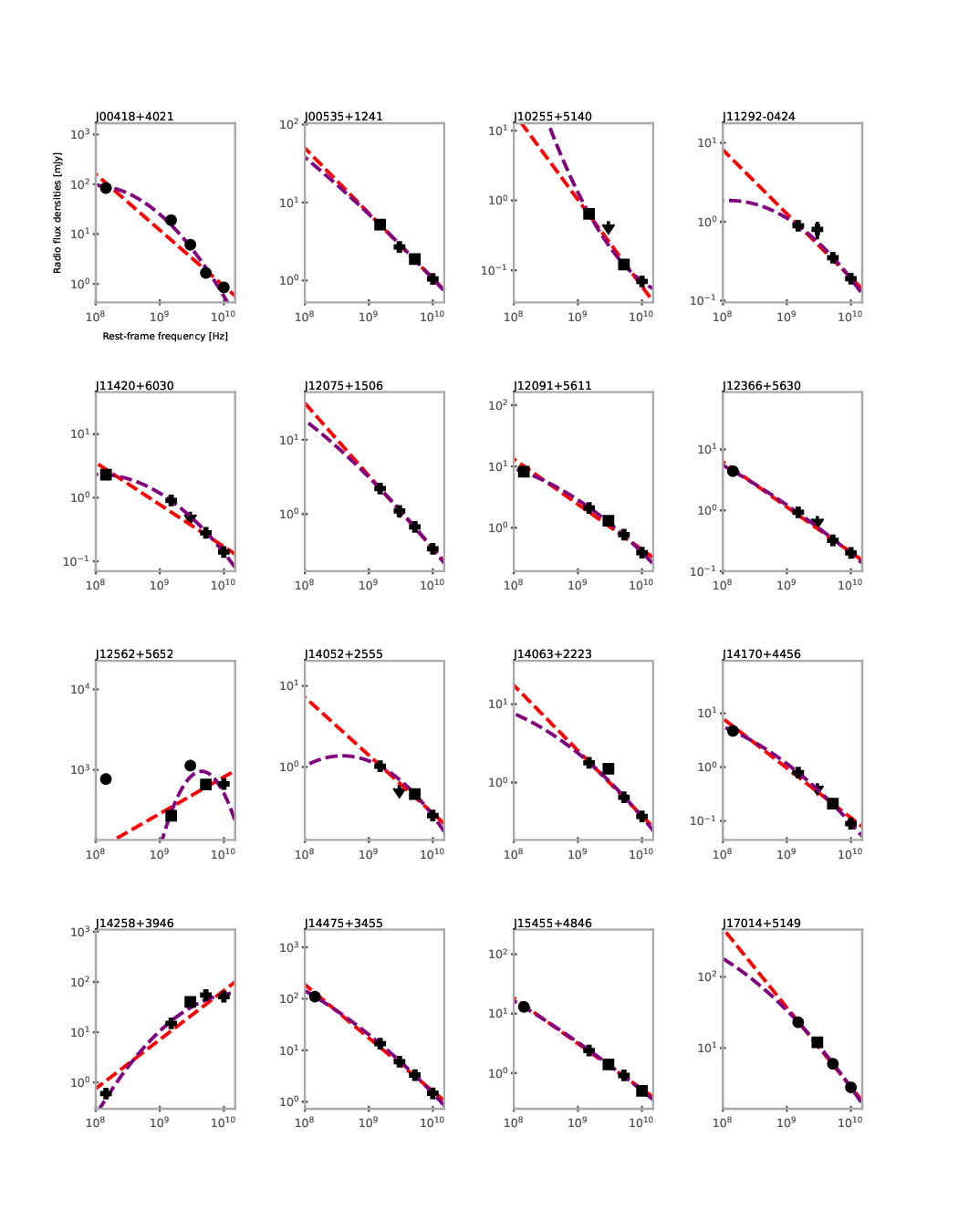}
    \caption{The radio spectra of the 16 sources detected in three bands or more. The dashed red line indicate the linear fit while the dashed purple line present the parabolic fit. Upper limits are shown but have been excluded during fitting procedure. The marker shape indicate the morphology of the source at the given frequency : circle for extended, square for intermediates and cross for compact sources.}
    \label{fig_spectrun}
\end{figure*}

\twocolumn

\section{Optical spectroscopy}\label{optical_spectro}

\begin{figure*}
\center
 \includegraphics[width=0.425\columnwidth]{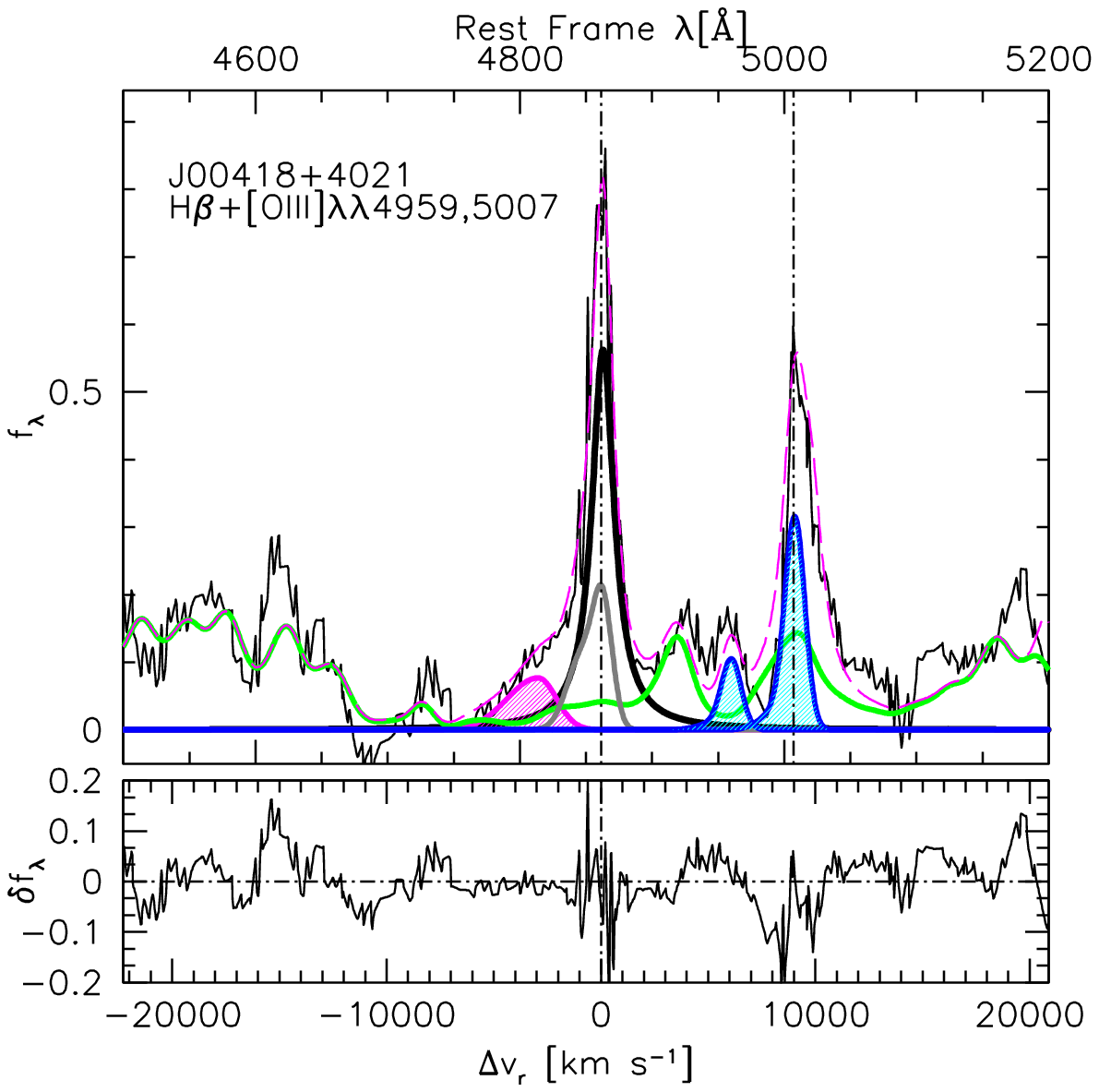} 
 \includegraphics[width=0.425\columnwidth]{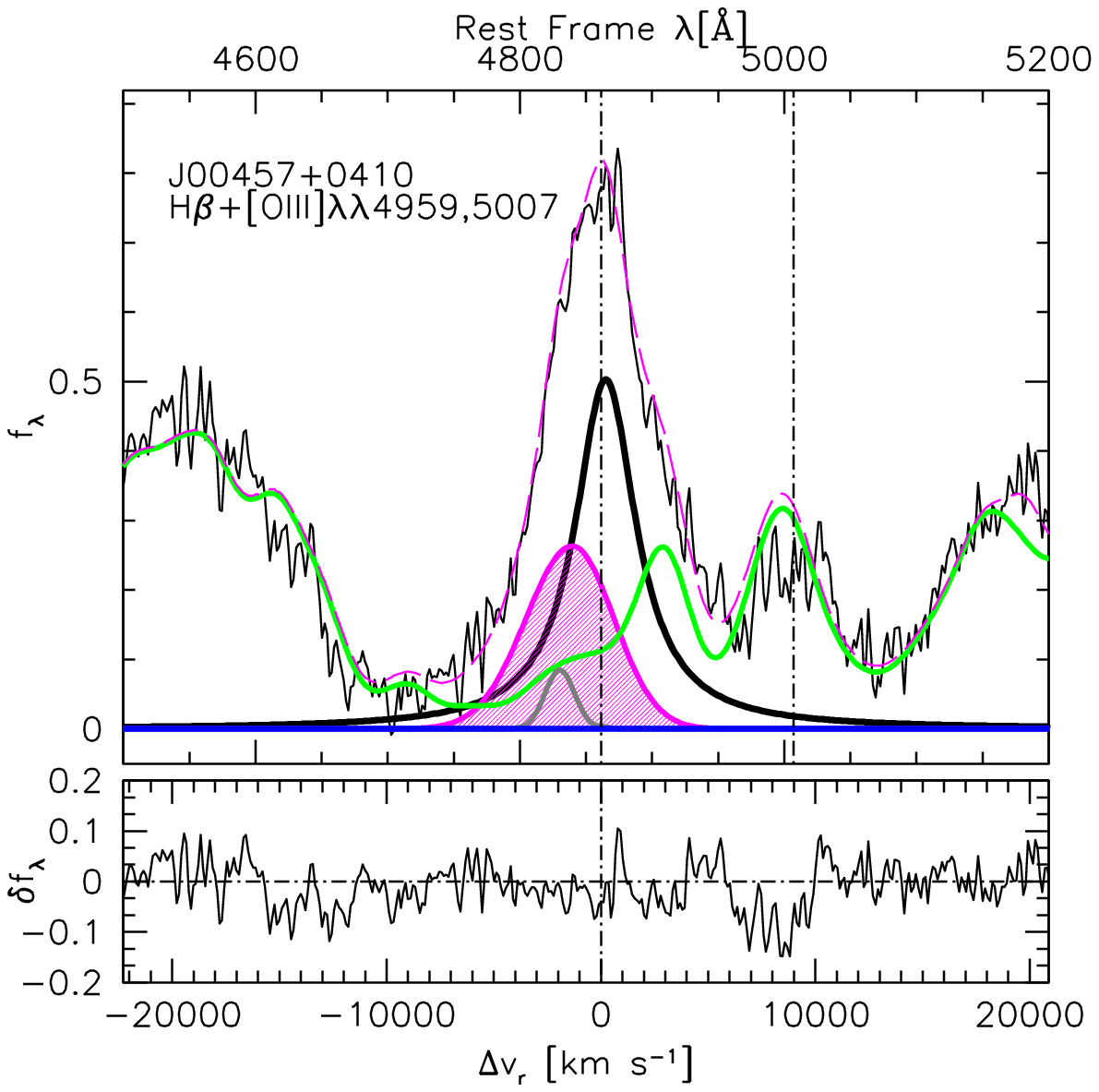}
 \includegraphics[width=0.425\columnwidth]{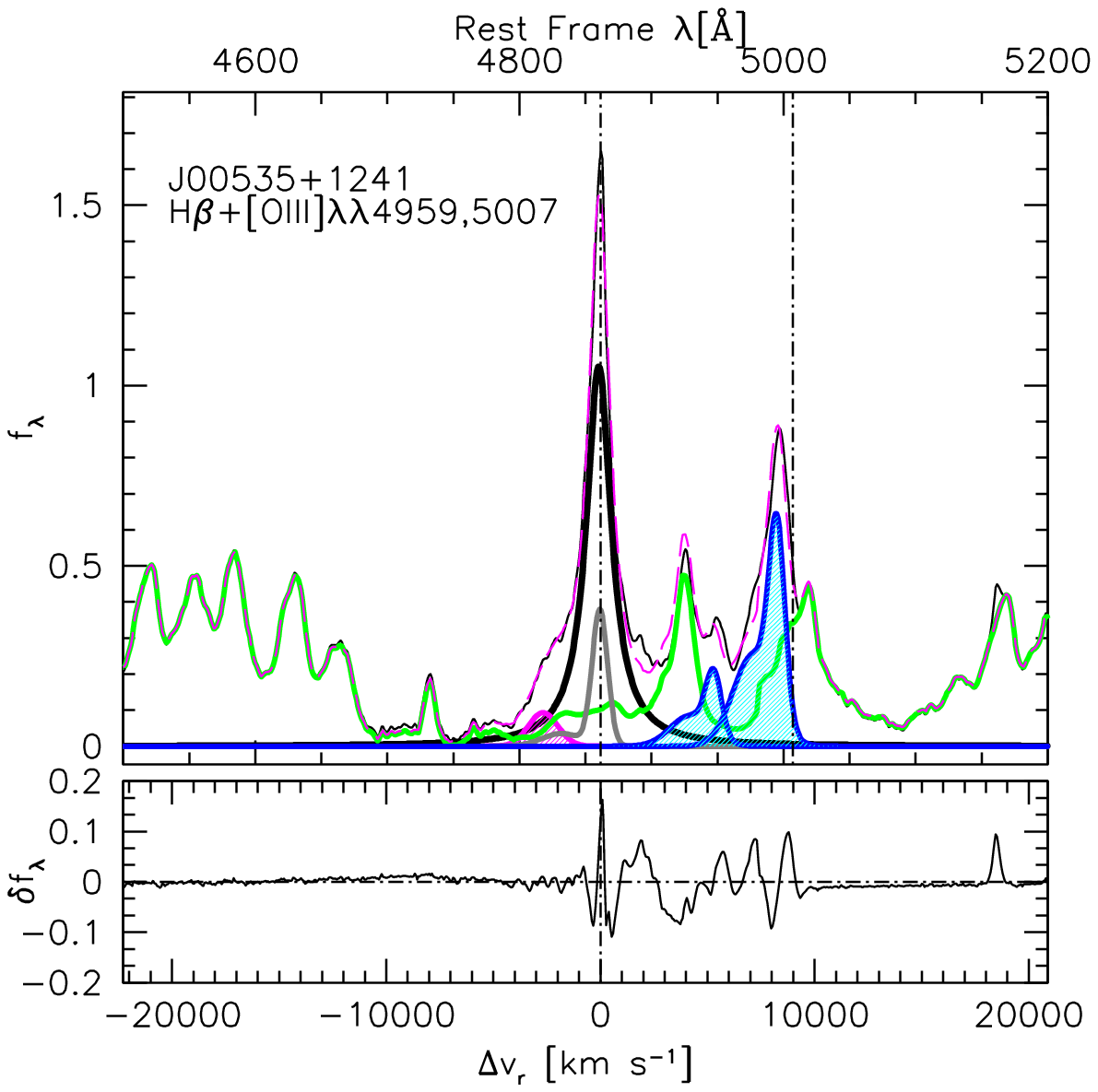}   
 \includegraphics[width=0.425\columnwidth]{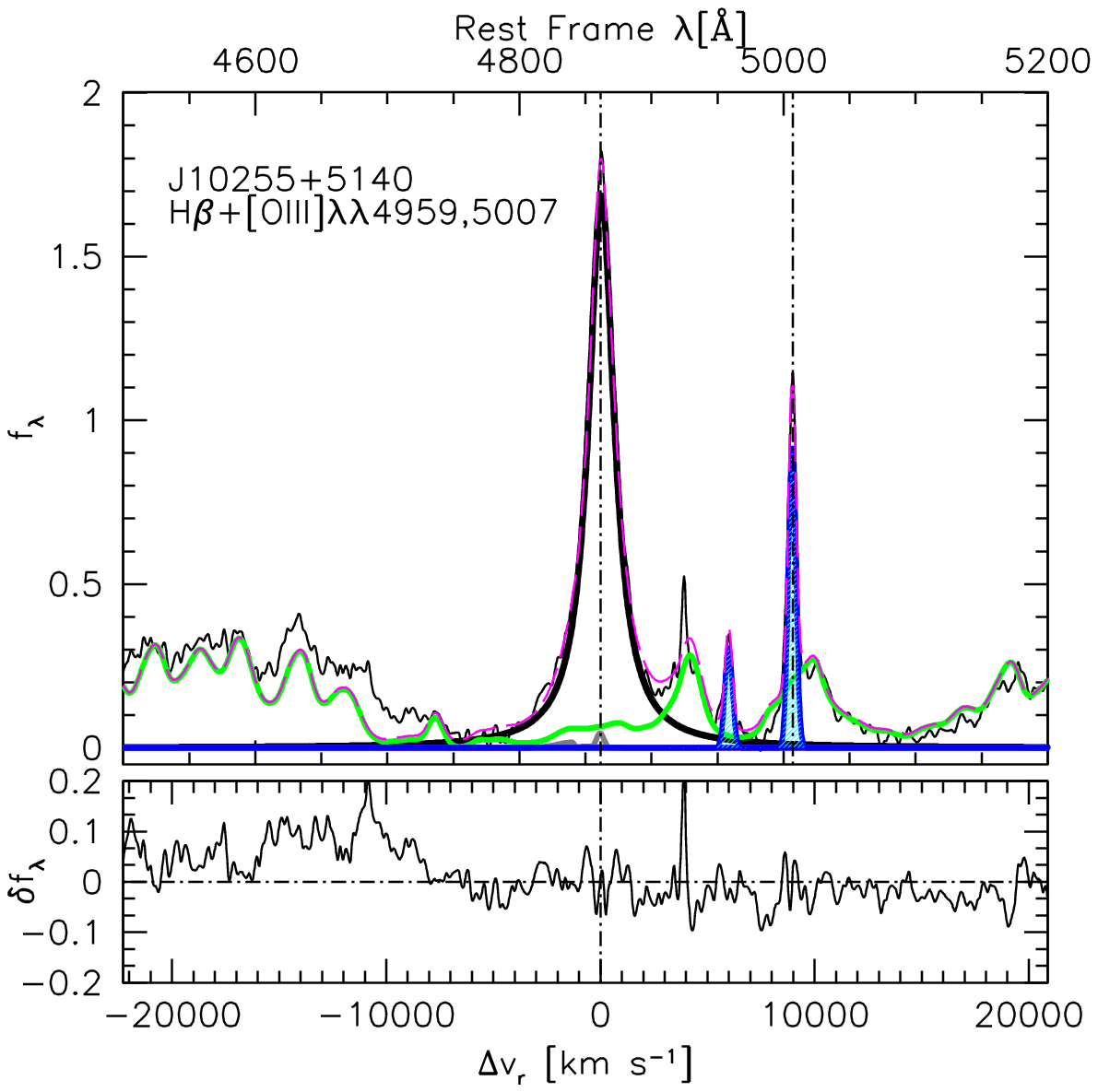}\\
 \includegraphics[width=0.425\columnwidth]{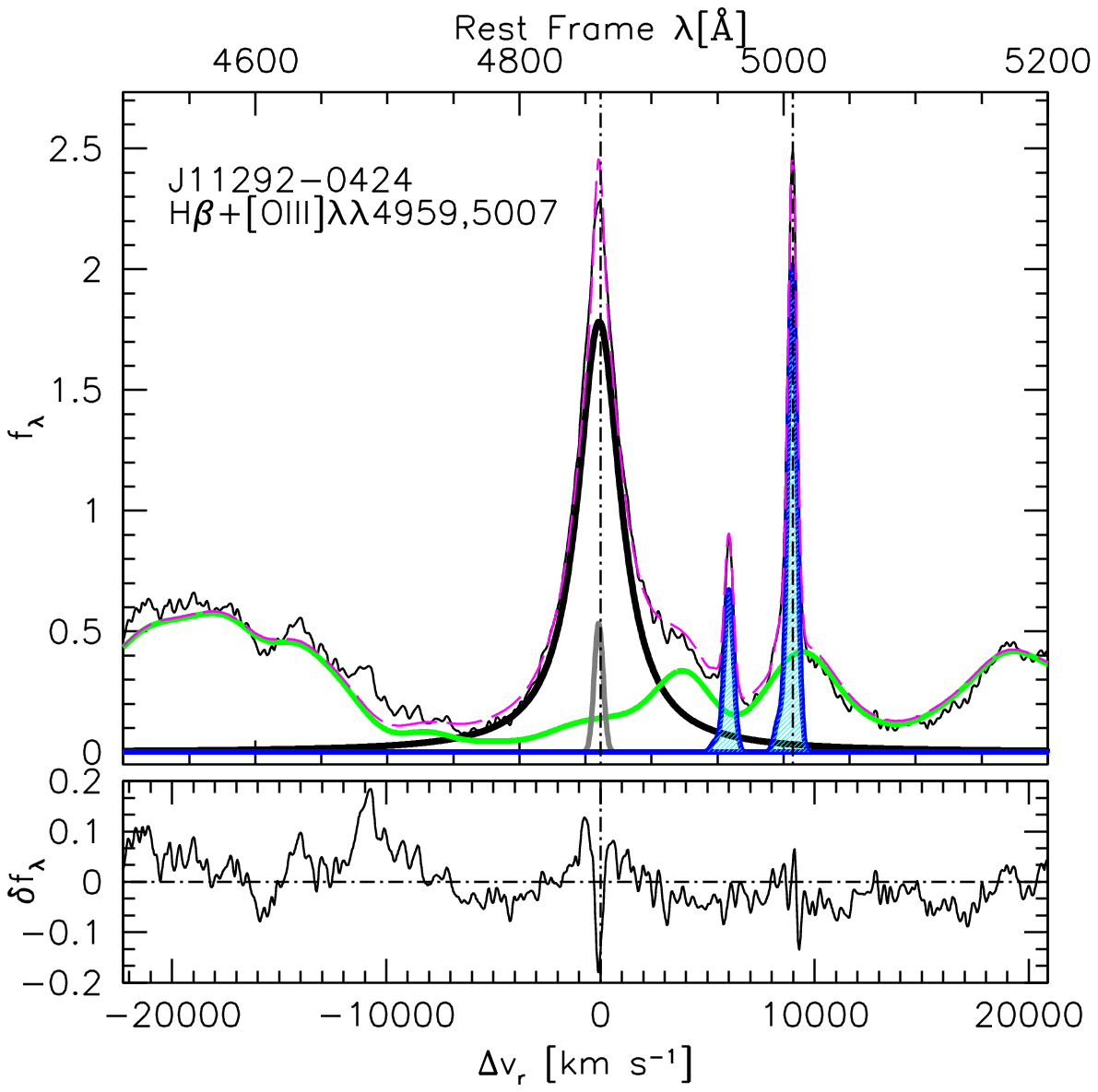}  
 \includegraphics[width=0.425\columnwidth]{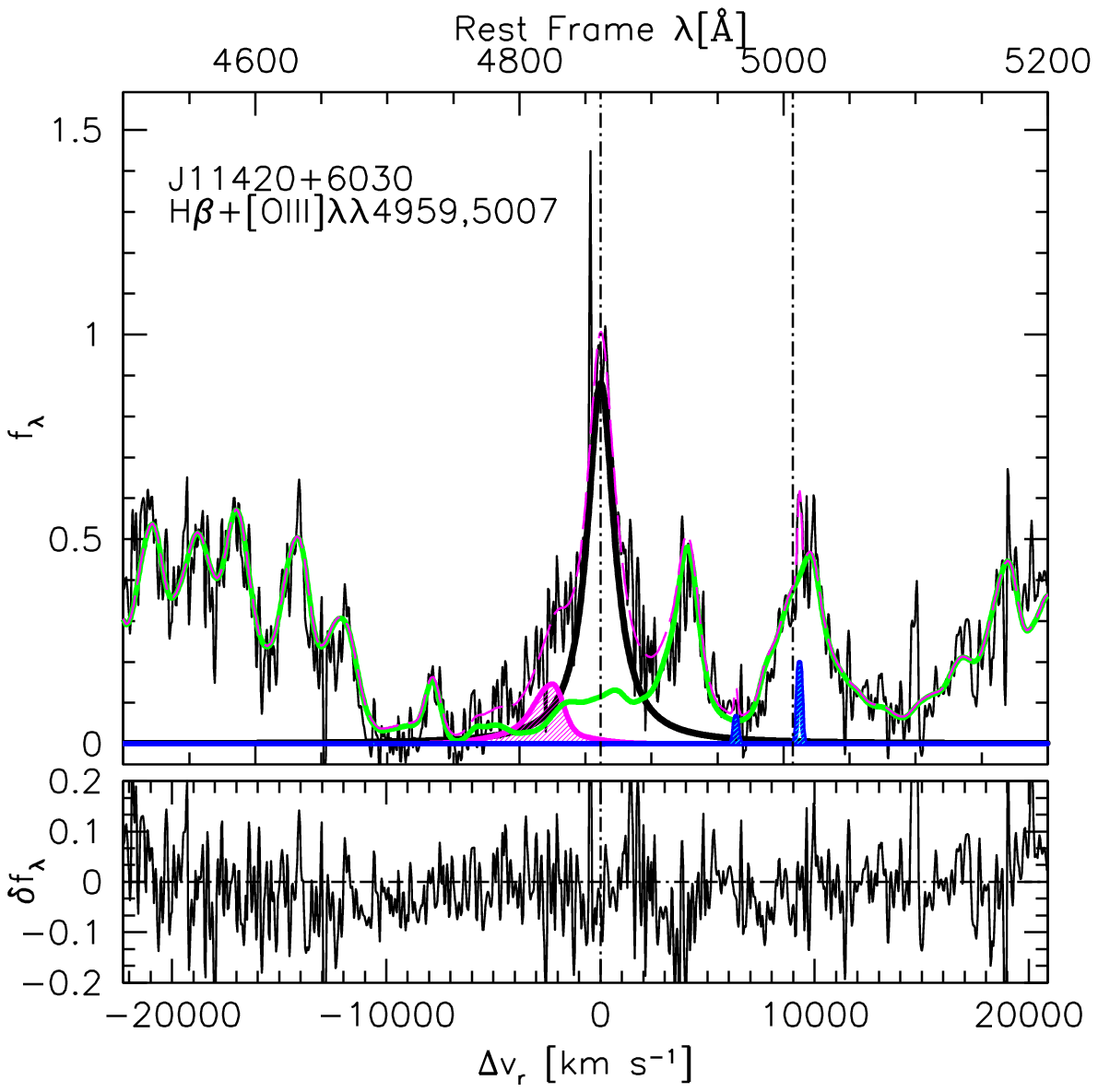}
 \includegraphics[width=0.425\columnwidth]{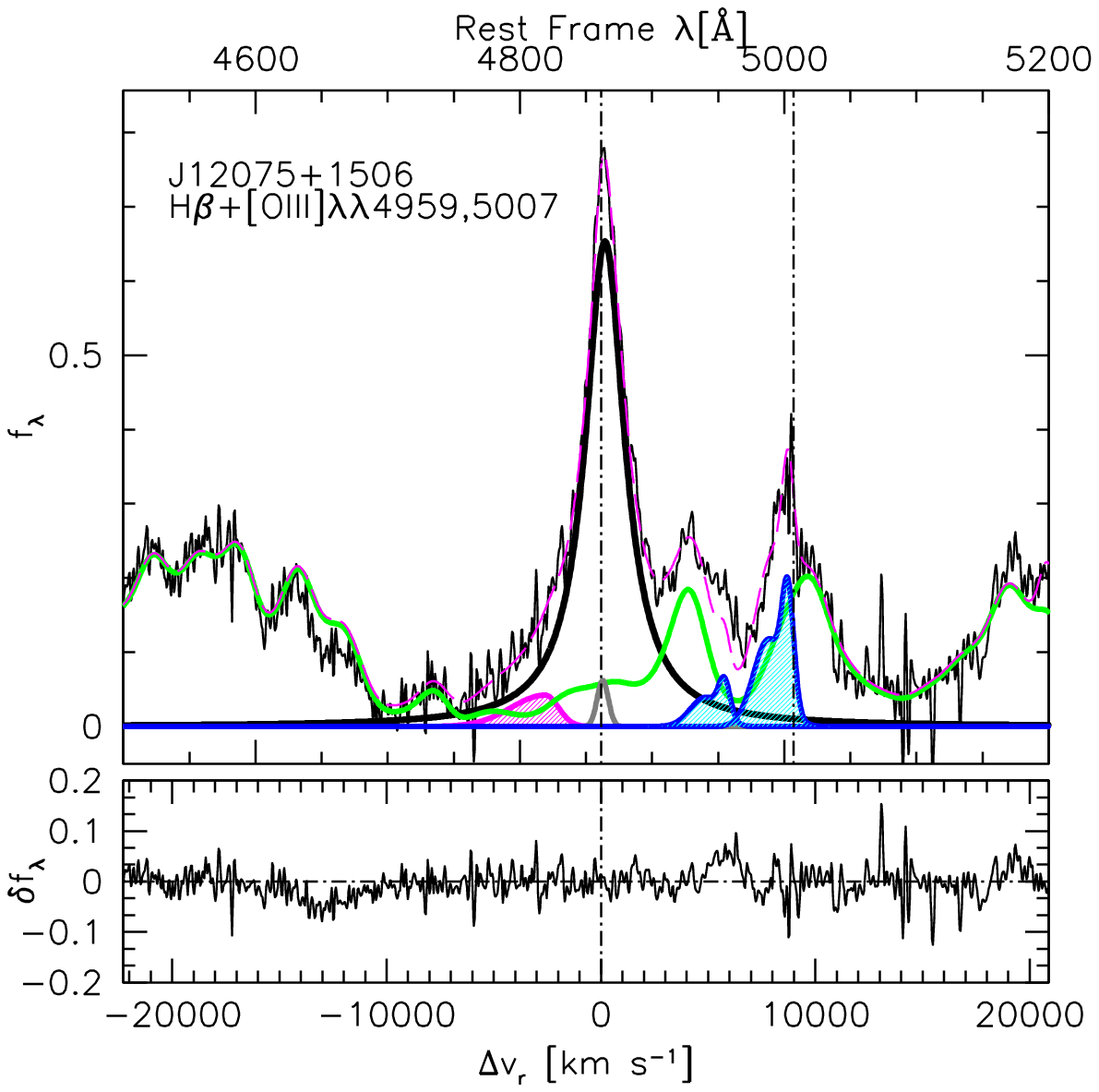} 
 \includegraphics[width=0.425\columnwidth]{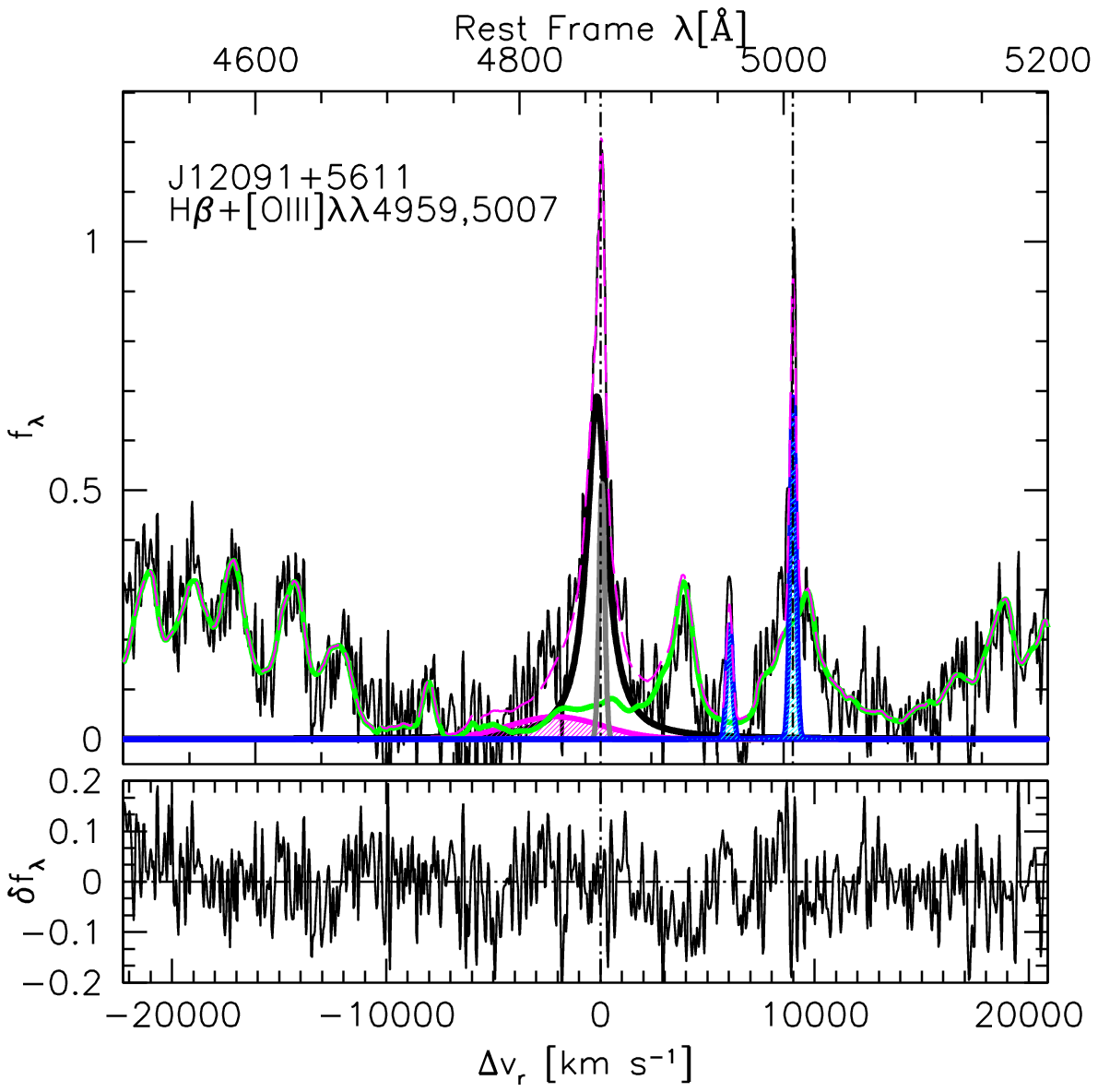}\\ 
 \includegraphics[width=0.425\columnwidth]{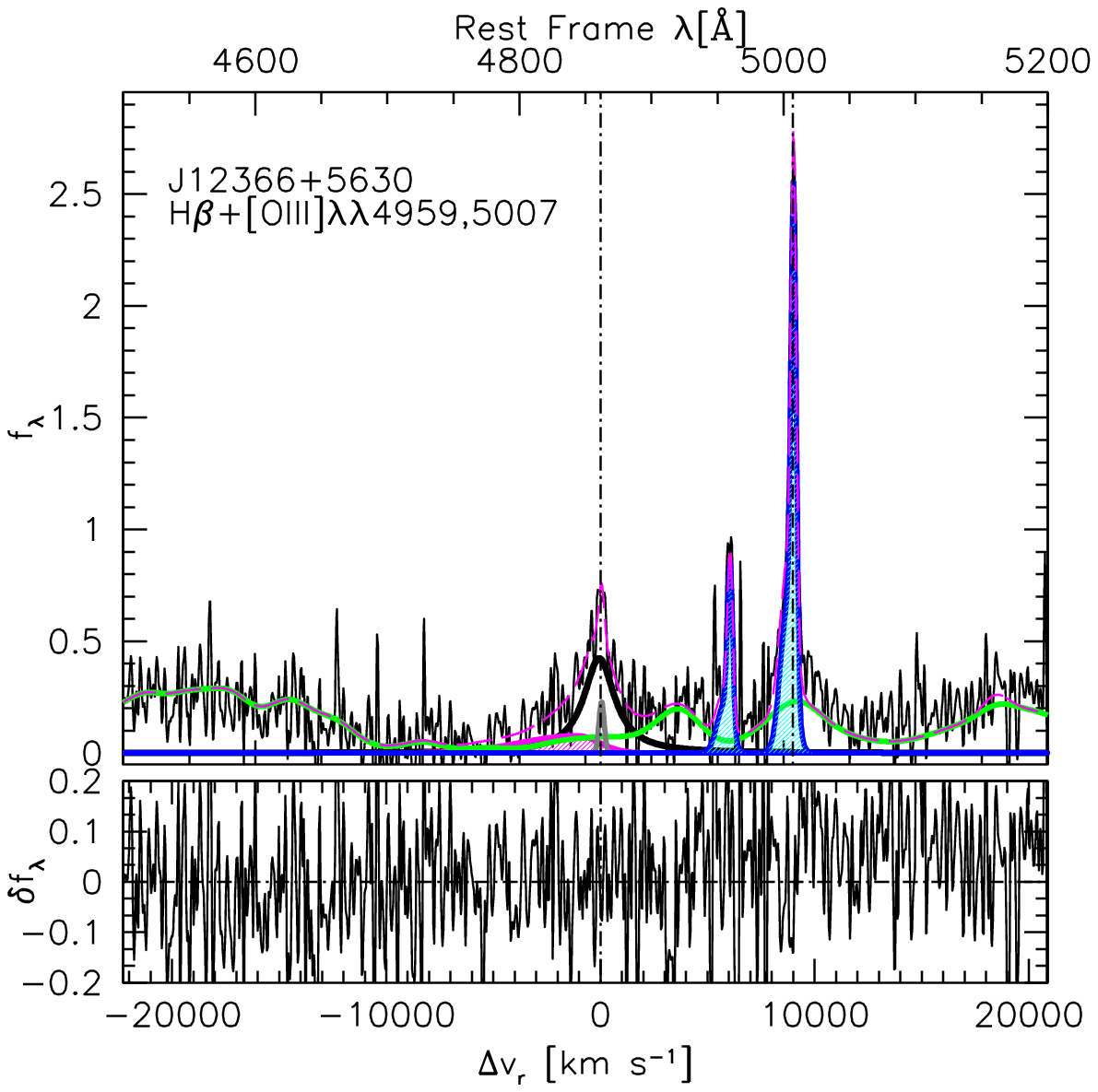}
 \includegraphics[width=0.425\columnwidth]{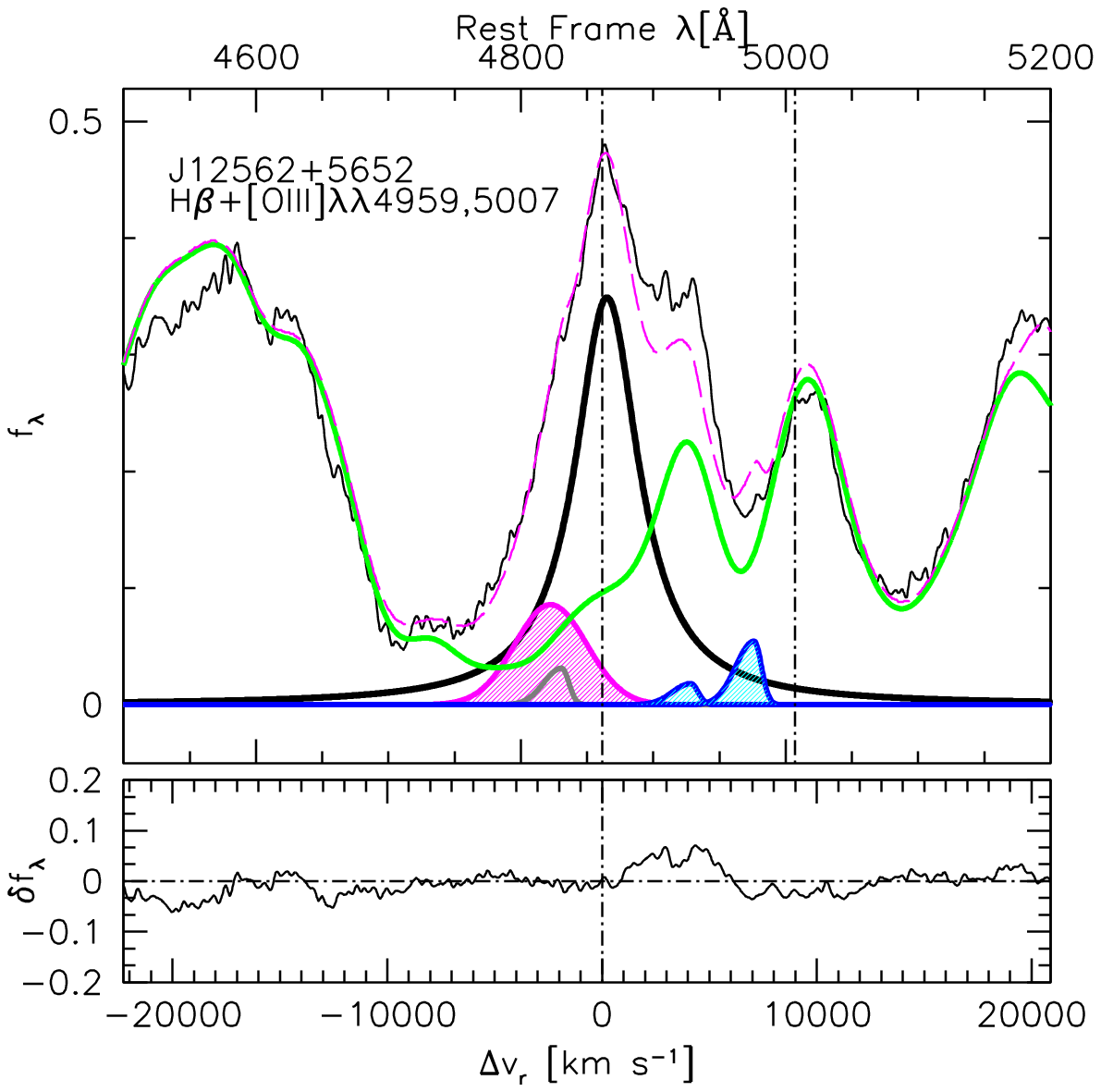}
 \includegraphics[width=0.425\columnwidth]{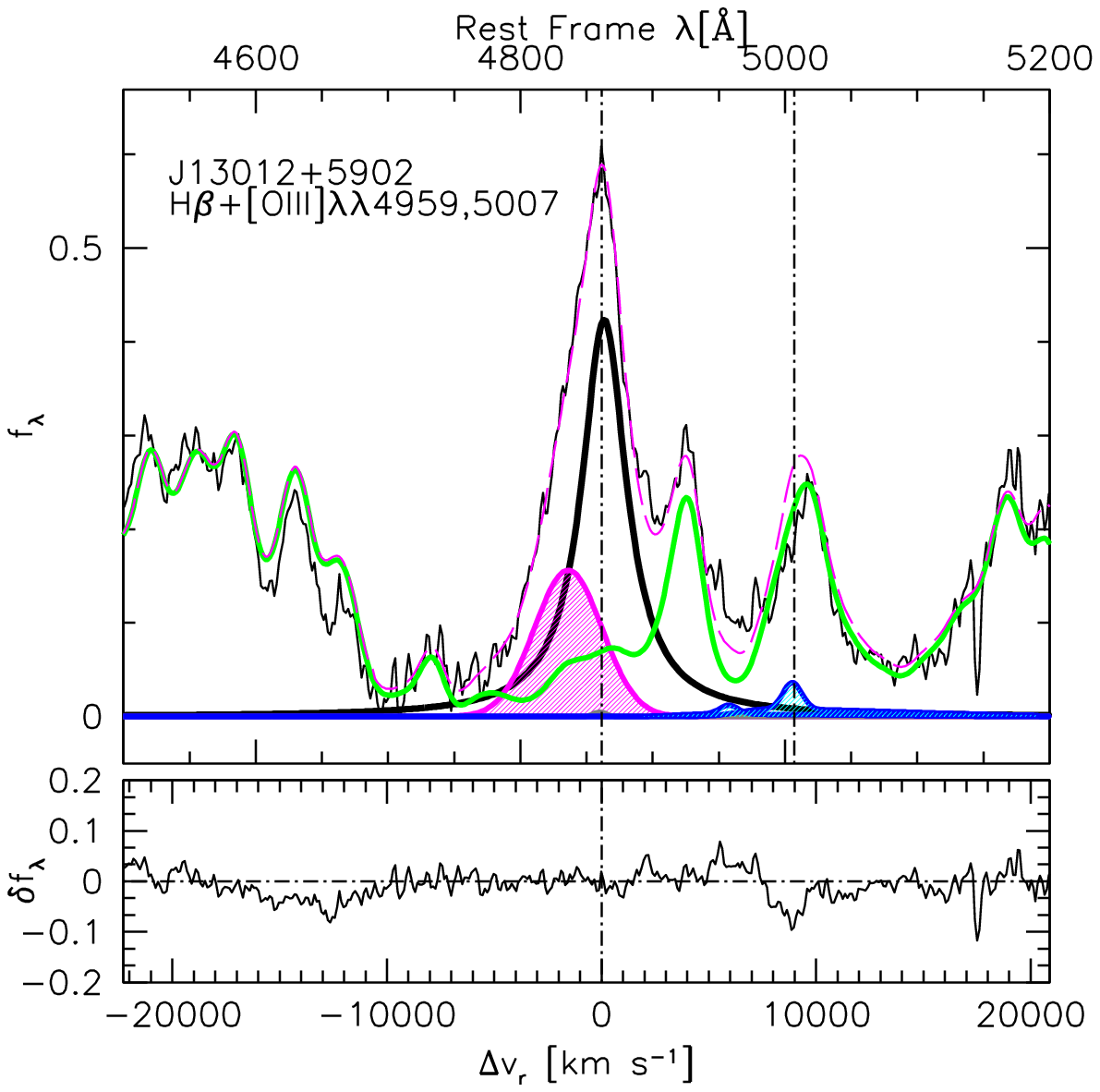}  
 \includegraphics[width=0.425\columnwidth]{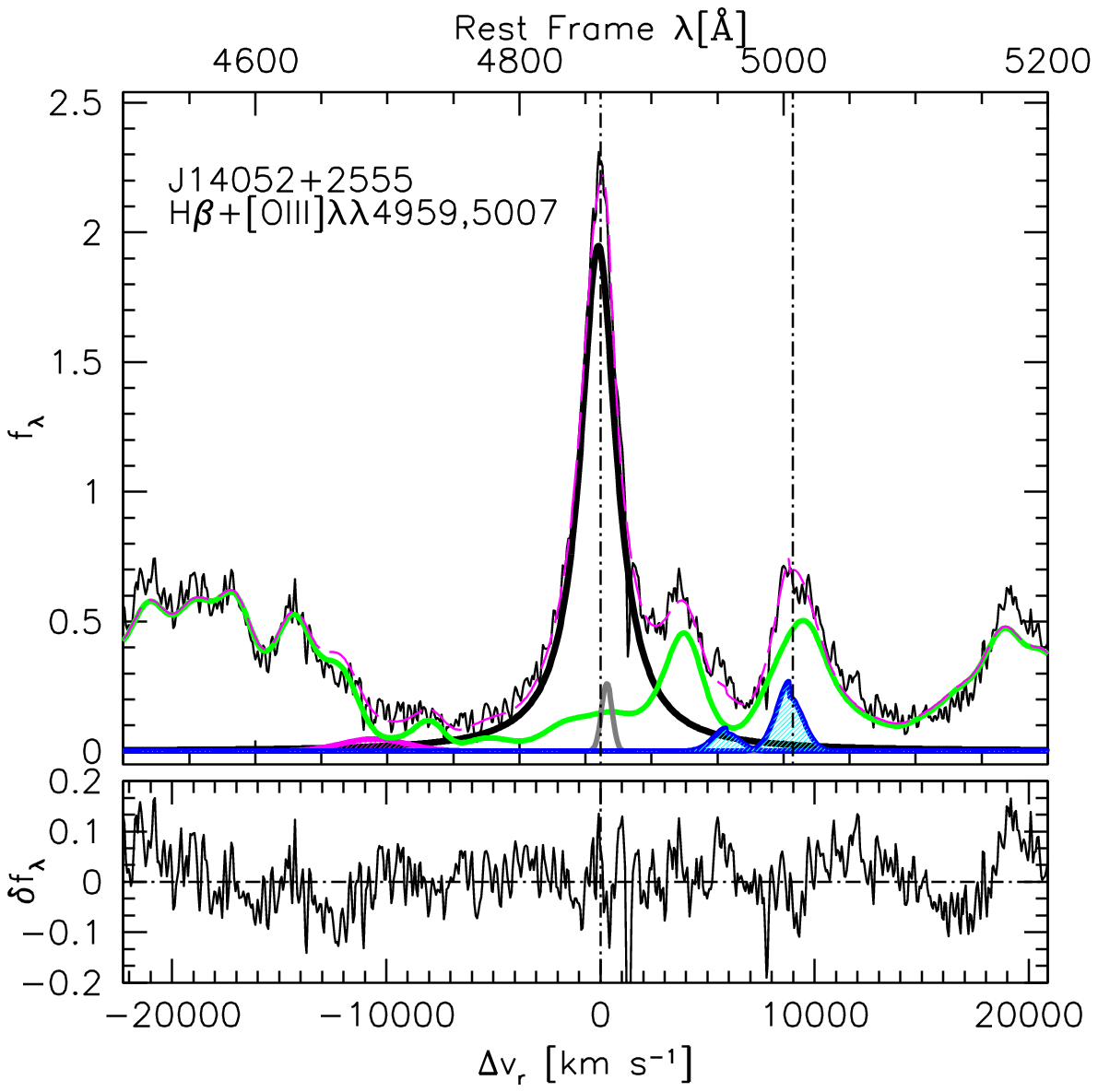}\\
 \includegraphics[width=0.425\columnwidth]{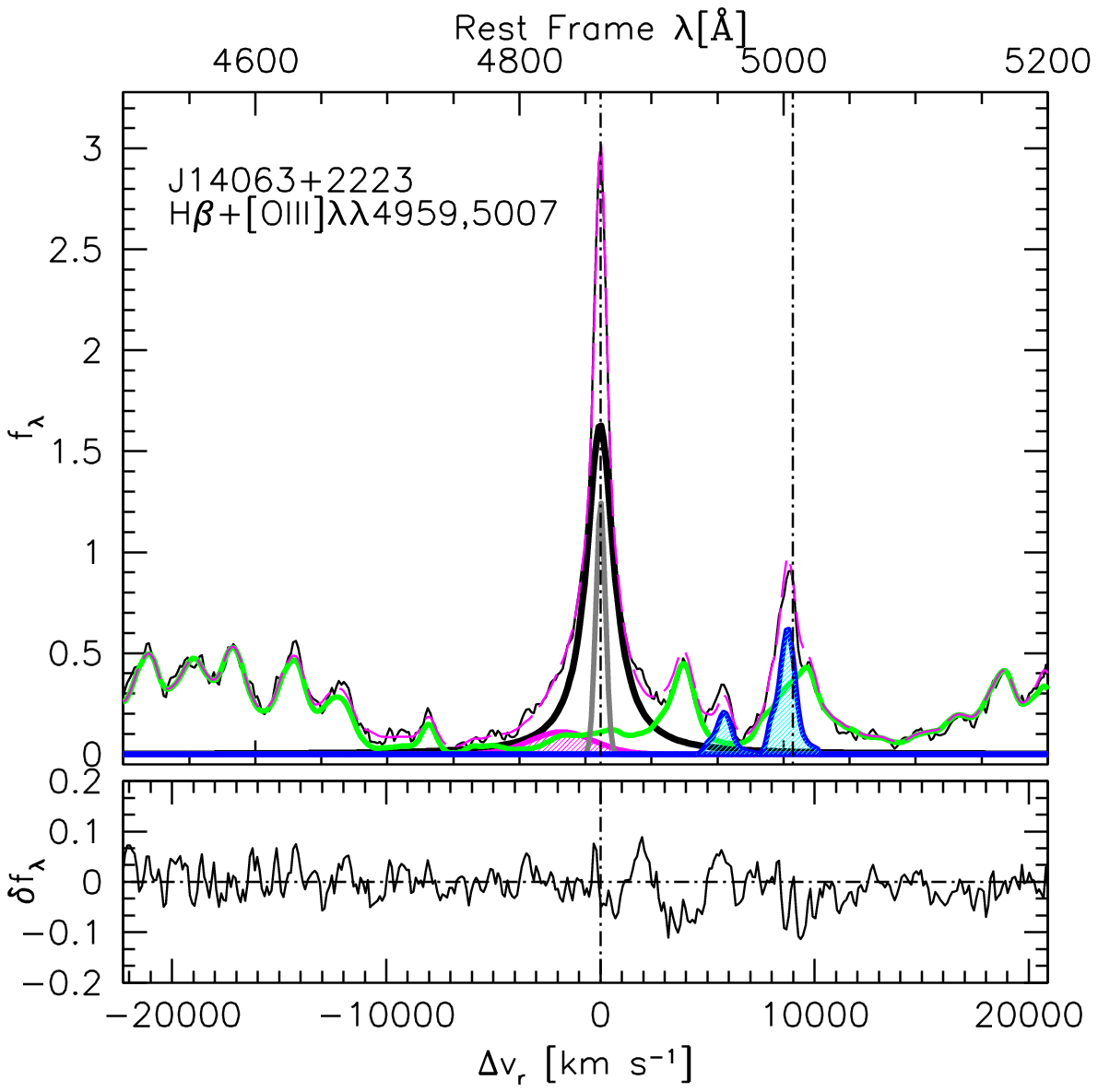}  
 \includegraphics[width=0.425\columnwidth]{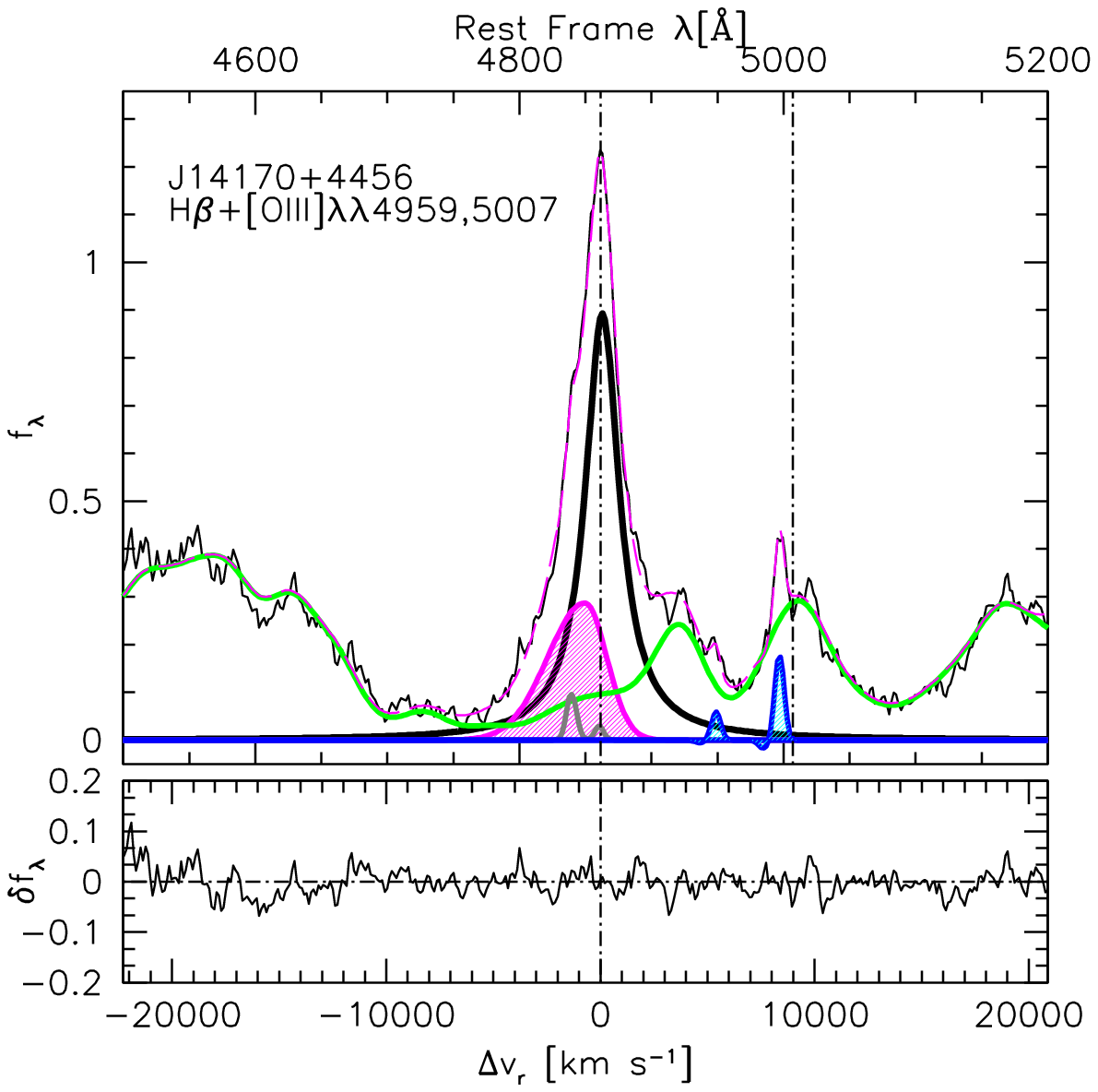}
 \includegraphics[width=0.425\columnwidth]{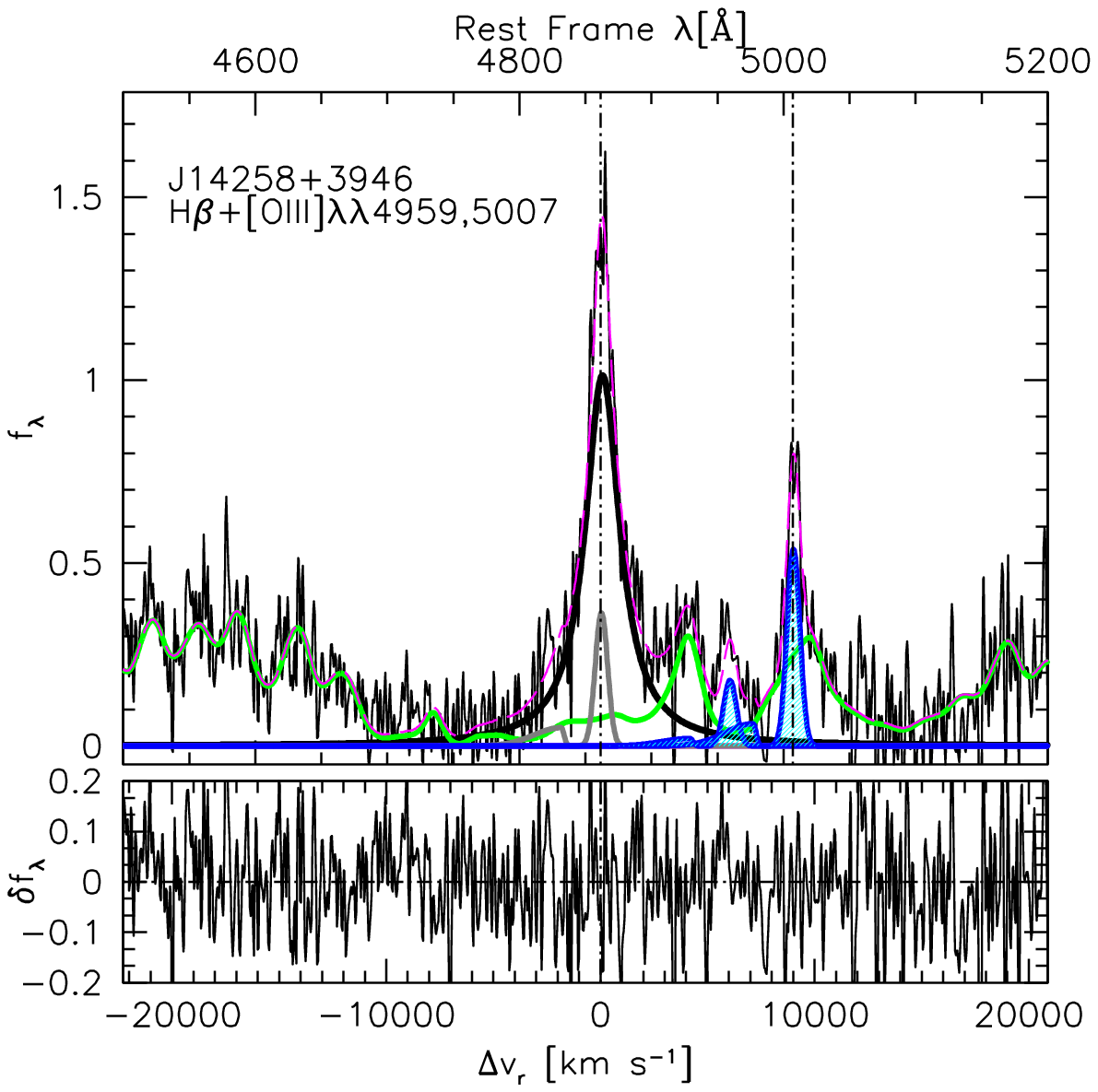}   
 \includegraphics[width=0.425\columnwidth]{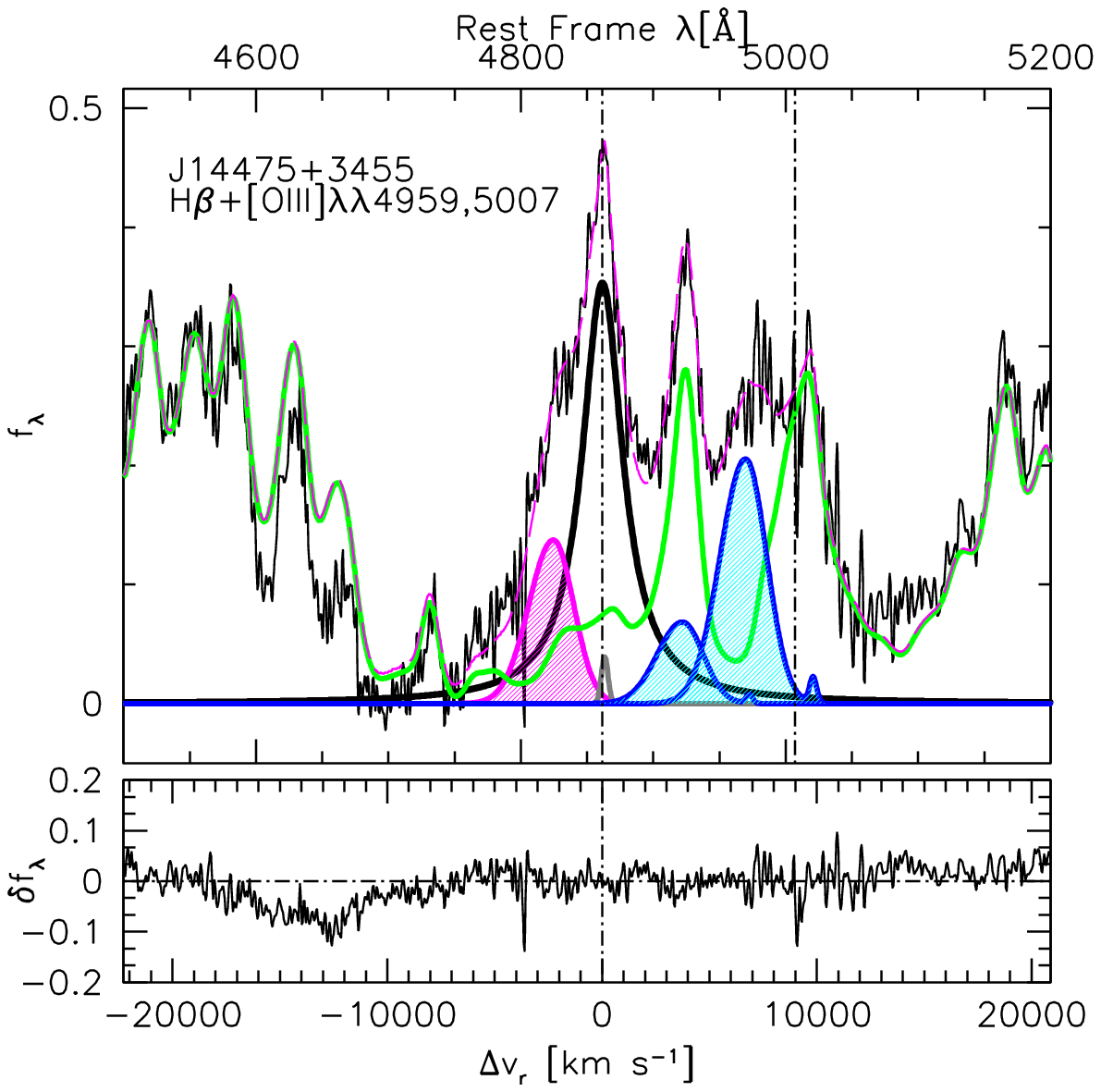}\\
 \includegraphics[width=0.425\columnwidth]{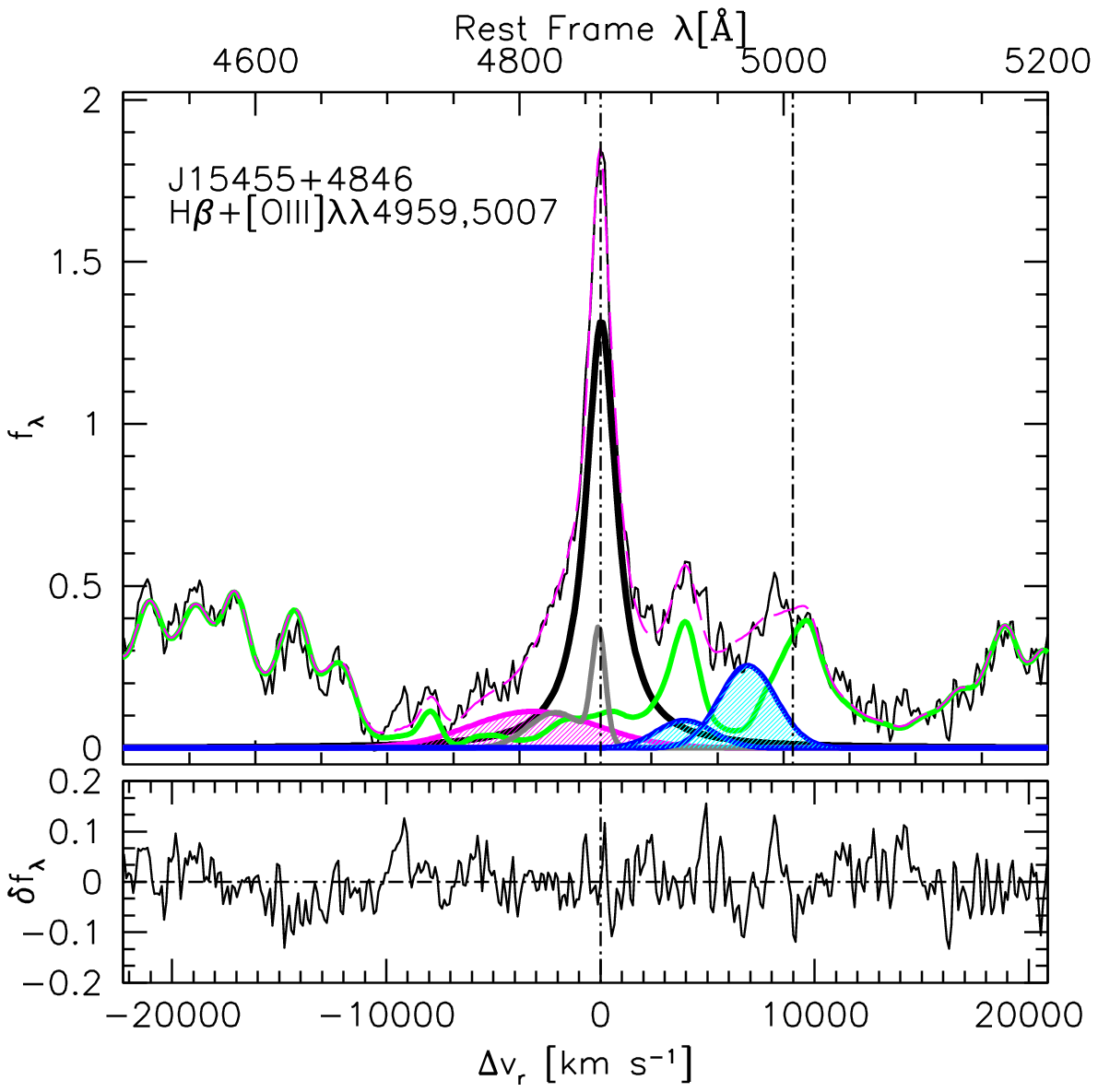} 
 \includegraphics[width=0.425\columnwidth]{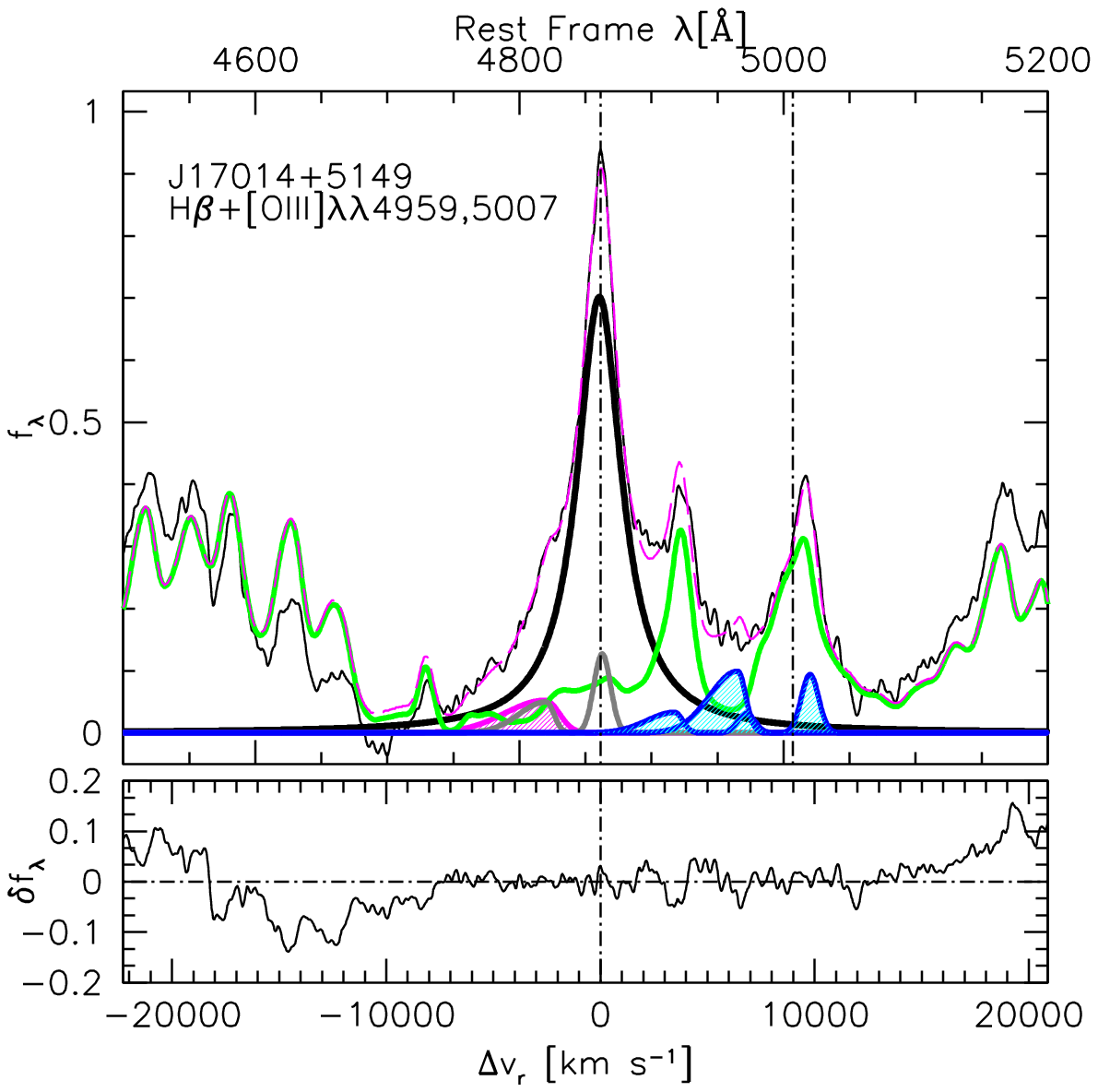} 
\caption{ Continuum-subtracted, optical spectra of the 18 targets in the \hb\ spectral range (upper panels). Abscissa is rest-frame wavelength (top scale) and radial velocity with respect to rest frame (bottom); ordinate is normalized specific flux, also in the rest frame. Thin black line: observed spectrum, thick black: broad component of \hb; thick blue shaded cyan: full profile of \oiiiopt; magenta: \hb\ broad blueshifted emission; grey: \hb\ narrow component; pale green: \feii\ template.  The dashed magenta line traces the full empirical model of the spectrum. Lower panels: observed spectrum minus empirical model. \label{fig:fosa}} 
\end{figure*}

Fig. \ref{fig:fosa} shows the results of a multicomponent nonlinear analysis of the \hb\ spectral range, after continuum subtraction. Several previous papers have successfully applied the same technique \citep{negrete_highly_2018}; here we recall only the salient points. The main components of the  model are a power law continuum, \feii\ emission according to a standard template \citep{boroson_emission-line_1992}, \hb\ emission well-fit by a Lorentzian profile \citep[see e.g.,][for a variety of approaches]{sulentic_average_2002,shapovalova_spectral_2012,wang_self-shadowing_2014,cracco_spectroscopic_2016,negrete_highly_2018,crepaldi_optical_2025}, and \oiiiopt\ emission. In sources where \feii\ is much weaker, the Lorentzian is found at a wavelength consistent with the one of the quasar rest frame;   small shifts ($\lesssim \pm 100$) are possible due to uncertainties in the systemic redshift ($\lesssim 50$ \kms), compounded with a slight mismatch originating in the fitting process.  The outflow emission originating from the broad-line region (BLR) is modeled as  a blueshifted skew Gaussian, on the basis of the trapezoidal profile most often observed in the \civ\ profile of xA sources \citep[e.g.,][]{zamanov_kinematic_2002,leighly_hubble_2004,martinez-aldama_extreme_2018}. The narrow-line region (NLR) emission is similarly modeled by separating a symmetric Gaussian and a skew Gaussian \citep{azzalini_class_1985}, usually shifted toward shorter wavelengths, and helpful to model the outflow component that is prominent in the \oiii\ emission \citep{zhang_blueshifting_2011}. The shaded areas in Fig. \ref{fig:fosa} emphasize the \hb\ outflow profile (magenta) and the full blueshifted profile of \oiiiopt\ (cyan). 
The \hb\ BLUE (shaded magenta) is barely detectable in several sources, but produce are remarkable blueward asymmetry especially in the sources with strong \feii\ emission. The \oiiiopt\ full profile (shaded cyan) is often dominated by the SBC shifted to the blue. The \oiii\ FWHM is large, and its equivalent width  is low, to the point that the line is, in several cases, barely detectable over the strong \feii\ emission. This result is in agreement with the basic tenet of the  Eigenvector 1 paradigm (the anticorrelation between the peak intensity of \oiii\ and \rfe, \citealt{boroson_emission-line_1992}).


Table~\ref{tab:meashb}  reports measurements  carried out on full profile of the \hb\ line,   as well as on the blueshifted component associated with the BLR outflows. The centroids, the asymmetry (AI) and kurthosis (KI) indexes  reported in the Tables are computed according to the definition of \citet{zamfir_detailed_2010}.  The column providing information on the strength of \feii\ emission reports the intensity ratio \rfe = \feiiq/\hb, where \feiiq\ is the intensity of the blend on the blue side of \hb\ (Fig.  \ref{fig:fosa}), that is the  parameter used for the selection of xA sources. The following columns report relative intensity normalized to the full broad  profile  of the \hb\ broad component (BC), symmetric and unshifted, and its FWHM. For the blueshifted broad component (BLUE), the relative contribution is listed along with  FWHM, peak shift and skew parameter  that describes the asymmetry of the skew normal function used to model its profile.   


Table~\ref{tab:measoiii} reports measurements  carried out on the \oiii\ line,  following a scheme analogous to Table~\ref{tab:meashb} for \hb. The centroids, AI and KI are reported for the full \oiii\ profile  following the spectrophotometric parameters. We distinguish a narrow component (NC) and a semi-broad component (SBC), with blueshifts ranging from almost 0 to a few thousands \kms. We report relative intensity for the NC and SBC, along with FWHM and Shift. For the \oiii\ SBC, the shift (defined as the mode of the skew normal distribution) is not representative of the bulk displacement of the line with respect to the rest frame. In this case, we provide also the centroid at half maximum $c(\frac{1}{2})$.

\subsection{Estimation of wind parameters}

\subsubsection{\oiiiopt}

The large blueshifts and blue-ward skewness of the \oiiiopt\ lines suggests the presence of an outflow for the emitting gas, possibly originating from nuclear winds, or from radiative feedback on the gas of the host galaxy bulge, where the receding side of the flow is obscured or absorbed \citep{zamanov_kinematic_2002,cresci_blowin_2015,singha_close_2022,kakkad_bass_2022}. The relations for line emitting gas mass, mass flow, thrust and kinetic power based on the BLUE component of \oiii\ are the ones provided by \citet{marziani_quasar_2017}, scaled to $n_\mathrm{H} = 10^4$ cm$^{-3}$ (Sect. \ref{radiowind}). They are standard relations \citep{cano-diaz_observational_2012,fiore_agn_2017}, obtained assuming optically-thin emission confined in a bipolar outflow within a solid angle 2$\cdot \Omega$. 

The immediate inference from the \oiii\ outflow parameters (Tab. \ref{tab:outkin}) computed from the data listed in Table \ref{tab:measoiii}  is that they show an especially strong dependence on luminosity, with the kinetic power reaching   tenth of percent of the bolometric luminosity ($\sim 0.1 - 0.2$\%) in the case of the most luminous sources.   This result is consistent with previous works  \citep{carniani_ionised_2015,fiore_agn_2017,bischetti_wissh_2017,vietri_wissh_2018,kakkad_super_2020}.  The mass outflow rate is always  $\dot{M} \lesssim 10$ M$_\odot$ yr$^{-1}$, with only a few luminous sources reaching $\dot{M} \sim5$ M$_\odot$ yr$^{-1}$. The low values of the wind parameters are also a consequence of the low equivalent width of the \oiii\ emission \citep{boroson_emission-line_1992,zamanov_kinematic_2002}. Apparently, even if the sources of the present sample are close to maximum radiative output per unit black hole mass, they are capable of inducing modest mechanical effects, especially at low-luminosity.  

\subsubsection{\hb}

An   estimation of the  wind parameters for \hb\ BLUE is less standard, as the mildly-ionized wind emission has been usually traced by the resonance high-ionization line of \civ\ \citep[e.g.,][]{marziani_most_2016,vietri_wissh_2018}. The blueshifted \hb\ emission is presumably associated with the BLR: spatial scales are smaller, and the electron densities are expected to be much higher with respect to the \oiii\ emitting regions, but are less well-known. It is customary to assume a standard value for the BLR electron density $n \sim 10^9 > 10^8$\  cm$^{-3}$ \citep{netzer_physics_2013}, even if the density of the low-ionization part of the BLR is much higher \citep{marziani_broad-line_2010,negrete_broad-line_2012,temple_fe_2020,temple_high-ionization_2021}, $\sim 10^{12} - 10^{13}$ cm$^{-3}$, and the outflow might be clumpy \citep[e.g.,][]{ward_aGN-driven_2024}. 
The \hb\ emissivity can be written as 
\(
j_{\text{H}\beta} = n^2 \alpha_{\text{H}\beta}(T) h \nu_{\text{H}\beta}
\)
where $n$ is the Hydrogen   density ($n_\mathrm{e} \sim n_\mathrm{H} \sim n$)
and \( \alpha_{\text{H}\beta}(T) \) is the effective recombination coefficient for \( \text{H}\beta \). At \( T_\mathrm{e} = 20,000 \, \text{K} \), \( \alpha_{\text{H}\beta} \approx 3.0 \times 10^{-14} \, \text{cm}^3 \, \text{s}^{-1} \) \citep{osterbrock_astrophysics_2006}, yielding an \hb\ emissivity $j_{\text{H}\beta} \sim 1.23\cdot 10^{-7} n_\mathrm{9}^2$ \ergss\ cm$^{-3}$. 
Therefore, the total mass of the line emitting gas is
\[ M_\mathrm{HII} \sim n \mu m_\mathrm{P} \frac{L(H\beta) \, \mathrm{BLUE}}{j_\mathrm{H\beta}} \sim 10^3 \frac{L_{44}(H\beta)  \mathrm{BLUE}}{ n_9} \mathrm{M}_\odot \, \] where we assume $n_\mathrm{H} = 10^9$ cm$^{-3}$, the molecular weight $\mu = 1.4$, and $m_\mathrm{P}$\ is the proton mass.  

Using the same approach applied to \oiii, we can compute the outflow’s dynamical parameters as traced by the \hb\ line. In this scenario, the gas emitting \hb\  is presumably still being accelerated by radiation pressure, thereby gaining momentum. An additional factor must be incorporated into these calculations, either by adopting a simplified model for an unbound gas subject to gravity and radiation pressure \citep{netzer_effect_2010,marziani_quasar_2017} or by applying an empirical recipe developed for the \civ\ line \citep{deconto-machado_exploring_2024}, assuming a final outflow velocity of $c(\frac{1}{2}) + 2\sigma$ for the blueshifted component.  We apply the simplest assumption including a factor $\kappa = 5$\ \citep{marziani_quasar_2017}. The resulting relations for line emitting mass, mass flow, thrust and kinetic power based on the BLUE component of \hb\ are:  
\begin{gather*}
    \dot{M} \sim 60 L_\mathrm{H\beta,44} v_{1000} r_\mathrm{0.1pc}^{-1} M_\odot \text{ yr}^{-1} \\
    \dot{M}v \sim 2\cdot 10^{36} k_5 L_\mathrm{H\beta,44} v_{1000}^2  r_\mathrm{0.1pc}^{-1}  \text{ g cm s}^{-2} \\
    \dot{\epsilon} \sim 5\cdot 10^{44} k^2_5 L_\mathrm{H\beta,44} v_{1000}^2  r_\mathrm{0.1pc}^{-1}  \text{ erg s}^{-1},
\end{gather*}
assuming $n_\mathrm{H} = 10^9$ cm$^{3}$, and velocity scaled to 1000 \kms.  

The results  obtained for the BLR outflow traced by the blueshifted excess of the \hb\ line, reported in Table \ref{tab:outkin}, are consistent with the ones for the \oiii\ outflow. In the \hb\  case, there is however no straightforward comparison with earlier results based on the same line: the BLUE component is faint and difficult to measure, especially in cases with strong \feii. Since \hb\ BLUE should be emitted in the same mildly-ionized outflow producing \civ\ and other UV high-ionization lines, a comparison can be made with the outflow parameters derived from \civ\ \citep{marziani_blue_2016,vietri_wissh_2018,deconto-machado_high-redshift_2023,deconto-machado_exploring_2024}.  The mass outflow rate is $\dot{M} \sim 1 \lesssim \mathrm{a \, few}$ M$_\odot$ yr$^{-1}$, with only J13012+5902 reaching $\dot{M} \sim 40$ M$_\odot$ yr$^{-1}$. Even if these values could be considered more as lower limit as anything else, the basic outcome is again  consistent with previous results, where only at the largest luminosities the kinetic power might become as large as to imply a significant feedback effect on the host galaxy \citep{di_matteo_energy_2005,ishibashi_active_2012,king_powerful_2015,harrison_kiloparsec-scale_2014,king_black_2016,laha_ionized_2020}.  The  implication is that, while these sources are mostly candidate super-Eddington quasars, their BLR winds do not necessarily produce the strongest feedback, as the feedback effect primarily depends on luminosity -- and thus on black hole mass -- if \lledd $\gtrsim 0.2$\ \citep{marziani_super-eddington_2025}: \lledd\ is constrained to be \lledd $\lesssim 1$ \citep{mineshige_slim-disk_2000}, and hence to vary within a small range. On the converse \mbh\ and AGN luminosity (to which  line luminosity is roughly proportional)   can be factors  $\sim$ 100 higher in the most luminous quasars at the cosmic noon. Only in those cases the mechanical power of the NLR and BLR  winds may reach values close to the ones needed for a global effect on the host galaxy \citep{deconto-machado_high-redshift_2023}.  \

\subsection{Accretion parameters} \label{accretion_parameter}


The black hole masses are computed assuming virial broadening for the emission lines: \mbh =$ f r_\mathrm{BLR} \mathrm{FWHM}^2 / G$,  where $f$\ is the virial factor and  $r_\mathrm{BLR}$\ the radius of the emitting region \citep[e.g.,][]{vestergaard_determining_2006,shen_mass_2013}.
The dependence of $f$\ on the viewing angle has been parameterized by \citet{mejia-restrepo_effect_2018} as $f \approx $ (FWHM/4550)$^{-1.17}$, and the \rb\ dependence on Eddington ratio by applying a correction to the standard scaling between \rb\ and optical luminosity at 5100 \AA\ \citep{du_radiusluminosity_2019,martinezaldama_scatter_2020}. 

{The main empirical correlation with Eddington ratio is the \rfe\ prominence parameter \citep[e.g.,][]{sulentic_eigenvector_2000,marziani_searching_2001,sun_dissecting_2015,du_fundamental_2016,panda_quasar_2019}, defined as the ratio between the fluxes of the \feii\ blend centered at 4570 \AA\ and \hb. The parameter \rfe\ is one of the two main parameters (along with FWHM \hb) defining the quasar Eigenvector 1 main  sequence \citep{boroson_emission-line_1992,sulentic_eigenvector_2000}. The classical scaling law linking BLR distance \rb\ from AGN continuum to luminosity \citep{bentz_radiusluminosity_2009} shows a dependence on Eddington ratio \citep{martinezaldama_scatter_2020}.    A tight relation between \rb\ and luminosity can be recovered once \rb\ is corrected, using the robust measurement of  \rfe\ in place of \lledd. The following relation \citep{du_radiusluminosity_2019}:}

\begin{equation}
   \log r_\mathrm{BLR} \approx 17.063 + 0.45  \log L_{5100,44} - 0.35 R_\mathrm{FeII}\ \mathrm{[cm],} \ 
\end{equation}

{yields:}

\begin{eqnarray}
 \log M_\mathrm{BH}(\mathrm{H}\beta) &\approx &{  5.220} + 0.45 \log L_{5100,44} + \nonumber\\
 & & 0.83 \log \mathrm{FWHM}  -  0.35  R_\mathrm{FeII} \,   \mathrm{[M}_\odot\mathrm{].} \label{eq:mbhml} 
\end{eqnarray}
 
Eq. \ref{eq:mbhml}  implies a significant increase of the \mbh\ values for the sources with the narrowest profile    with respect to the scaling law of \citet{vestergaard_determining_2006}, providing estimates in closer agreement with the ones obtained from more robust methods \citep{donofrio_correlation_2024}. The systematic \mbh\ uncertainty is estimated by computing the dispersion between the \mbh\ values computed from Eq. \ref{eq:mbhml} and three different scaling laws, of them two based on the \hb\ line width \citep{vestergaard_determining_2006,shen_sloan_2024}, and one on the \hb\ luminosity \citep{greene_estimating_2005}:   $\delta \log$ \mbh $\approx 0.17$, with a negligible bias between \mbh\ from Eq. \ref{eq:mbhml} and the average of the other scaling laws, $\delta \log$ \mbh $\approx - 0.05\pm 0.20$. The uncertainty associated with the scaling law is still   dominating with respect to the statistical uncertainties in line FWHM and in continuum flux measurement, $\lesssim 10$\%\ at 1$\sigma$ confidence level.   

The bolometric correction (BC) was computed accounting for its dependence on  luminosity \citep{runnoe_updating_2012}. We utilize the  BC parameterization by \citet{netzer_bolometric_2019} that are fairly consistent, at low luminosity, with the ones described for highly accreting quasars: for Eddington ratios in the range $\log$ \lledd\ $\sim -0.5 - 0.0$, at $\log \lambda L_\lambda$(5100\AA)$ \sim  43 -  44$ [erg s$^{-1}$]  {the bolometric correction has been estimated between  16 and 23 \citep{garnica_spectral_2025}} and 10 and 18.8 \citep{ferland_state---art_2020}. If the Garnica et al. and the \citet{netzer_bolometric_2019} bolometric corrections are applied to the present sample, the systematic difference in bolometric luminosity $L$ is negligible, $\approx$ 12 \%. Results for optical and  bolometric luminosity, and for \mbh\ and \lledd\ are reported in Table \ref{tab:valuesmbh}. 

Table~\ref{table_uncertainties} reports statistical and systematic linear fractional uncertainties for the accretion parameters.
The main source of uncertainty is the systematic uncertainty associated with the different scaling laws employed for \mbh\ estimates. The uncertainty is apparently reduced with respect to earlier works \citep{vestergaard_determining_2006}, but is still affecting heavily the \mbh\ and \lledd\ estimates. If systematic and statistical errors are quadratically propagated (column Total in Table~\ref{table_uncertainties}), the uncertainties reach values $\gtrsim$ 70 \%\ for the \lledd, $\delta \log $\lledd $\sim \pm 0.32$, an aspect that should be considered in any correlation analysis involving \lledd.

\begin{figure*}
\center
\includegraphics[width=0.9\textwidth]{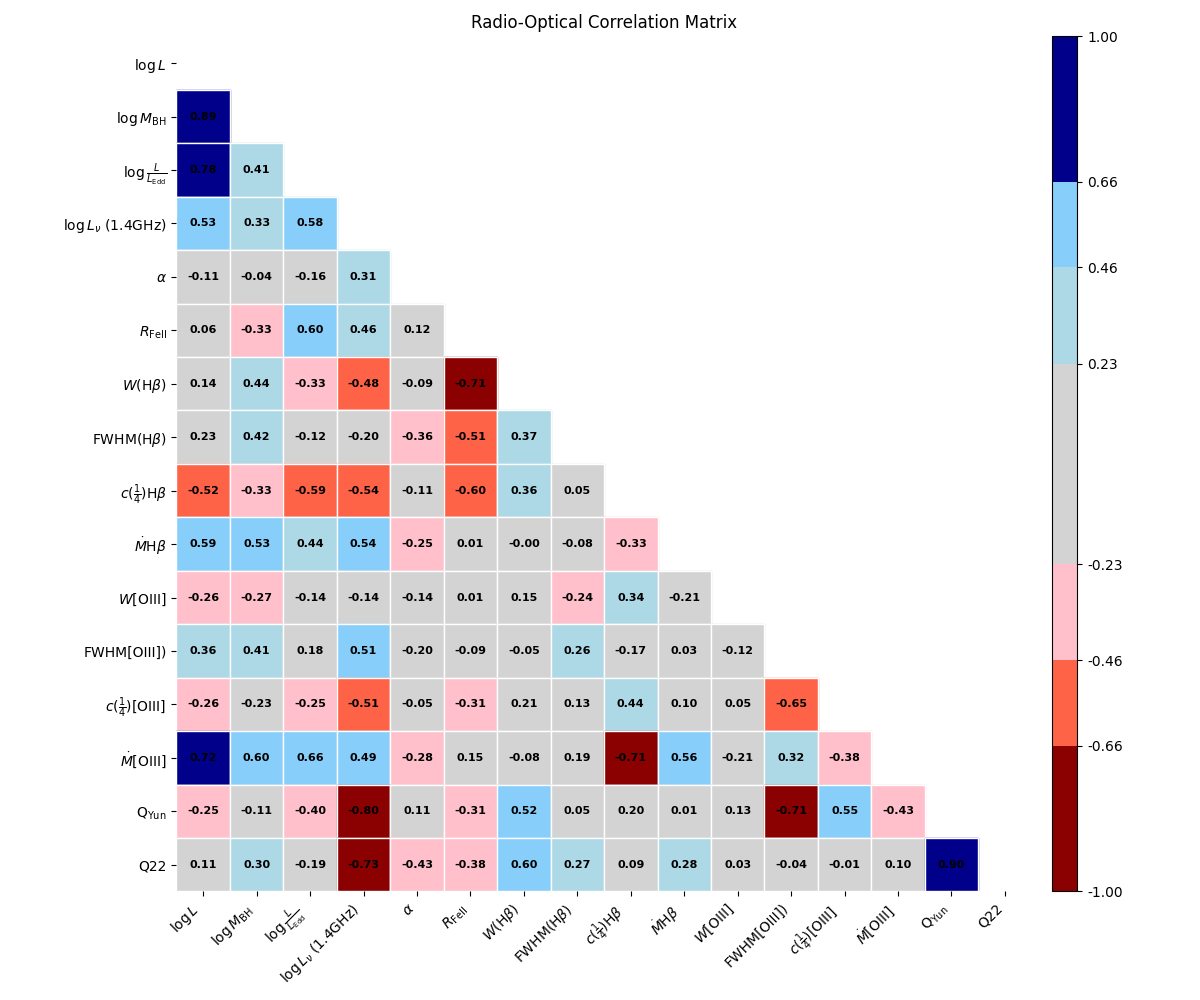}
    \caption{Correlation matrix (Pearson's) involving the main optical (\hb\ and \oiii), radio and  IR derived parameters. The limits in the color ranges  correspond to a statistical significance of $\pm 1,2,3 \sigma$\ for the Pearson's correlation coefficient computed for 18 objects. The parameters considered are in this order, the bolometric luminosity $L$, the black hole mass \mbh, the Eddington ratio \lledd, the radio power at 1.4 GHz, the radio spectral index $\alpha$, \feii\ prominence  ratio $R_\mathrm{FeII}$, the equivalent with of \hb\ (broad components), its FWHM and centroid displacement at one-half peak intensity $c(\frac{1}{2})$, the mass outflow rate measured from blueshift emission of \hb, $\dot{M}$; the equivalent width, FWHM, $c(\frac{1}{2})$, $\dot{M}$ \ of the \oiii\ line; the $Q$\ parameters defined according to \citet{yun_radio_2001} and to Eq. \ref{eq:q22}.}
    \label{fig:corrmatrix}
\end{figure*}

\onecolumn

\begin{sidewaystable*}
\setlength{\tabcolsep}{1.2pt}
\caption{Analysis of the \hb\ and \feiiq\ emission features}
\label{tab:meashb}
\begin{tabular}{@{\extracolsep{3pt}}l c c c c c c c c c c c c c c c c c c c c c c c@{}} 
\hline\hline 
&& \multicolumn{9}{c}{\hb\ full broad profile (BC + BLUE)}&& \multicolumn{1}{c}{\feiiq} && \multicolumn{2}{c}{\hb\ BC}  && \multicolumn{3}{c}{\hb\ BLUE} \\
\cline{3-11 }\cline{13-13} \cline{15-16} \cline{18-21}
Object & $f_{\lambda,5100}$ &  \multicolumn{1}{c}{Flux} & \multicolumn{1}{c}{EW} &  \multicolumn{1}{c}{FWHM} &  \multicolumn{1}{c}{c($\frac{1}{4}$)} & \multicolumn{1}{c}{c($\frac{1}{2}$)} & \multicolumn{1}{c}{c($\frac{3}{4}$)} & \multicolumn{1}{c}{c($\frac{9}{10}$)} & \multicolumn{1}{c}{AI} & \multicolumn{1}{c}{KI}  
&&    \multicolumn{1}{c}{\rfe} && \multicolumn{1}{c}{F/F(\hb)} & FWHM  && 
  \multicolumn{1}{c}{F/F(\hb)}  &\multicolumn{1}{c}{FWHM}& \multicolumn{1}{c} {Shift}  & \multicolumn{1}{c} {Skew}  &  \\

\multicolumn{1}{c}{(1)}& \multicolumn{1}{c}{(2)} & \multicolumn{1}{c}{(3)} & \multicolumn{1}{c}{(4)} & \multicolumn{1}{c}{(5)} & \multicolumn{1}{c}{(6)} &\multicolumn{1}{c}{(7)} & \multicolumn{1}{c}{(8)} & \multicolumn{1}{c}{(9)} & \multicolumn{1}{c}{(10)} & \multicolumn{1}{c}{(11)} && \multicolumn{1}{c}{(12)} & &\multicolumn{1}{c}{(13)} &\multicolumn{1}{c}{(14)} &&\multicolumn{1}{c}{(15)} &\multicolumn{1}{c}{(16)} &  \multicolumn{1}{c}{(17)} & \multicolumn{1}{c}{(18)}       \\
\hline
 J00418+4021       	&	3.59E-16	&	7.7E-15	&	23.2	&	1247	&	87	&	96	&	103	&	90	&	-0.002	&	0.336	&&	1.32	&&	0.8344	&	1257	&&	0.166	&	3412	&	-2928	&	0.594	\\
 J00457+0410       	&	7.03E-16	&	4.6E-14	&	67.6	&	4405	&	-668	&	-452	&	-162	&	-53	&	-0.175	&	0.355	&&	1.18	&&	0.6664	&	3413	&&	0.334	&	5099	&	-1332	&	0.898	\\
 J00535+1241       	&	5.77E-15	&	2.3E-13	&	41.0	&	1412	&	-83	&	-61	&	-60	&	-59	&	-0.019	&	0.327	&&	1.83	&&	0.9354	&	1408	&&	0.065	&	1629	&	-2676	&	1.010	\\
 J10255+5140       	&	1.44E-15	&	9.1E-14	&	60.1	&	1480	&	48	&	41	&	41	&	41	&	0.006	&	0.339	&&	0.78	&&	0.9970	&	1463	&&	0.003	&	3634	&	-533	&	0.605	\\
 J11292-0424       	&	1.56E-15	&	1.7E-13	&	115.9	&	2455	&	-55	&	-76	&	-53	&	-60	&	0.002	&	0.333	&&	0.95	&&	1.0000	&	2454	&&	0.000	&	\ldots	&	\ldots	&	\ldots	\\
 J11420+6030       	&	2.75E-16	&	1.0E-14	&	41.7	&	1542	&	-625	&	-73	&	-78	&	-75	&	-0.293	&	0.243	&&	2.12	&&	0.9290	&	1522	&&	0.071	&	4370	&	-2934	&	1.013	\\
 J12075+1506       	&	1.20E-15	&	4.7E-14	&	66.4	&	2253	&	-11	&	98	&	112	&	95	&	-0.051	&	0.313	&&	1.08	&&	0.9515	&	2238	&&	0.049	&	3775	&	-2598	&	0.393	\\
 J12091+5611       	&	1.11E-16	&	2.7E-15	&	22.9	&	1186	&	-319	&	-265	&	-258	&	-266	&	-0.050	&	0.314	&&	2.09	&&	0.8363	&	1141	&&	0.164	&	5069	&	-2095	&	1.010	\\
 J12366+5630       	&	1.43E-16	&	3.5E-15	&	25.8	&	1973	&	-529	&	-289	&	-227	&	-201	&	-0.213	&	0.288	&&	2.10	&&	0.7662	&	1753	&&	0.234	&	6200	&	-1158	&	0.392	\\
 J12562+5652       	&	1.12E-14	&	4.3E-13	&	34.7	&	4480	&	-545	&	-321	&	-14	&	72	&	-0.165	&	0.316	&&	1.93	&&	0.8460	&	3599	&&	0.154	&	3947	&	-2393	&	1.010	\\
 J13012+5902       	&	1.91E-14	&	7.1E-13	&	56.8	&	3313	&	-601	&	-370	&	-89	&	-6	&	-0.215	&	0.319	&&	1.37	&&	0.7323	&	2523	&&	0.268	&	3712	&	-1551	&	1.000	\\
 J14052+2555       	&	2.26E-14	&	2.3E-12	&	105.7	&	2054	&	-94	&	-94	&	-94	&	-94	&	0.000	&	0.333	&&	1.02	&&	1.0000	&	2053	&&	0.000	&	\ldots	&	\ldots	&	\ldots	\\
 J14063+2223       	&	5.80E-15	&	4.1E-13	&	58.9	&	1633	&	-132	&	-43	&	-24	&	-19	&	-0.077	&	0.301	&&	1.10	&&	0.9053	&	1528	&&	0.095	&	3522	&	-1800	&	1.000	\\
 J14170+4456       	&	1.02E-14	&	6.1E-13	&	66.0	&	2270	&	-448	&	-174	&	-59	&	3	&	-0.220	&	0.314	&&	1.17	&&	0.7320	&	1929	&&	0.268	&	4098	&	-721	&	0.586	\\
 J14258+3946       	&	1.07E-16	&	5.5E-15	&	49.4	&	1977	&	-148	&	13	&	2	&	-4	&	-0.093	&	0.368	&&	1.10	&&	1.0000	&	1975	&&	0.000	&	\ldots	&	\ldots	&	\ldots	\\
 J14475+3455       	&	1.77E-15	&	3.7E-14	&	26.9	&	3862	&	-953	&	-863	&	-168	&	-122	&	-0.291	&	0.260	&&	1.97	&&	0.9981	&	2298	&&	0.002	&	2572	&	-2957	&	0.459	\\
 J15455+4846       	&	4.87E-16	&	3.5E-14	&	73.4	&	1794	&	-77	&	10	&	27	&	10	&	-0.052	&	0.310	&&	1.06	&&	0.8097	&	1723	&&	0.190	&	6858	&	-3164	&	1.051	\\
 J17014+5149       	&	2.62E-15	&	1.3E-13	&	46.9	&	2582	&	-329	&	-78	&	-71	&	-69	&	-0.105	&	0.296	&&	1.19	&&	0.9436	&	2528	&&	0.056	&	4497	&	-2564	&	0.335	\\
 \hline
\end{tabular}
\begin{tablenotes}
   \item[*] {\bf Notes.} 
All measured quantities refer to rest-frame. Col. 1: Object coordinate name.  Col. 2: specific flux at 5100 \AA\ in units of erg\, s$^{-1}$ cm$^{-2}$ \AA$^{-1}$. Cols. 3:  flux in units of   erg\, s$^{-1}$\,  cm$^{-2}$.  Col. 4:   equivalent width in  \AA. Cols. 5:  FWHM of the \hb\ full broad profile in \kms; Cols. 6-9: profile centroids at fractional peak intensity in \kms. Col. 10: asymmetry index; Col. 11: kurtosis index. See \citealt{zamfir_detailed_2010} for definition. Col. 12: ratio \rfe, see text for definition. Col. 13: fractional intensity of the \hb\ BC; Col. 14: FWHM \hb\ BC in \kms; Col. 15: fractional intensity of the blueshifted broad component BLUE; Col. 16: FWHM in \kms; Col. 17: peak shift (mode) of BLUE in \kms. Col. 18: skew of BLUE. 
\end{tablenotes}
\end{sidewaystable*}

\begin{sidewaystable*}
\hspace{-1cm}
\caption{Profile analysis for [OIII]$\lambda$5007\AA \label{tab:measoiii}} 
\scalebox{0.98}{
\setlength{\tabcolsep}{2.5pt}
\begin{tabular}{l c c c r r r r c c c c r r c c c c c c} 
\hline\hline 

\multicolumn{1}{c}{Object} & \multicolumn{9}{c}{Full   profile  (NC + SBC)} && \multicolumn{3}{c}{[O{\sc iii}] NC } && \multicolumn{5}{c}{[O{\sc iii}] SBC} \\ 
\cline{2-10} \cline{12-14} \cline{16-20}
& Flux  & \multicolumn{1}{c}{EW} &  \multicolumn{1}{c}{FWHM} & \multicolumn{1}{c}{c($\frac{1}{4}$)} & \multicolumn{1}{c}{c($\frac{1}{2}$)} & \multicolumn{1}{c}{c($\frac{3}{4}$)} & \multicolumn{1}{c}{c($\frac{9}{10}$)} & AI & KI && F/F([O{\sc iii}]) & \multicolumn{1}{c}{FWHM} & \multicolumn{1}{c}{Shift} && \multicolumn{1}{c}{F/F([O{\sc iii}])} & \multicolumn{1}{c}{FWHM} &   \multicolumn{1}{c}{c($\frac{1}{2}$)} &  \multicolumn{1}{c}{Shift} & \multicolumn{1}{c}{Skew}\\
\multicolumn{1}{c}{(1)} & \multicolumn{1}{c}{(2)} & \multicolumn{1}{c}{(3)} & \multicolumn{1}{c}{(4)} & \multicolumn{1}{c}{(5)} & \multicolumn{1}{c}{(6)} & \multicolumn{1}{c}{(7)} & \multicolumn{1}{c}{(8)} & \multicolumn{1}{c}{(9)} & \multicolumn{1}{c}{(10)} && \multicolumn{1}{c}{(11)} & \multicolumn{1}{c}{(12)} & \multicolumn{1}{c}{(13)} && \multicolumn{1}{c}{(14)} & \multicolumn{1}{c}{(15)} & \multicolumn{1}{c}{(16)} & \multicolumn{1}{c}{(17)} & \multicolumn{1}{c}{(18)}\\
 \hline
 J00418+4021       	&	2.31E-15	&	7.00	&	1094	&	12	&	52	&	53	&	44	&	-0.040	&	0.446	&&	0.89	&	1051	&	65	&&	0.11	&	1641	&	-868	&	-875	&	0.97		\\
 J00457+0410       	&	0.00E+00	&	0.00	&	\ldots	&	\ldots	&	\ldots	&	\ldots	&	\ldots	&	\ldots	&	\ldots	&&	\ldots	&	\ldots	&	\ldots	&&	\ldots	&	\ldots	&	\ldots	&	\ldots	&	\ldots		\\
 J00535+1241       	&	1.01E-13	&	18.36	&	1095	&	-1359	&	-823	&	-763	&	-750	&	-0.475	&	0.251	&&	0.44	&	849	&	-718	&&	0.56	&	2158	&	-1766	&	-1819	&	1.00		\\
 J10255+5140       	&	9.45E-15	&	6.62	&	419	&	-24	&	-30	&	-20	&	-29	&	0.017	&	0.508	&&	1.00	&	402	&	-30	&&	0.00	&	\ldots	&	\ldots	&	\ldots	&	\ldots		\\
 J11292-0424       	&	2.88E-14	&	20.50	&	501	&	-41	&	-31	&	-30	&	-30	&	-0.032	&	0.439	&&	0.95	&	486	&	-20	&&	0.05	&	445	&	-644	&	-669	&	0.97		\\
 J11420+6030       	&	2.02E-16	&	0.82	&	\ldots	&	\ldots	&	\ldots	&	\ldots	&	\ldots	&	\ldots	&	\ldots	&&	1.00	&	206	&	303	&&	0.00	&	\ldots	&	\ldots	&	\ldots	&	\ldots		\\
 J12075+1506       	&	5.98E-15	&	8.89	&	1556	&	-979	&	-778	&	-389	&	-382	&	-0.537	&	0.242	&&	0.32	&	617	&	-254	&&	0.68	&	1609	&	-1176	&	-1163	&	1.02		\\
 J12091+5611       	&	4.34E-16	&	3.94	&	321	&	-62	&	-62	&	-62	&	-60	&	-0.010	&	0.480	&&	0.96	&	307	&	10	&&	0.04	&	2335	&	-268	&	-224	&	1.01		\\
 J12366+5630       	&	2.74E-15	&	21.89	&	328	&	-104	&	-44	&	-35	&	-43	&	-0.218	&	0.334	&&	0.56	&	282	&	-21	&&	0.44	&	750	&	-322	&	-286	&	1.01		\\
 J12562+5652       	&	1.49E-14	&	1.21	&	1358	&	-2272	&	-2135	&	-2047	&	-1960	&	-0.324	&	0.449	&&	0.00	&	\ldots	&	\ldots	&&	1.00	&	1959	&	-2135	&	-1881	&	0.40		\\
 J13012+5902       	&	3.26E-14	&	2.85	&	1220	&	-115	&	-86	&	-75	&	-103	&	-0.011	&	0.364	&&	0.30	&	1005	&	-86	&&	0.70	&	2916	&	1823	&	-1262	&	5.25		\\
 J14052+2555       	&	1.36E-13	&	6.31	&	1243	&	-185	&	-201	&	-282	&	-276	&	0.094	&	0.295	&&	0.72	&	1500	&	-336	&&	0.28	&	3000	&	-3043	&	-2376	&	0.10		\\
 J14063+2223       	&	5.96E-14	&	9.20	&	803	&	-358	&	-261	&	-238	&	-231	&	-0.181	&	0.399	&&	0.56	&	675	&	-215	&&	0.44	&	495	&	-426	&	-912	&	4.85		\\
 J14170+4456       	&	1.58E-14	&	1.61	&	494	&	-601	&	-604	&	-604	&	-609	&	0.021	&	0.413	&&	1.00	&	494	&	-607	&&	0.000	&	\ldots	&	\ldots	&	\ldots	&	\ldots		\\
 J14258+3946       	&	8.91E-16	&	8.47	&	681	&	-54	&	-41	&	-39	&	-46	&	-0.017	&	0.451	&&	0.79	&	683	&	30	&&	0.21	&	3000	&	-2541	&	-1824	&	0.10		\\
 J14475+3455       	&	1.62E-14	&	9.64	&	2472	&	-2454	&	-2416	&	-2363	&	-2335	&	-0.069	&	0.465	&&	0.02	&	372	&	819	&&	0.98	&	2718	&	-2416	&	-2264	&	0.83		\\
 J15455+4846       	&	6.63E-15	&	14.77	&	3006	&	-2036	&	-2055	&	-2036	&	-2055	&	0.009	&	0.465	&&	0.00	&	730	&	778	&&	1.00	&	3000	&	-2055	&	-2088	&	1.02		\\
 J17014+5149       	&	1.29E-14	&	5.08	&	5209	&	-1606	&	-1407	&	-1227	&	-1100	&	-0.169	&	0.747	&&	0.31	&	916	&	778	&&	0.69	&	4820	&	-3054	&	-2529	&	0.28		\\
\hline
\end{tabular}
}
\vspace{0.05truecm}

{\raggedright \textbf{Notes}.\textit{}
All measured quantities refer to rest-frame. 
Col. 1: Object coordinate name.  
Cols. 2: \oiii\  flux in units of  ergs\, s$^{-1}$\,  cm$^{-2}$.  
Col. 3:   equivalent width in  \AA. 
Cols. 4:  FWHM of the \oiii\ full  profile in \kms; 
Cols. 5-8: profile centroids at  $\frac{1}{4}, \frac{1}{2}, \frac{3}{4}, 0.9$\ fractional peak intensity in \kms. 
Col. 9: asymmetry index; 
Col. 10: kurtosis index, as for \hb.  
Col. 11: fractional intensity of the \oiii\ NC; 
Col. 12: FWHM \oiii\ NC in \kms; 
Col. 13: peak shift of the \oiii\ NC in \kms.  
Col. 14: fractional intensity of the blueshifted semi-broad component SBC; 
Col. 15: FWHM of SBC in \kms; 
Col. 16: SBC centroid at  half maximum in \kms; 
Col. 17: SBC peak shift (mode)  in \kms. 
Col. 18: SBC skew.     \par} 
\label{tab:valueso3}
\end{sidewaystable*}

\begin{table} 
    \centering\tabcolsep=1pt
    \caption{Mass outflow rates and kinetic powers}
    \begin{tabular}{lccccc}\hline\hline
Object    & \multicolumn{2}{c}{\hb} &&  \multicolumn{2}{c}{\oiii}  \\ \cline{2-3}\cline{5-6}
& $\dot{M}$ & $\dot{\epsilon}$ &&   $\dot{M}$ & $\dot{\epsilon}$\\
  & [M$_\odot$ yr$^{-1}$] & [erg s$^{-1}$]  && [M$_\odot$ yr$^{-1}$] & [erg s$^{-1}$]\\
\hline    
J00418+4021   &  0.0251      &  1.16E+41    &&  0.001  &  2.11E+39  \\        
J00457+0410   &  3.3547      &  7.05E+42    &&   \ldots  &    \ldots  \\           
J00535+1241   &  0.1866      &  7.87E+41    &&  0.233  &  1.31E+42  \\    
J10255+5140   &  0.0004      &  3.32E+38    &&   \ldots  &   \ldots \\    
J11292-0424   &  0.0000      &  0.00E+00    &&  0.003  &  5.39E+39  \\    
J11420+6030   &  1.9106      &  8.84E+42    &&   \ldots  &   \ldots \\    
J12075+1506   &  2.9473      &  1.21E+43    &&  1.194  &  4.49E+42  \\    
J12091+5611   &  0.1955      &  6.46E+41    &&  0.0005  &  3.88E+38  \\    
J12366+5630   &  0.5111      &  9.33E+41    &&  0.087  &  8.94E+40  \\    
J12562+5652   &  0.3743      &  1.41E+42    &&  0.038   &  2.60E+41  \\    
J13012+5902   &  69.596      &  1.70E+44    &&  4.924   &  2.87E+43  \\    
J14052+2555   &  0.0000      &  0.00E+00    &&  1.947  &  1.90E+43  \\    
J14063+2223   &  0.8687      &  2.47E+42    &&  0.070   &  9.60E+40  \\    
J14170+4456   &  1.8542      &  2.11E+42    &&   \ldots  &   \ldots \\    
J14258+3946   &  0.0000      &  0.00E+00    &&  0.063   &  5.14E+41  \\    
J14475+3455   &  9.7100      &  4.53E+43    &&  7.902  &  6.11E+43  \\    
J15455+4846   &  3.6547      &  1.82E+43    &&  1.196  &  7.86E+42  \\    
J17014+5149   &  1.8024      &  7.29E+42    &&  1.34  &  1.31E+43  \\    
\hline
    \end{tabular}
    \label{tab:outkin}
\end{table}

\begin{table}
\centering\tabcolsep=2pt
\caption{Accretion parameters} 
\begin{tabular}{l c c c c } 
\hline
\multicolumn{1}{c}{Object} &  \multicolumn{1}{c}{$\log \lambda L_\lambda$(5100\AA) } & \multicolumn{1}{c}{$\log L_\mathrm{bol}$}   
 & \multicolumn{1}{c}{$\log$\mbh} & \multicolumn{1}{c}{$\log L_\mathrm{bol}/L_\mathrm{Edd}$}  \\ 
 \hline
 J00418+4021       	&	43.33	&	44.67	&	7.03	&	-0.54	\\
 J00457+0410       	&	44.99	&	45.99	&	8.19	&	-0.37	\\
 J00535+1241       	&	44.38	&	45.51	&	7.37	&	-0.03	\\
 J10255+5140       	&	43.53	&	44.82	&	7.36	&	-0.72	\\
 J11292-0424       	&	43.84	&	45.07	&	7.63	&	-0.73	\\
 J11420+6030       	&	45.04	&	46.03	&	7.59	&	0.27	\\
 J12075+1506       	&	45.71	&	46.57	&	8.39	&	0.00	\\
 J12091+5611       	&	44.31	&	45.45	&	7.17	&	0.11	\\
 J12366+5630       	&	44.74	&	45.79	&	7.51	&	0.10	\\
 J12562+5652       	&	44.36	&	45.49	&	7.66	&	-0.35	\\
 J13012+5902       	&	46.59	&	47.27	&	8.73	&	0.36	\\
 J14052+2555       	&	45.81	&	46.65	&	8.43	&	0.04	\\
 J14063+2223       	&	44.79	&	45.83	&	7.83	&	-0.18	\\
 J14170+4456       	&	45.17	&	46.14	&	8.07	&	-0.10	\\
 J14258+3946       	&	44.38	&	45.50	&	7.74	&	-0.42	\\
 J14475+3455       	&	45.80	&	46.64	&	8.13	&	0.33	\\
 J15455+4846       	&	44.86	&	45.89	&	7.92	&	-0.21	\\
 J17014+5149       	&	45.34	&	46.27	&	8.23	&	-0.14	\\ \hline
\end{tabular}
\tablefoot{Col. 1: Object coordinate name. 
Col. 2: decimal logarithm of luminosity at 5100 \AA\  in units  erg\ s$^{-1}$. Col. 3: bolometric luminosity  in units  erg\ s$^{-1}$, from $\log \lambda L_\lambda$(5100\AA) applying the bolometric correction by \citet{netzer_bolometric_2019}.  Col. 4 is the logarithm of black hole mass in units of M$_\odot$. Col. 5: Eddington ratio.} 
\label{tab:valuesmbh}
\end{table}

\begin{table}
\begin{center}
\caption{Statistical and systematic linear fractional uncertainties for the accretion parameters}
\label{table_uncertainties}
\begin{tabular}{lcccccc}\hline
Parameter    & Statistical & Systematic & Total  \\ \hline
FWHM	& 0.100&	\ldots&	0.100  \\ 
$\lambda L_\lambda$(5100\AA)	& 0.200&	\ldots&	0.200	 \\ 
\mbh\	    & 0.224&	0.602&	0.642	 \\
BC	    &\ldots&	 0.346&	0.346	 \\
$L$	&0.200	&0.346	   &0.399	\\
\lledd\	& 0.224 &	0.694 &	0.729 \\ \hline
\end{tabular}
\end{center}
\end{table}

\end{appendix}

\end{document}